\documentclass[aps,pre,showpacs,twocolumn,superscriptaddress,longbibliography]{revtex4-2}
\usepackage[pdftex]{graphicx}
\usepackage[pdftex,bookmarks,colorlinks,breaklinks]{hyperref} 
\usepackage{amsmath,amssymb}
\hypersetup{linkcolor=red,citecolor=blue,filecolor=blue,urlcolor=blue} 
\hypersetup{pdftoolbar=true,pdfpagemode=UseNone,pdfstartview=FitH,colorlinks=true,extension=}

\providecommand*{\dos}{\ensuremath{\rho}}
\providecommand*{\intdos}{\ensuremath{I}}
\providecommand*{\corr}{\ensuremath{\dos_{+1,-1}}}
\providecommand*{\ncorr}{\ensuremath{\intdos_{+1,-1}}}
\providecommand*{\pf}{\ensuremath{p_{\mathrm{f}}}}
\providecommand*{\pr}{\ensuremath{p_{\mathrm{r}}}}
\providecommand*{\nr}{\ensuremath{n_{\mathrm{r}}}}

\begin{document}
\graphicspath{{./},{./figs/},{./figs/coupling/}}

\title[Microwave studies of the three chiral ensembles]{
  Microwave studies of the three chiral ensembles\\in chains of coupled
  dielectric resonators}
\author{M.~Richter}
\email{martin.richter@nottingham.ac.uk}
\affiliation{School of Mathematical Sciences, University of Nottingham, Nottingham NG7 2RD, United Kingdom}
\affiliation{School of Science and Technology, Nottingham Trent University, Nottingham NG11 8NS, United Kingdom}
\affiliation{Universit\'{e} C\^{o}te d'Azur, CNRS, Institut de Physique de Nice (INPHYNI), 06108 Nice, France, EU}
\author{A.~Rehemanjiang}
\affiliation{Fachbereich Physik der Philipps-Universit\"at Marburg, D-35032 Marburg, Germany}
\author{U.~Kuhl}
\email{ulrich.kuhl@unice.fr}
\affiliation{Universit\'{e} C\^{o}te d'Azur, CNRS, Institut de Physique de Nice (INPHYNI), 06108 Nice, France, EU}
\author{H.-J.~St\"ockmann}
\email{stoeckmann@physik.uni-marburg.de}
\affiliation{Fachbereich Physik der Philipps-Universit\"at Marburg, D-35032 Marburg, Germany}

\date{\today}
\begin{abstract}
  Random matrix theory has proven very successful in the understanding of the spectra of chaotic systems. 
  Depending on symmetry with respect to time reversal and the presence or absence of a spin 1/2 there are three ensembles, the Gaussian orthogonal (GOE), Gaussian unitary (GUE), and Gaussian symplectic (GSE) one. 
  With a further particle-antiparticle symmetry the chiral variants of these ensembles, the chiral orthogonal, unitary, and symplectic ensembles (the BDI, AIII, and CII in Cartan's notation) appear which are the main point of interest in this paper. 
  Following a recently published work on chiral random matrix ensembles and their experimental realizations (Rehemanjiang et al., PRL 124, 116801 (2020)), this is achieved by using dielectric cylinders placed between two parallel aluminium plates. 
  These cylinders act as microwave resonators which are used to create tight-binding chains of finite length up to $N=5$. 
  The different ensembles are achieved by using different types of couplings: 
  For the orthogonal case spatial proximity is used, for the unitary case microwave circulators are used, and for the symplectic case a combination of circulators and cables is used to create the necessary symmetry. 
  In all cases the predicted repulsion behavior between positive and negative eigenvalues for energies close to zero are verified by a comparison with theory taking the finite size of the systems into account. 
  We will show that the difference to the expected universal behavior is given by logarithmic corrections only. 
  These corrections stem from the Hamiltonians having zero entries in their off-diagonal blocks. 
\end{abstract}

\maketitle

\section{Introduction}

Symmetries and universality classes belong to the most fundamental building blocks of modern physics. 
Universality can be seen whenever complex systems are investigated. 
These are systems whose simple constituents, when put together, give rise to a much richer behavior than the simple building blocks would admit. 
This very general observation leads to the applicability of the same methods to a wide range of problems ranging from number theory~\cite{kea00} over quantum physics~\cite{stoe07b,haa01b}, electronic transport problems~\cite{fei09} and WiFi-telecommunications~\cite{tul04,cou11} to financial mathematics~\cite{sch10b,mun10} and delay in bus and train times~\cite{krb02,jag17}. 
One of the first and therefore most prominent examples were early experiments on nuclear scattering where the statistical properties of measured resonances of compound nuclei could be described using random matrices~\cite{dys62c,sel86a,mit10}. 
These comparisons show that symmetries of the systems under consideration play an important role for the universality class of the system. 
Early on, the presence or absence of (generalized) time-reversal symmetries were taken into account and gave rise to the so called three-fold way~\cite{dys62c} of \textit{Gaussian Orthogonal Ensemble}, \textit{Gaussian Unitary Ensemble}, and \textit{Gaussian Symplectic  Ensemble}. 
If these symmetries are taken into account correctly, then the universal features of the spectra can be reproduced to great detail by the corresponding random matrix ensembles~\cite{cas80,boh84c}. 
This correspondence between the spectral properties of complex quantum systems and random matrix theory has been proven on the basis of semiclassical arguments~\cite{muel04}.

In random matrix theory the Hamiltonian of a system is substituted by a matrix $H$ with randomly, usually Gaussian, distributed matrix elements $H_{nm}$ subject to constraints depending on symmetry. 
An immediate constraint results from the hermiticity of the Hamiltonian, $H_{mn}=H_{nm}^*$. 

Most important in this context is time-reversal symmetry. 
Denoting the time-reversal operator by $T$, time-reversal symmetry of the Hamiltonian means
\begin{equation} \label{intro01}
    HT=TH \,. 
\end{equation}
For systems without spin 1/2 the time-reversal operator is just given by the complex-conjugate, $T={\cal K}$, whence follows $T^2=1$. 
For systems with spin 1/2 the time-reversal operator has to be replaced by $T={\cal K}\tau_y$, where $\tau_{y}$ is a quaternion, with $T^2=-1$, see Eq.~\eqref{eq:definition_tau_y} later. 

The three options ($T^2=1$, no $T$, $T^2=-1$) give rise to the three classical random matrix ensembles, the orthogonal, the unitary and the symplectic one, respectively.

A new symmetry appears in the context of particles and anti-particles.
Let $C$ be the charge-conjugation operator. 
This operator is anti-unitary and the symmetry with respect to an exchange of particles and anti-particles means 
\begin{equation} \label{intro02}
  HC=-CH\,,
\end{equation}
where again there are the three possibilities: $C^2=1$ and $C^2=-1$, and no $C$. 
It follows that, if $E_n$ is an eigenvalue of $H$, then the same is true for $-E_n$, including the possibility of eigenvalues at $E=0$. 
If both $C$ and $T$ exist and obey Eqs~(\ref{intro01}) and~(\ref{intro02}), respectively, then
\begin{equation} \label{eq:chiral_sym}
  HCT + CTH = 0
\end{equation}
follows automatically. 
Note that $CT$ in this equation is unitary as both $C$ an $T$ are anti-unitary. 

All possible combinations of ($T^2=1$, no $T$, $T^2=-1$) and ($C^2=1$, no $C$, $C^2=-1$) yield a total of nine different random matrix ensembles. 
Together with the last remaining option no $T$, no $C$, but $CT$, one finally ends up with the {\em ten-fold way} \cite{zir96}. 

From the very beginning there had been the interest in experimental demonstrations of the predictions of random matrix theory, where the distribution of spacings of adjacent levels played a dominating role. 
For the Gaussian orthogonal ensemble there is an abundant number of realizations, including systems not quantum-mechanical in origin (for example Refs.~\cite{boh91b,rey14} for vibro-acoustics, but see also Sec.~3.2 of Ref.~\cite{stoe99}). 
For the Gaussian unitary ensemble the number of realizations is still small, the first of them being performed in microwave billiard systems \cite{so95,sto95b} and later in microwave networks \cite{law10}. 
The Gaussian symplectic ensemble usually is associated with spin 1/2 systems, but Joyner et al.~\cite{joy14} proposed an alternative way to realize the necessary condition $T^2=-1$ in a graph with a peculiarly designed geometry. 
The proposal was experimentally realized in our group in a microwave network \cite{reh16,reh18}. 

For the new ensembles systematic experimental studies still seem to be missing, though there are a lot of studies of systems showing chiral symmetry (see Ref.~\cite{bee15} for a review). 
It is the aim of the present work to provide more details of our recent microwave study~\cite{reh20} and several figures or sub-figures are taken from it. 
At its center stand the three chiral relatives of the classical ensembles, the chiral orthogonal (chiOE), the chiral unitary (chiUE), and the chiral symplectic (chiSE) ensemble. 
More specifically, these are the three ensembles where $H\cdot CT = CT\cdot H$ and the corresponding presence or absence of $T$ and $C$, respectively, see Table~\ref{tab:chiral_ensemble}. 

\begin{table}[tb]
  \begin{center}
    \begin{tabular}{c|c|c|c|}
      & \,chiOE\, & \,chiUE\, & \,chiSE\, \\\hline
      Cartan notation & BDI & AIII & CII \\\hline
      $\alpha$ & 0 & 1 & 3 \\
      $\beta$ & 1 & 2 & 4 \\
      $T^2$ & +1 & $\times$ & -1\\
      $C^2$ & +1 & $\times$ & -1\\
    \end{tabular}
    \caption{\label{tab:chiral_ensemble} 
      Parameters $\alpha$, $\beta$ for the chiral orthogonal, unitary, and symplectic ensembles, respectively. 
      In the second line the Cartan notations for the three ensembles are given. 
      The last rows shows the values of the squared operators $T^2$ and $C^2$. 
      The chiral UE has no such operators which commute or anti-commute with $H$ (indicated by '$\times$'). 
      However, all ensembles fulfill $H\cdot CT = CT\cdot H$. 
    }
  \end{center}
\end{table}

The studies concentrate on the eigenvalues close to $E=0$. 
It is here where the positive eigenvalues feel the existence of their negative partners, resulting in a possible repulsion of eigenvalues. 
The ensembles studied by us are not Gaussian. 
This leads to a specific, non-universal (i.e.~$N$-dependent) behavior of the densities of state and a modification of the expected repulsion behavior as compared to random matrix theory (RMT).

\section{The chiral relatives of the classical ensembles}

The first extension of the classical ensembles had been studied by Verbarschoot and coworkers, who looked for the consequences of particle-antiparticle symmetry in the context of the Dirac equation \cite{ver93}. 
Here the proximity of the electron and the positron states close to $E=0$ generates an oscillatory modulation in the ensemble averaged density of states. 

For the observation of such a symmetry the existence of particles and anti-particles is not really needed. 
Sufficient is a system consisting of two subsystems I and II with interactions only between members of I an II, but no internal interactions within I or II. 
The Hamiltonian for such a situation is given by
\begin{equation} \label{eq:chiham}
    H=\left(\begin{array}{cc}
        0 & A \\
        A^\dag & 0
      \end{array}\right)\,,
\end{equation}
where the off-diagonal blocks contain the interaction between the subsystems. 
Just like in the traditional counter parts GOE, GUE, and GSE, the entries of $A$ are given by real, complex, and quaternion real numbers for the three ensembles chiOE, chiUE, and chiSE, respectively,
\begin{eqnarray*}
  a_{nm}&=&\left\{
            \begin{array}{ll}
              \rm{real} & {\rm chiOE} \\
              {\rm complex} & {\rm chiUE} \\
              {\rm quaternion\hspace{0.5em}real}\quad & {\rm chiSE}.
            \end{array}\right. 
\end{eqnarray*}

For all the ensembles, the chiral symmetry is immediately evident from the structure of $H$ by realizing that the unitary operator
\begin{eqnarray}
  J &=& \left(
        \begin{array}{cc}
          {\bf 1}_n & 0 \\
          0 & -{\bf 1}_m \\
        \end{array}
  \right)
\end{eqnarray}
anti-commutes with $H$,
\begin{equation} \label{eq:structural_chiral_sym}
  J H=- H J\,,
\end{equation}
where ${\bf 1}_n$ is the unit matrix of rank $n$.

For sake of completeness we want to give explicit expressions for $T$ and $C$ for all three chiral ensembles: 
For both chiOE and chiUE we can set $T = {\cal K}$ and $C = {\cal K}J$ such that $TC=J$ in both cases and Eq.~\eqref{eq:chiral_sym} holds via Eq.~\eqref{eq:structural_chiral_sym}. Furthermore, we get the necessary relations~\eqref{intro01} and~\eqref{intro02} for the chiOE and $T^2 = C^2 = J^2 = +{\bf 1}_{n+m}$. 
The explicit forms of $T$ and $C$ for the chiSE follow from writing the quaternionic elements of $A$ in terms of \(2\times 2\) blocks of the form
\begin{align} \label{eq:quaternions_via_complex_nums}
  a_{nm}
  \nonumber
         &= {(a_{nm})}_0 \cdot 1\,\, + {(a_{nm})}_x\cdot\tau_x + \\
         &\hphantom{=}\,\,\,{(a_{nm})}_y\cdot\tau_y + {(a_{nm})}_z\cdot\tau_z
  \\
  &= \left(
             \begin{array}{cc}
               {a'}_{nm} & {a''}_{nm} \\
               -{a''}_{nm}^* & {a'}_{nm}^*
             \end{array}
  \right)
\end{align}
with complex numbers ${a'}_{nm}$ and ${a''}_{nm}$. 
With this we can define
\begin{eqnarray}
  J_{\pm} &=& \left(
        \begin{array}{cc}
          {\bf \tilde{1}}_n & 0 \\
          0 & \pm{\bf \tilde{1}}_m \\
        \end{array}
  \right)\,,
\end{eqnarray}
where ${\bf \tilde{1}}_n$ is a $2n\times 2n$ matrix with $n$ $2\times 2$ matrices
\begin{equation} \label{eq:definition_tau_y}
  \tau_y = \left(
    \begin{array}{cc}
      0 & -1 \\ 1 & 0
    \end{array}\right)
\end{equation}
on its diagonal. 
From this we can set $T = KJ_{-}$ and $C = KJ_{+}$. This fulfills the chiral symmetry Eq.~\eqref{eq:chiral_sym} via the unitary matrix
\begin{equation}
  \label{eq:TCchiSE}
  TC = \left(
    \begin{array}{cc}
      -{\bf 1}_n & 0 \\
      0 & {\bf 1}_m \\
    \end{array}
  \right)
\end{equation}
and we have $T^2 = C^2 = -{\bf 1}_{2n + 2m}$ as needed. 

Assuming $n \geq m$, the characteristic polynomial of $H$ for all three chiral ensembles is given by 
\begin{eqnarray}
  \nonumber
    \chi(E)&=& \left|
               \begin{array}{cc}
                 E\cdot{\bf 1}_n & -A \\
                 -A^\dag & E\cdot{\bf 1}_m \\
               \end{array}
             \right|
  \\&=&
  \nonumber
             E^{-m\phantom{n}}\left|
               \begin{array}{cc}
                 E\cdot{\bf 1}_n & -A \\
                 0 & E^2\cdot{\bf 1}_m - A^\dag A \\
               \end{array}
             \right|
  \\&=&
        \label{eq:chi}
    E^{n-m}\left|E^2\cdot{\bf 1}_m-A^\dag A\right|\,,
\end{eqnarray}
where we multiplied the lowest $m$ rows with $E$ and added $A^\dagger\cdot 1^{\mathrm{st}}\,\mathrm{row}$ to them in the second equation. 
Equation~(\ref{eq:chi}) has a number of important consequences: 
\begin{itemize}
\item[(i)] For $n>m$ there are $\nu=n-m$ eigenvalues \mbox{$E_0^{(i)}=0$, ($i=1, \dots, \nu$)}.
They are {\em topologically protected}; i.e., they do not depend on the interaction between the two subsystems and can only be destroyed by lifting the chiral symmetry. 
In other words, they are protected as long as the Hamiltonian is of shape~\eqref{eq:chiham} for arbitrary sub-matrices $A$. 
\item[(ii)] All other eigenvalues appear in pairs $E_n$ and \mbox{$E_{-n}=-E_n$  ($n=1,2,\dots)$}. 
\item[(iii)] For energies far from $E=0$ the statistical properties of the chiral ensembles approach those of the classical ones \item[(iv)] The eigenvalues close to zero, $E_1, E_2, E_3, \dots$, feel the proximity of their partners $E_{-1}, E_{-2}, E_{-3}, \dots$. 
\end{itemize}
The last point results in a universal oscillatory modulation in the density of states $\dos (E)$ and a possible repulsion of the eigenvalues close to $E=0$ \cite{iva02},
\begin{equation} \label{eq:rep}
  \dos (E) \sim |E|^{\alpha+\nu\beta}\,,
\end{equation}
where $\alpha, \beta$ are given in Table~\ref{tab:chiral_ensemble}. 
Parameter $\beta$ is the universality index known already from the classical ensembles, and $\alpha=\beta -1$ for their chiral relatives. 
For reference, Table~\ref{tab:chiral_ensemble} shows the Cartan notations for the three ensembles, used as a mathematical classification scheme of the ten universal ensembles~\cite{bee15}.

The study of these ensembles, in particular the eigenvalue repulsion behavior close to $E=0$ is the objective of the present work. 
Everything addressed in the present section is well-known to experts working in the field, but we considered it appropriate to summarize the essential features of the chiral ensembles here to make the access easier to readers not familiar with them. 
For more details see the review by Beenakker~\cite{bee15} which served as the main basis for the present introduction.

\section{The set-up}

\begin{figure}
    \parbox{\linewidth}{
    \vspace*{-2ex}
    \hspace*{-.1\linewidth}
    \includegraphics[width=1.09\linewidth]{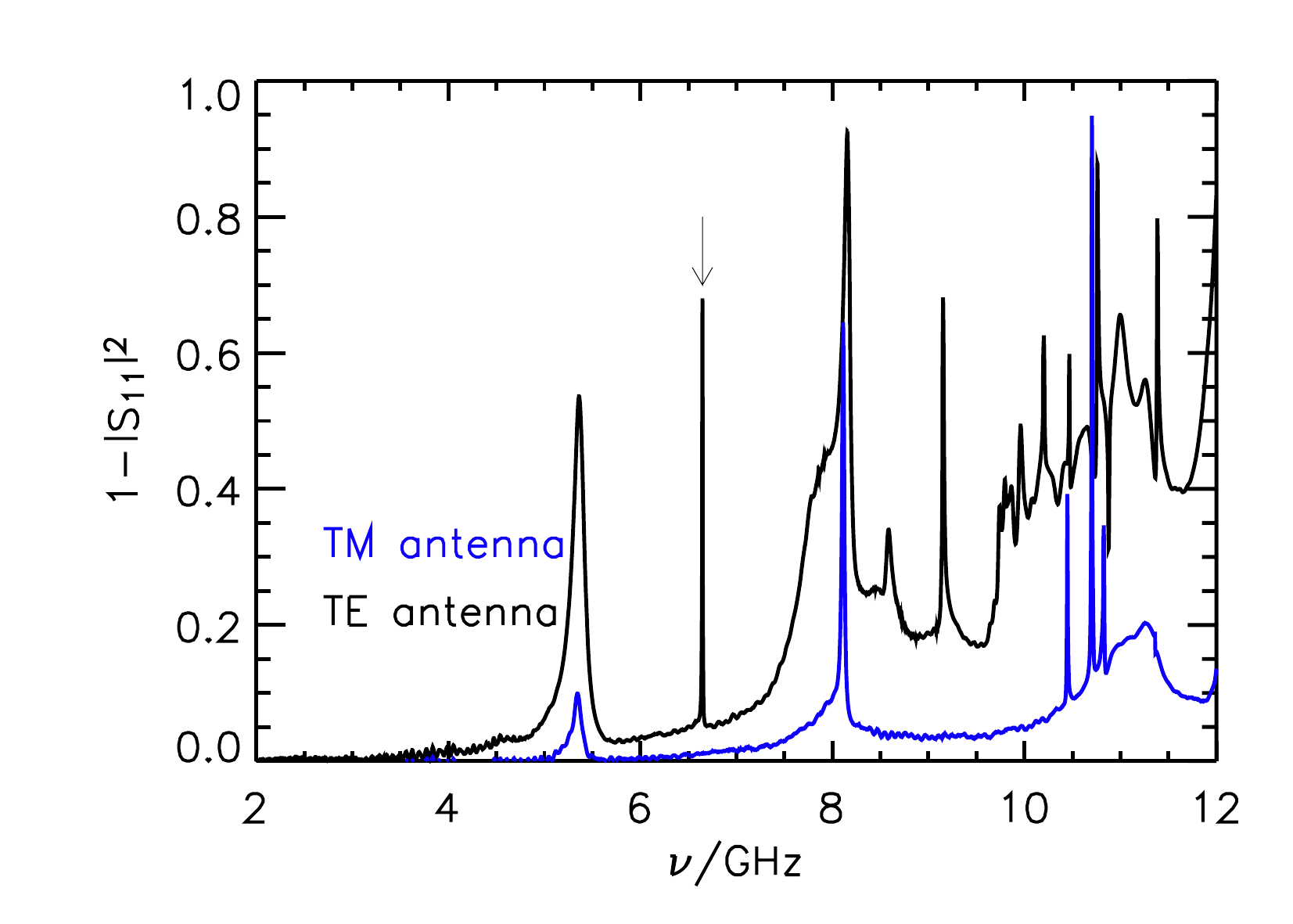}\\[0pt]
    \hspace*{-.43\linewidth}\raisebox{.50\linewidth}[0pt][0pt]{\includegraphics[width=0.22\linewidth]{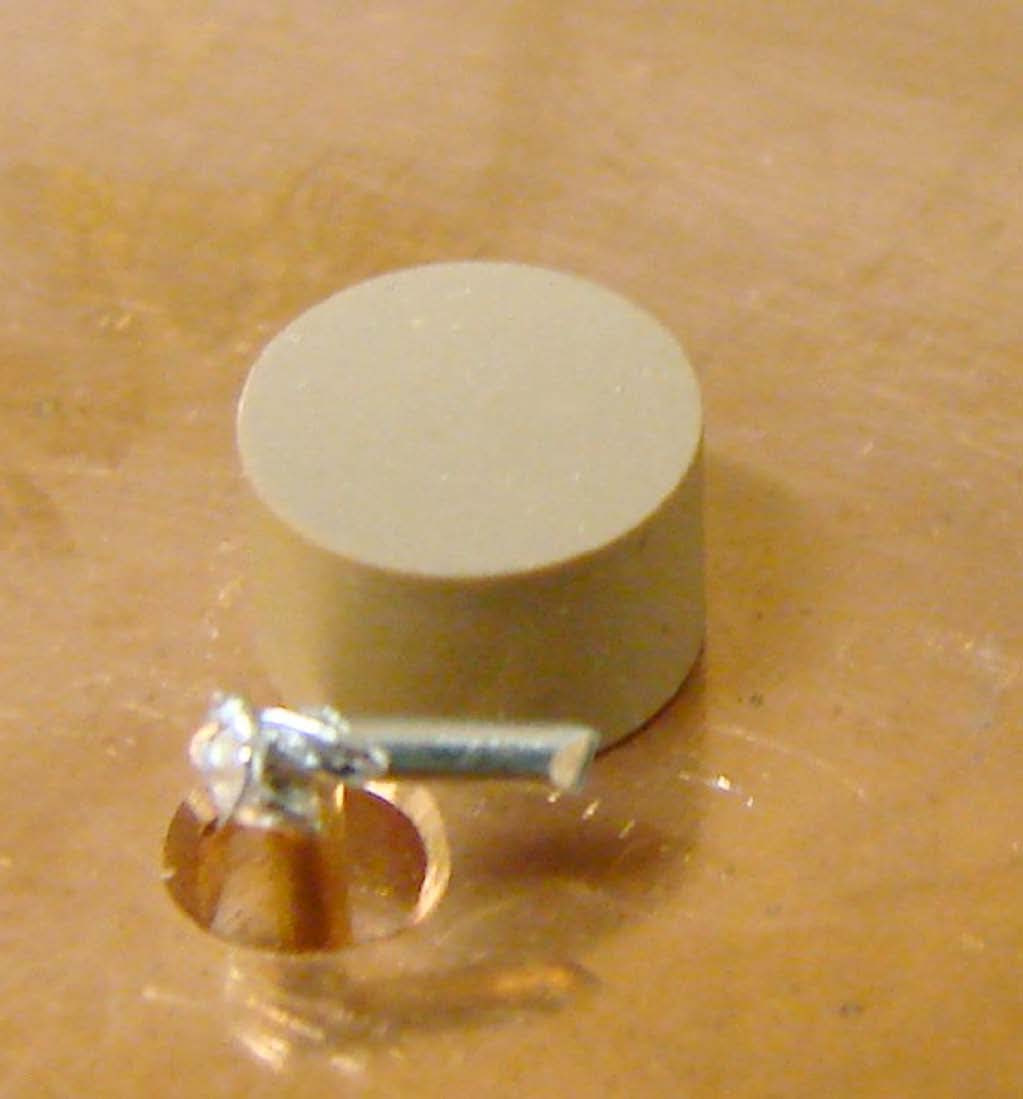}}\\[-5ex]
    }
    \caption{ \label{fig:TETM} 
      Reflection spectrum of a single disk measured with 
      (i) a monopole antenna exciting the TM resonances (blue), and 
      (ii) with a bent antenna (see insert) exciting the TE resonances as well (black).
      The arrow indicates the lowest TE mode, i.e., the one used in the experiment 
      (see Ref.~\cite{kuh10a} for further details). 
      The inset shows a resonator (height $5\,\mathrm{mm}$, diameter $7.6\,\mathrm{mm}$, index of refraction $n \approx 6$). 
  }
\end{figure}

The main building blocks for the experimental realization of the chiral ensembles are dielectric cylinders ($h = 5\mathrm{mm}$, $r = 3.8\mathrm{mm}$) with an index of reflection of $n\approx 6$ (see inset in Fig.~\ref{fig:TETM}). 
They are used to create a tight-binding system which is a very simple realization of the chiral symmetry above, Eq.~\eqref{eq:chiham}. 
The cylinders are placed between aluminium bottom and top plates, the latter one being removed for the photograph. 
There are two types of resonance modes, the transverse magnetic (TM) mode with the electric field $\vec{E}$ parallel and the magnetic field $\vec{B}$ perpendicular to the cylinder axis, and the transverse electric modes (TE) with the roles of $\vec{E}$ and $\vec{B}$ interchanged. 
For the experiments a vector network analyzer (Agilent 8720ES) was used allowing for a measurement of microwave reflection and transmission amplitudes. 
The TM modes can be excited by a monopole antenna parallel to the cylinder axis. 
Figure \ref{fig:TETM} shows a corresponding reflection spectrum showing a large number of resonances. 
To excite the TE modes a bent antenna has been used (see inset of Fig.~\ref{fig:TETM}) or a loop antenna on top of the cylinder. 
A reflection spectrum obtained with the bent antenna is shown in Fig.~\ref{fig:TETM} too. 
The TM resonances are seen again, since the bent antenna also contains a vertical part. 
But there are a number of additional resonances, the TE modes. 
For the experiments the lowest TE mode was used, marked by an arrow in the figure. 
It is very sharp compared to the TM resonances for the following reason: 
For a cylindrical geometry with top and bottom parallel to each other, the wave number $k$ may be decomposed as 
\begin{equation}
    k^2=k_\perp^2+k_\parallel^2\,,
\end{equation}
where $k_\parallel$ is the in-plane component, and $k_\perp$ is the perpendicular component. 
Component $k_\parallel$ governing the in-plane propagation is hence given by 
\begin{equation} \label{eq:k}
    k_\parallel=\sqrt{k^2-k_\perp^2} \,. 
\end{equation}
Within the disk the perpendicular component is given by $k_\perp=\pi/(nd)$ and outside by $k_\perp=\pi/(d)$, where $d$ is the height of the disk, and where it is assumed that top and bottom plate are in direct contact with the disk \cite{kuh10a}. 
Hence, there is a $k$ range
\begin{equation}\label{eq:krange}
  \frac{\pi}{nd} < k < \frac{\pi}{d}
\end{equation}
where the waves are standing within the cylinder, and evanescent outside. 
With $k=2\pi\nu/c$, where $c$ is the velocity of light and $\nu$ the microwave frequency, this corresponds to upper lower and upper frequency limits of $5$ and $30\,\mathrm{GHz}$. 
TE modes within this range, in particular the TE$_0$ mode, thus correspond to bound states, in contrast to the TM modes, which are just resonances. 
This is exactly what is needed for the realization of a tight-binding system and explains the small line width of this mode. 

In reality, the distance between the two plates is larger than the height of the cylinders, namely $11\,\mathrm{mm}$ as compared to $5\,\mathrm{mm}$. 
Therefore, there is a gap of $6\,\mathrm{mm}$ between the top plate and the top face of the cylinder allowing for an insertion of loop antennas used in part of the experiments. 
As a consequence, $n$ in Eq.~(\ref{eq:krange}) has to be replaced by an effective $n_{\rm{eff}}$. 

Bringing two cylinders together, the mutual overlap of the evanescent tails generates a coupling.  The situation can by phenomenologically described by a $2\times 2$ Hamiltonian
\begin{equation} \label{eq:ham}
  H = \left(\begin{array}{cc} \nu_0+\frac{\Delta\nu}{2} & a \\ a & \nu_0-\frac{\Delta\nu}{2}\end{array}\right)\,,
\end{equation}
where $a$ is the coupling constant between the two cylinders and $\nu_0\pm\Delta\nu/2$ are the TE$_0$ eigenfrequencies, different from each other due to fabrication tolerance and contact imperfections on the metallic surface. 
For the cylinders used in the experiments the mean eigenfrequency was $\nu_0=6.880\,\mathrm{GHz}$ with a spread $\Delta\nu$ of approximately $3\,\mathrm{MHz}$, corresponding to 20\,\% of the linewidth. 

\begin{figure}
  \begin{center}
    \includegraphics[width=\linewidth]{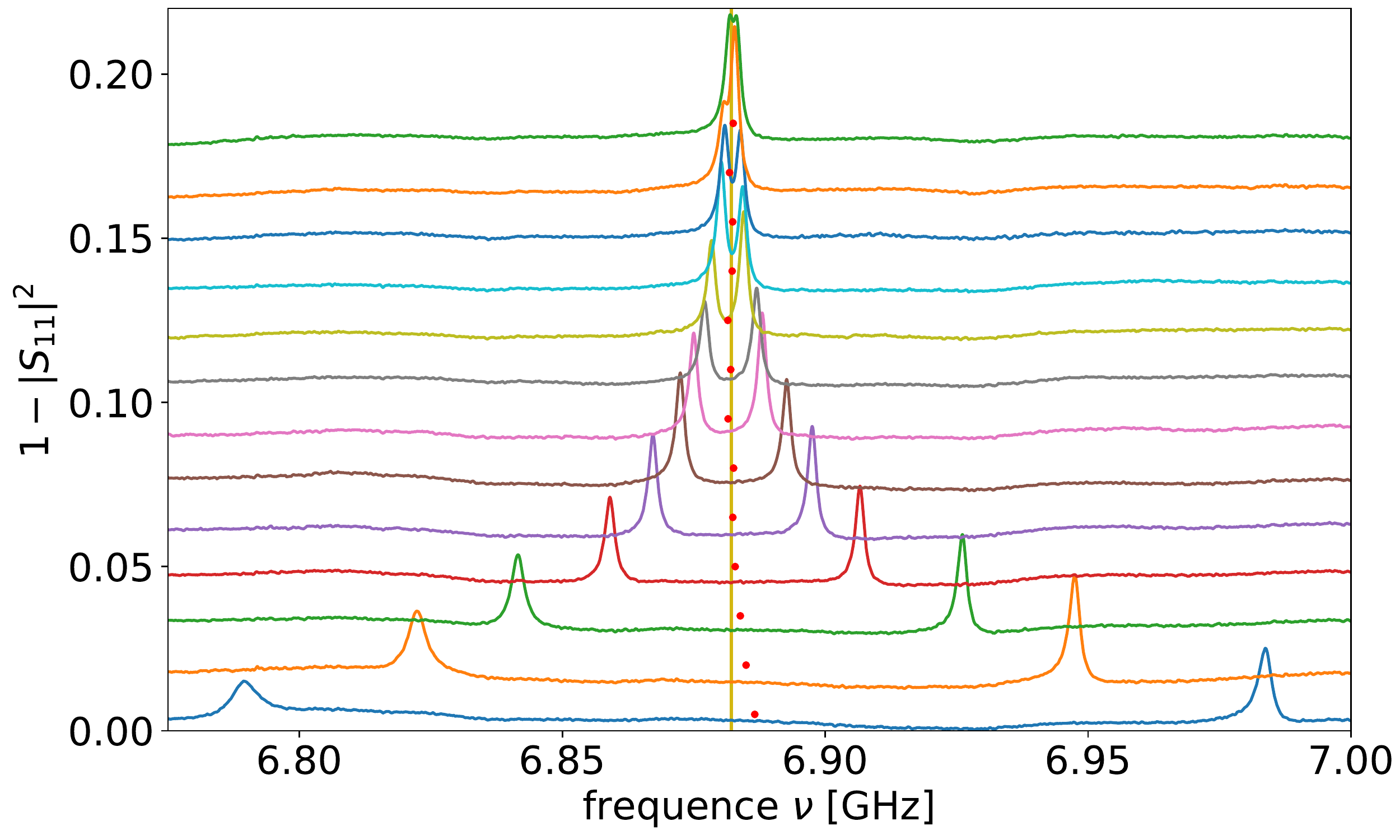}
    \caption{ \label{fig:2disks}
      Spectra of a two disk system in dependence of their distance $d$, used to establish a relation between the distance and the coupling constant $a(d)$. 
      The measurements for different distances $d$ are offset along the ordinate for better visibility. 
      The lowest curve corresponds to a distance of $1\,\mathrm{mm}$ between the two disks.
      Adjacent curves differ by $\Delta d = 0.5\,\mathrm{mm}$.
      The curve formed by the red dots shows the mean value of the extracted resonance frequencies.
    }
  \end{center}
\end{figure}

The eigenfrequencies of the two-disk systems resulting from Hamiltonian~\eqref{eq:ham} are given by
\begin{equation}
  \label{eq:split_eigenfreqs}
  \nu_\pm=\pm\sqrt{|a|^2+\frac{(\Delta\nu)^2}{4}} \,. 
\end{equation}
Figure~\ref{fig:2disks} shows spectra for the two-disks system in dependence of their distance $d$ thus yielding the coupling constant in dependence of $d$, $a=a(d)$.
For the shown example the eigenfrequency of the two disks had been identical within the limits of experimental uncertainty, i.e., $\Delta\nu=0$ and $\nu_\pm=\pm|a|$. 
Such two-disks measurements have been used for a calibration of the distance in terms of the coupling constants, $d(a)$. 
More specifically, we used the known solution for the electric field in a cylindrical resonator to fit this relationship from the experimental calibration. 
The expression used for the fit was~\cite{kuh10a} 
\begin{eqnarray} \label{eq:definition_coupling_from_splitting}
  \frac{1}{2}(\nu_+ - \nu_-) = a 
  &=& \int\!\mathrm{d}^2r\,\psi^{*}_1 \hat{H}\psi_2\\
  \label{eq:fit_func_coupling}
  &=& a_0\left\vert K_0\left[\gamma(r_D +\frac{1}{2}d)\right]\right\vert^2\,,
\end{eqnarray}
see Fig.~\ref{fig:chiOE_spectra} and Appendix~\ref{sec:two-reson-coupl} for details of the measurements and extracted two-resonator couplings. 
We used these fits to map the wanted Gaussian distributions of coupling constants $a$ onto a corresponding distribution of distances $d(a)$ by inverting the obtained fit.

Dielectric disk systems have been used repeatedly in our group, among others also for the realization of a microwave analogue of graphene \cite{bar13a}. 
One reason for the popularity of the latter had been the similarity of the density of states of graphene and a Dirac system with a zero mass term. 
Both have a triangular gap in their densities of states stemming from a linear energy-momentum dispersion relation at the K and K$^\prime$ points in the Brillouin zone. 
Microwave graphene at first sight thus seems to be the best candidate for a realization of the chiral ensembles. 
However, there are a number of disadvantages:

(i) To build a microwave graphene lattice 250 to 300 disks are needed. 
But due to the triangular gap of the density of states at the so-called Dirac point, only five to at most ten resonances are expected to show the signatures of the chiral ensembles, making this approach highly inefficient. 

(ii) In a graphene lattice the next-nearest neighbor distance exceeds the nearest neighbor distance only by a factor of $\sqrt{3}$. 
Next-nearest neighbor contributions can therefore not be neglected. 
They appear in the diagonal blocks of Hamiltonian (\ref{eq:chiham}) and spoil the chiral symmetry. 

\begin{figure}
  \includegraphics[width=0.7\columnwidth]{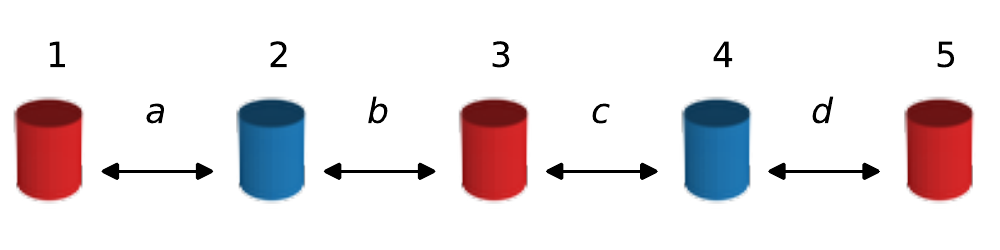}\\
  \includegraphics[width=0.7\columnwidth]{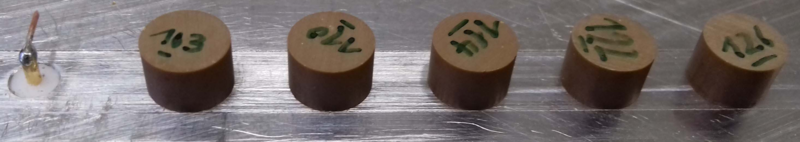}
  \caption{ \label{fig:schematic_chain}
    Top: Schematic representation of the tight-binding chain. 
    The two subsystems are indicated by red and blue (odd and even sites).
    Arrows indicate the nearest-neighbor couplings $a, b, c,$ and $d$. 
    Bottom: Experimental setup of the chiOE chain with bent-antenna. 
    The top plate of the setup was removed for the picture. 
  }
\end{figure}

To avoid these drawbacks, we use a different, much simpler system, namely the linear chain, see Fig.~\ref{fig:schematic_chain}. 
This, however, has the following consequence: In RMT, the off-diagonal $A$ matrices in Eq.~\eqref{eq:chiham} are completely filled with Gaussian random numbers. 
Due to the nearest-neighbor coupling in the linear chain of length $N$, the matrices $A$ become more and more sparse for increasing $N$. 
It is therefore necessary to investigate how this increasing sparsity will affect the RMT behavior. 

\section{Theory and Simulations}
\label{sec:theory}

In order to show that the experiment conducted does comply with the prediction for chiral symmetric systems, we need to derive predictions for the density of states and, as will later be shown, the correlation function for the eigenvalues with the smallest magnitude.

\subsection{Ensemble averaged density of states for $N=1,\dots,5$}
\label{sec:ensemble-aver-dens}

Starting point is the linear chain of resonators shown in Fig.~\ref{fig:schematic_chain}. 
The lower figure shows the disks used in the experiment next to an antenna at the left. 
The upper figure shows a schematic version with the two subsystems I and II indicated in red (odd sites) and blue (even sites). 
The only coupling (indicated by arrows) is assumed to be between adjacent sites, i.e., there is no internal interaction within I or II, a precondition for the existance of a chiral symmetry, see Eq.~\eqref{eq:chiham}. 
For $N\le 5$ the eigenvalues of the Hamiltonian describing the linear chain can be calculated analytically in terms of solutions of quadratic equations. 
For $N=3$, e.g., and only nearest neighbor interactions taken into account, the Hamiltonian reads, with the sequence 1, 3, 2 of rows and columns,
\begin{equation}\label{eq:ham3}
  H=\left(
    \begin{array}{ccc}
      \cdot & \cdot & a \\
      \cdot & \cdot & b \\
      a^* & b^* & \cdot \\
    \end{array}
  \right)\,,
\end{equation}
where $a$ is the coupling between sites 1 and 2 and $b$ is the couplings between sites 2 and 3. 
This makes the chiral symmetry of Eq.~\eqref{eq:chiham} clearly visible. 
The characteristic polynomial~\eqref{eq:chi} reads
\begin{eqnarray}
  \chi(E) &=& E\left|E^2- (|a|^2  + |b|^2)\right|\,,
\end{eqnarray}
hence, we have the eigenvalues
\begin{eqnarray}
  E_{0} &= 0,\quad  E_{\pm1} &= \pm\sqrt{|a|^2  + |b|^2}. 
\end{eqnarray}
For the chiral orthogonal ensemble, couplings $a$ and $b$ are real Gaussian variables following a distribution
\begin{align} \label{eq:gauss_distribution}
  p(a) &= \frac{1}{\sqrt{2\pi}}\mathrm{e}^{-a^2 / 2}
\end{align}
where a variance of one for $a$ has been assumed. 
With Eq.~\eqref{eq:gauss_distribution} the distribution of the non-trivial eigenvalues is given by a $\chi$ distribution with $f = 2$ degrees of freedom. 
For the chiral unitary case, $a$ and $b$ are complex Gaussian variables and the distribution has $f = 4$ degrees of freedom. 
Therefore, we get Wigner-like distributions for the positive eigenvalue $E_{+1}$
\begin{align} \label{eq:dos_N3_chGOE_chGUE_partial}
  \dos^{\beta}_{+1}(E) & \sim {E^{{2\beta - 1}}}\cdot\mathrm{e}^{-E^2 / 2},
\end{align}
and $\dos^{\beta}_{+1}$ is zero for negative arguments. 
For sake of readability, we omit the superscript $\beta$ of $\dos_{+1}$ in the following.
With this, the complete density of state reads
\begin{equation} \label{eq:dos_N3_chGOE_chGUE}
  \dos_{N=3}(E) = \frac{1}{3}\left(\rule[-1.0ex]{0pt}{2.5ex} \delta(E) + \dos_{+1}(|E|)\right)\,. 
\end{equation}
This equation also holds for the symplectic case, $\beta = 4$, where the entries $a$ and $b$ in Eq.~\eqref{eq:ham3} are given by quaternions. 

Due to the symmetry of the spectrum, the correlation between \(E_{+}\) and \(E_{-}\) is given by
\begin{equation} \label{eq:corr_func_Nthree}
  \corr(E) = \frac{\dos_{+}(E / 2)}{2}
\end{equation}
and the integrated correlation function follows as
\begin{equation} \label{eq:ncorr_func_Nthree}
  \ncorr(E) = \intdos_{+1}(E / 2)
\end{equation}
where \(\intdos_{+1}\) is the cumulative distribution function for a $\chi$ distribution with $f = 2\beta$ degrees of freedom. 
The expressions for $N=4$ and $N=5$ are more complicated and are given in Appendix~\ref{sec:density-states-n=4}. 
For a graphical representation of the theoretical curves for $N = 1, \dots, 5$, see the dashed lines in Figs.~\ref{fig:chiGOE_hist_TP},~\ref{fig:chiGUE_hist}, and~\ref{fig:chiGSE}. 

\subsection{Repulsion behavior at small energies}
\label{sec:repulsion-behavior}

One of the most helpful aspects of RMT is the ability to predict universal behavior in experimental data. 
The paradigmatic example is the universal level repulsion of adjacent energy levels for systems with a chaotic classical limit. 
In the chiral ensembles there is in addition a repulsion between the positive and negative eigenvalues with an exponent of $\alpha^\prime= \alpha+\nu \beta$, see Eq.~\eqref{eq:rep}.
For the systems considered here, the behavior is different in as so far as the results depend on $N$. 
This is apparent from the curves show in the comparison with the experimental results, see later.
The cases for $N=2$ and $N=3$ yield full off-diagonal blocks $A$ as can be seen in Eq.~\eqref{eq:ham3}. 
Therefore, they show the expected RMT behavior. 
Chains with $N \geq 4$ show deviations from the RMT prediction. 
This is due to the sparsity of the off-diagonal blocks $A$ already mentioned above. 

Using the analytic expressions for $N=2$ (trivial), $N=3$ (shown above), $N=4$, and $N=5$ (shown in Appendix~\ref{sec:density-states-n=4}) we can determine the repulsion behavior and compare with the general case of Eq.~\eqref{eq:rep}. 
For $N=3$ the scaling behavior of \(\dos{} \sim E^{\alpha'}\) and therefore \corr{} follows directly from Eq.~\eqref{eq:dos_N3_chGOE_chGUE_partial} for all values of $\beta$. 
The derivation of the repulsion exponent for $N=4$ and $N=5$ is presented in Appendix~\ref{sec:density-states-n=4}. 
For $N = 4$ and $\beta = 1$ the density of states \dos{} shows a logarithmic divergence at $E = 0$ instead of a constant behavior. 
An overview of the obtained repulsion exponents for the correlation function, \(\corr\sim E^{\alpha'}\), is given in Tab.~\ref{tab:repulsion_exponents_ncorr}. 
\begin{table}[tb]
  \centering
  \begin{tabular}{|@{\hspace{2ex}}c@{\hspace{2ex}}|@{\hspace{2ex}}c@{\hspace{2ex}}|@{\hspace{2ex}}c@{\hspace{2ex}}|@{\hspace{2ex}}l@{\hspace{2ex}}|c|}\hline
    N & $\beta$ & \(\dos_\mathrm{RMT}\) & \multicolumn{1}{c|}{\(\dos_\mathrm{linear\, chain}\)} & formula \\
    \hline
    2 & 1  & \(\sim |E|^0\) & \(\sim |E|^0\) &  \\
      & 2  & \(\sim |E|^1\) & \(\sim |E|^1\) &  \\
      & 4  & \(\sim |E|^3\) & \(\sim |E|^3\) &  \\
    \hline
    3 & 1  & \(\sim |E|^1\) & \(\sim |E|^1\) & \eqref{eq:dos_N3_chGOE_chGUE_partial} \\
      & 2  & \(\sim |E|^3\) & \(\sim |E|^3\) & \eqref{eq:dos_N3_chGOE_chGUE_partial} \\
      & 4  & \(\sim |E|^7\) & \(\sim |E|^7\) & \eqref{eq:dos_N3_chGOE_chGUE_partial} \\
    \hline
    4 & 1  & \(\sim |E|^0\) & \(\sim |E|^0\cdot (1 + \mathrm{const}\cdot\ln{|E|})\) & \eqref{eq:pf_N4_beta1_small_f} \\
      & 2  & \(\sim |E|^1\) & \(\sim |E|^1\cdot (1 + \mathrm{const}\cdot\ln{|E|})\) & \eqref{eq:pf_N4_beta2_small_f} \\
      & 4  & \(\sim |E|^3\) & \(\sim |E|^3\cdot (1 + \mathrm{const}\cdot\ln{|E|})\) & \eqref{eq:pf_N4_beta4_small_f} \\
    \hline
    5 & 1  & \(\sim |E|^1\) & \(\sim |E|^1\cdot (1 + \mathrm{const}\cdot\ln{|E|})\) & \eqref{eq:pf_N5_beta1_small_f} \\
      & 2  & \(\sim |E|^3\) & \(\sim |E|^3\cdot (1 + \mathrm{const}\cdot\ln{|E|})\) & \eqref{eq:pf_N5_beta2_small_f} \\
      & 4  & \(\sim |E|^7\) & \(\sim |E|^7\cdot (1 + \mathrm{const}\cdot\ln{|E|})\) & \eqref{eq:pf_N5_beta4_small_f} \\
    \hline
  \end{tabular}
  \caption{ \label{tab:repulsion_exponents_ncorr}
    Behavior of the density of states \dos{} close to $E=0$. 
    The leading expressions for the linear chains are identical to Eq.~\eqref{eq:rep} indicated in the third column. 
    A comparison with the full curves is shown in Fig.~\ref{fig:rep_ncorr_small_energies}. 
    Details on the cases $N=4$ and $N=5$ are given in Appendix~\ref{sec:repulsion_exponents_appendix} specifying the form of the logarithmic corrections.
    For $N=5$, $\beta=4$ an integral expression for $\rho(E)$ can be found in the appendix [Eq.~\eqref{eq:pf_N5_beta4}], but we have not been able to elaborate the asymptotic behavior. 
  }
\end{table}

One can see that the leading exponents are indeed as expected from Eq.~\eqref{eq:rep} for all cases up to logarithmic corrections for $N = 4$ and $N=5$. 

\begin{figure}
  \includegraphics[width=\columnwidth]{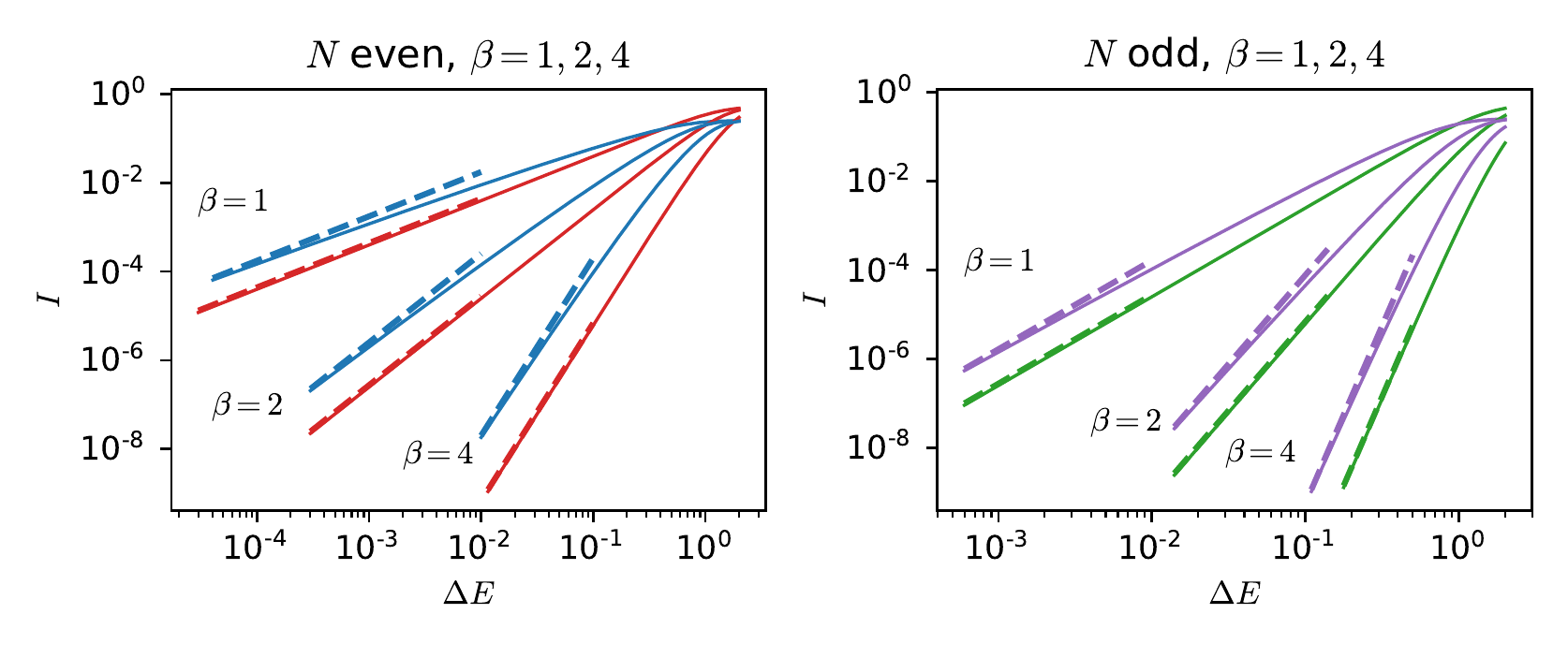}
  \caption{ \label{fig:rep_ncorr_small_energies}
    Repulsion behavior of the integrated correlation function $\ncorr$ for small energies. 
    Left: Curves for $N = 2$ (red), $N = 4$ (blue). 
    Right: Curves for $N = 3$ (green), $N = 5$ (violet). 
    The upper, middle, and lower pair of curves correspond to $\beta = 1, 2, 4$, respectively as indicated. 
    The dashed lines indicate the scaling $\Delta E^{\alpha' + 1}$. 
    The value of $\alpha'$ is taken from the scaling behavior of $\dos_\mathrm{RMT}$ in Tab.~\ref{tab:chiral_ensemble}. 
  }
\end{figure}

Fig.~\ref{fig:rep_ncorr_small_energies} shows a comparison of the analytical curves for the integrated curves $\ncorr$ and the expected scaling $\sim E^{\alpha' + 1}$ following from Eq.~\eqref{eq:rep}. 
Note that the logarithmic corrections are too small to be visible.

\section{Results}

\subsection{The chiral orthogonal ensemble}
\label{sec:chir-orth-ensemble}

\begin{figure}
  \begin{center}
    \includegraphics[width=\linewidth]{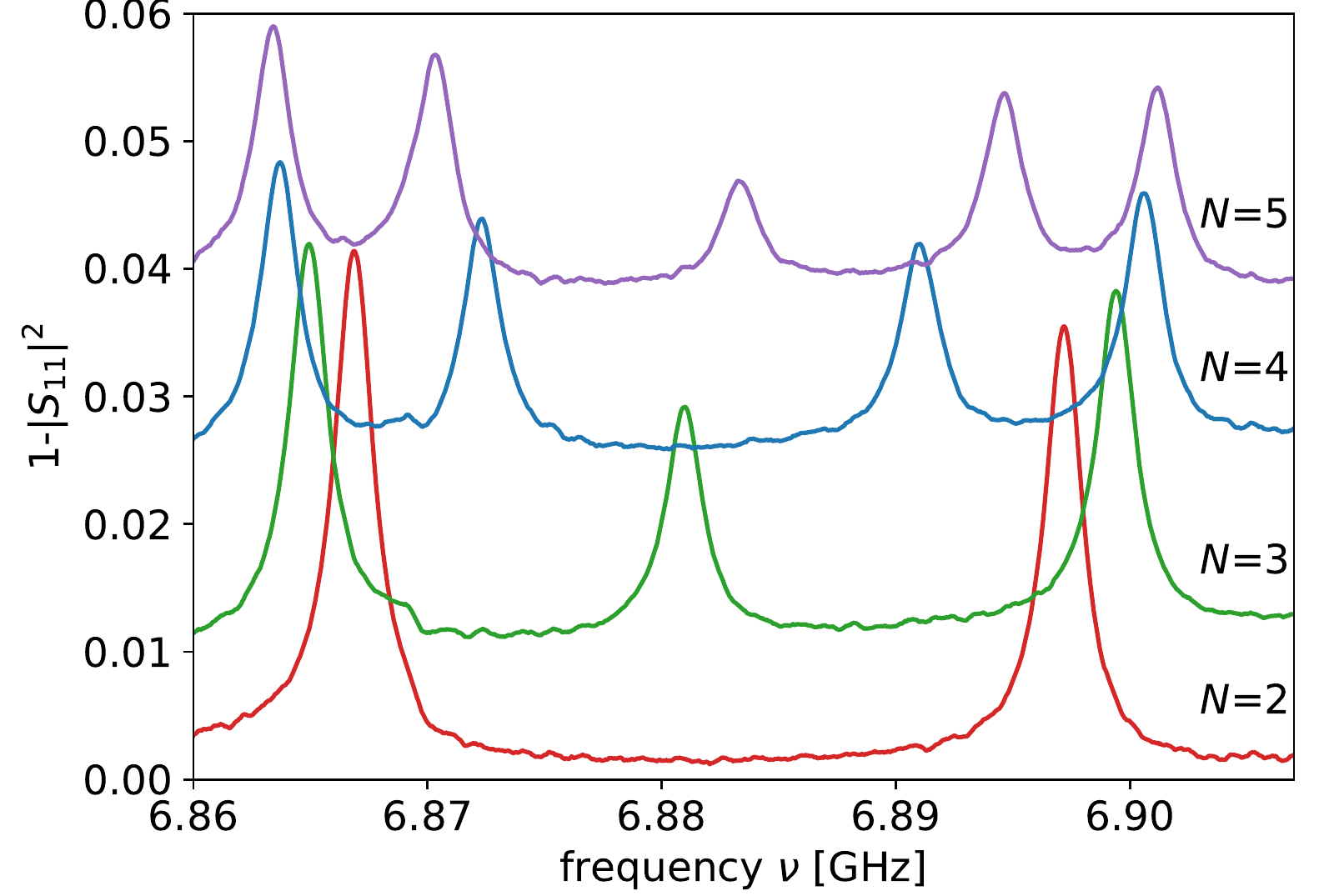}\\[2ex]
    \includegraphics[width=\linewidth]{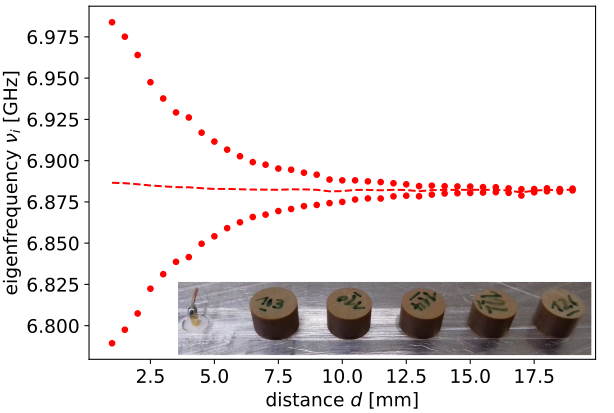}\\
    \caption{ \label{fig:chiOE_spectra}
      Spectra for a linear chain with $N$ = 2, 3, 4 and 5 dielectric cylinders. 
      Bottom: Eigenfrequencies for a two resonator system in dependence of their distance, used to calibrate the coupling constant $a$ in terms of the distance. 
      The dashed line denotes the center of gravity of the two eigenfrequencies. 
      The inset shows the used set-up. 
      The top plate has been removed for the photograph. 
      (See also Fig.~1 in \cite{reh20}.)
    }
  \end{center}
\end{figure}

For the realization of the chiOE the set-up shown in Fig.~\ref{fig:schematic_chain}~(bottom) had been used. 
Up to five disks have been placed in a row. 
Reflection spectra have been obtained with a bent antenna placed close to the leftmost disk.
Figure~\ref{fig:chiOE_spectra} shows typical reflection spectra for $N=2,\dots, 5$ disks obtained by removing one disk after the other from the right.
The chiral symmetry is clearly evident from the spectra: 
Firstly, the spectrum is symmetric around $\nu_0$. 
Secondly, for odd $N$ there is a resonance at $\nu_0$. 
The length $N=5$ showed to be a practical limit for the experiments. 
The extraction of the resonance positions are performed by searching local reflection minima and perform a fit of a Lorentzian around it and in case of overlapping resonances a multi-Lorentzian fit. 
Depending on the coupling strength of the antenna and on the wave function amplitude at the resonator to which the antenna is coupled (always at the end of the chain to avoid zeros of the wave function) we might miss one resonance. 
For larger $N$ there had been an increasing tendency of missing resonances which eventually became intolerable around $N >5$. 

We could not compensate this as the antenna coupling poses two problems: 
On one hand it should be strong to get resonances with a good signal-to-noise ratio, on the other hand it should be weak to avoid a detuning and broadening of the resonances due to the presence of the antenna. 
As a compromise we chose a distance of \(10\,\mathrm{mm}\) of the antenna from the next disk, where the resonance depths amounted to about 50 percent of the reflection signal, and the resonance shifts due to the coupling were of about 10 percent of the line width. 

Altogether spectra of $500$ realizations had been obtained. 
For the mutual distances between the disk a distribution had been used mapping onto a Gaussian distribution of coupling constants, as was explained in the preceding section.
More specifically, we used the distribution
\begin{equation} \label{eq:distr_couplings_chOE}
  p(a)=\frac{1}{\sqrt{2\pi\sigma^2}}e^{-\frac{a^2}{2\sigma^2}}\,, 
\end{equation}
for the couplings and the inverse of the corresponding fits from Eq.~\eqref{eq:fit_func_coupling}. 
For more details on the two-resonator measurements, see Appendix~\ref{sec:two-reson-coupl}. 
We chose $\sigma = 36.3\,\mathrm{MHz}$ as this allowed us to use a range of $d$ values from Fig.~\ref{fig:chiOE_spectra}~(bottom) leading to a good representation of the tails of the Gaussian couplings from Eq.~\ref{eq:distr_couplings_chOE}. 
This sigma led to a mean value of the distance of $d = 5\,\mathrm{mm}$ between adjacent resonators.

\begin{figure}
  \begin{center}
    \includegraphics[width=\linewidth]{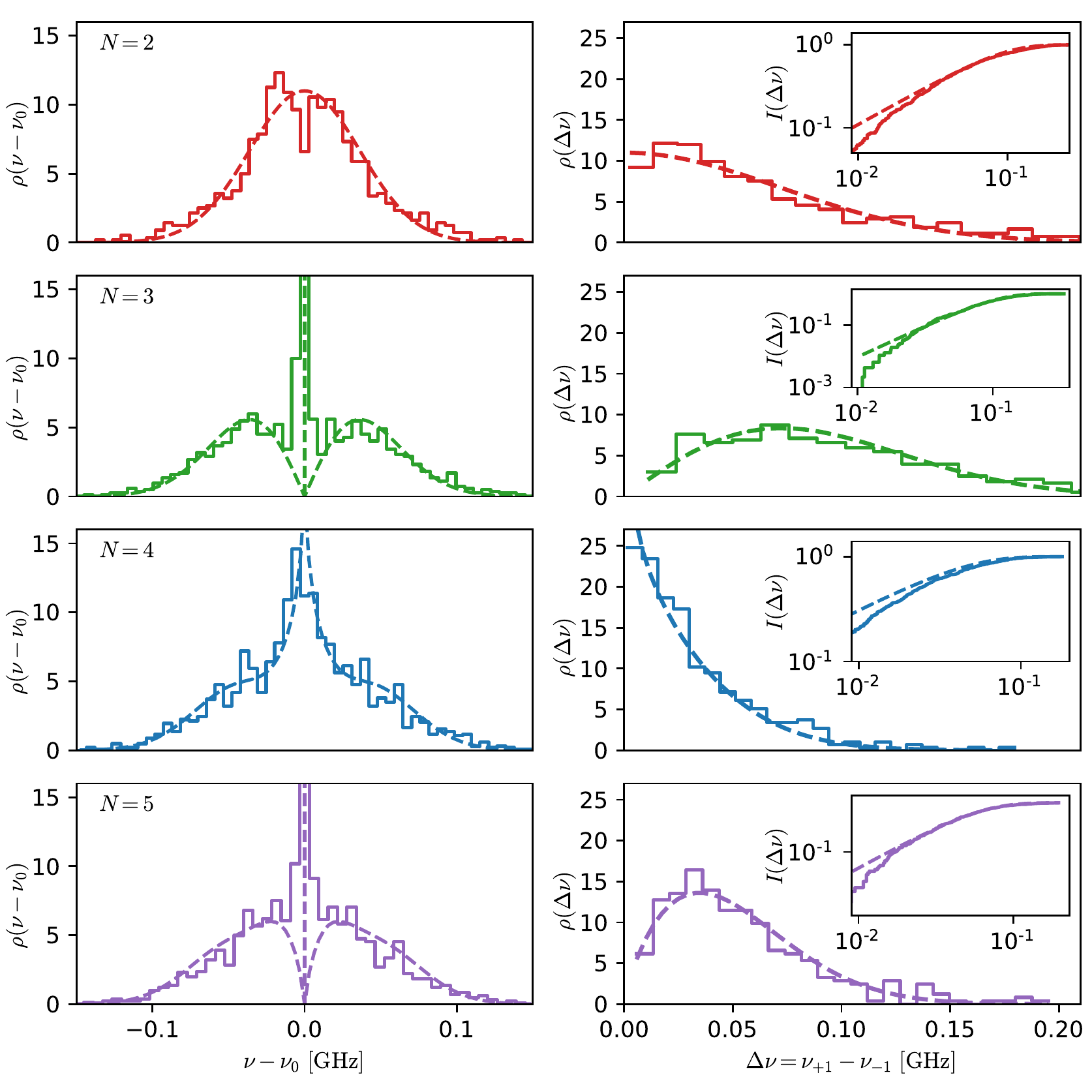}
    \caption{\label{fig:chiGOE}\label{fig:chiGOE_hist_TP}
      Left: Ensemble averaged density of states of the chiOE for $N=2$, $3$, $4$, and $5$.
      The dashed lines correspond to the theoretical expectation.
      Right: Two-point correlation function $\corr(\Delta\nu)$ for the chiOE.
      The dashed lines show the theoretical prediction from Eq.~\eqref{eq:dos_general_3dim_with_abbrevs}.
      The inset shows the integrated two-point correlation function $\ncorr(\Delta\nu)$ in a log-log plot together with the theoretical prediction, Eq.~\eqref{eq:cdfTPcorrfunc}. 
      (See also Fig.~2 in \cite{reh20}.)
    }
  \end{center}
\end{figure}

Figure~\ref{fig:chiGOE_hist_TP}, left panel, shows the results for the ensemble averaged density of states $\langle\dos(E)\rangle$ for linear chains of lengths $N=2,3,4,5$.
For odd $N$ a peak at $E=0$ is predicted, and a linear repulsion of the eigenenergies from the energy zero, see Eq.~(\ref{eq:rep}). 
For even $N$ there should be no repulsion \cite{bee15}. 
All these features are found in the experiment. 
The dashed lines correspond to the analytical expressions for $\langle\dos(E)\rangle$, see section \ref{sec:theory}. 
A good overall agreement is found, with two exceptions: 
(i) The eigenfrequencies of the disks are not identical, but differ by some MHz, thus spoiling the chiral symmetry slightly. 
This results in a hole in the distribution at $E=0$ for $N=2$, and to a smaller extent also for $N=4$. 
(ii) For short distances between the disks the coupling constant depends very sensitive on the distance, making a reliable realization of the Gaussian distribution in the tails problematic. 
Furthermore, there is a natural cut-off, corresponding to the direct contact of the disk. 
This explains the significant deviations of the tails of the experimental distributions from the theoretical ones. 
Fortunately, all these drawbacks, though distorting the details of the distributions, do not influence the repulsion behavior at $E=0$. 

The density of states is not optimal to compare the experimental results with theory: 
Each non-zero element in the diagonal block of the Hamiltonian~(\ref{eq:chiham}) destroys the chiral symmetry. Perturbation of the chiral symmetry result in two imperfections: 
(i) a shift of the center of gravity of the spectrum, 
(ii) a left-right asymmetry between the ``electron'' and the ``positron'' part of the spectrum. 
For the linear chain with an odd number of elements this has in particular the consequence that the zero energy peak is smeared out. 
Therefore, we studied another quantity, the two-point correlation function $\corr(E)$ giving the probability density to find an energy distance $E$ between the states +1 and -1. 
As this uses the difference between two energies, perturbations resulting in a shift of the absolute frequency across different realizations drop out. 
Note that the correlation $\corr(E)$ is, up to a factor of two, identical with the ensemble average density $\dos_{+1}(E)$ of state 1, i.e., the smallest positive energy.

Figure~\ref{fig:chiGOE_hist_TP}, right panel, shows these $\corr$, together with the random matrix expectation (dashed). 
To accentuate the repulsion behavior, in addition the integrated pair correlation function $\ncorr(E)=\int_0^{E} dE'\,\corr(E')$ is shown in the inset in a log-log plot. 

\subsection{The chiral unitary ensemble}
\label{sec:chir-unit-ensemble}

\begin{figure}
  \begin{center}
    \includegraphics[width=\linewidth]{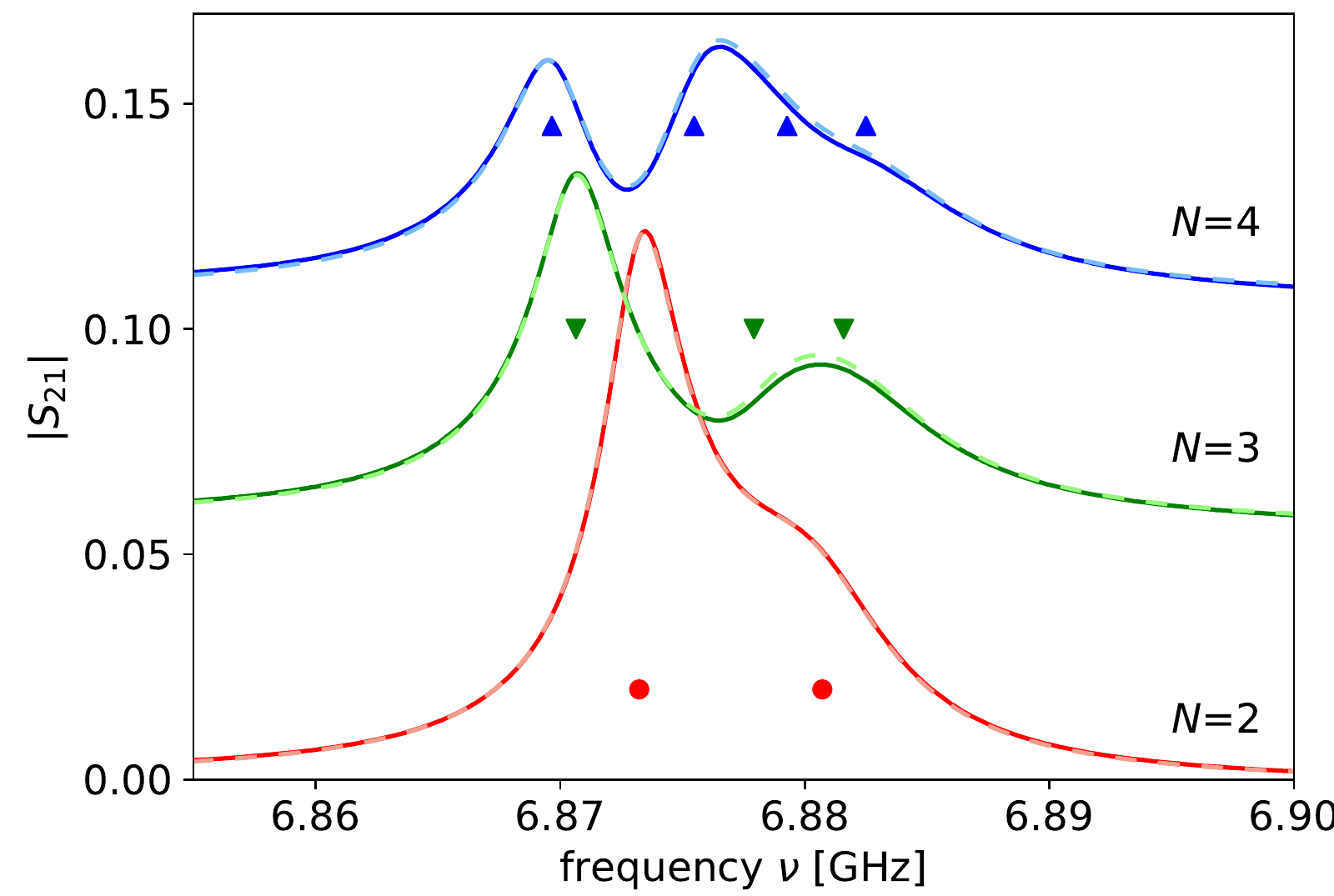}\\[2ex]
    \includegraphics[width=\linewidth]{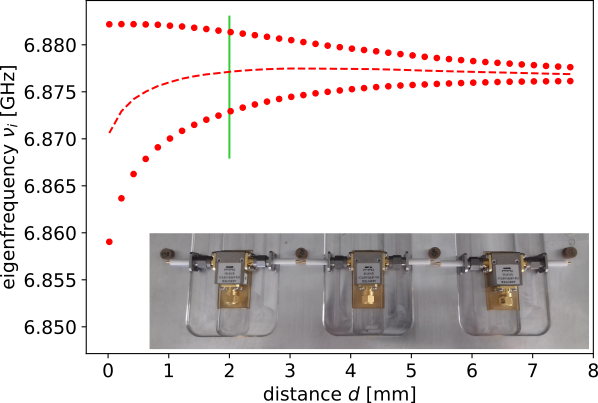}\\
    \makebox[0pt][l]{%
      \hspace*{.163\linewidth}%
      \raisebox{.335\linewidth}[0pt][0pt]{
        \includegraphics[viewport={40pt 72pt 170pt 149pt},clip,width=0.31\linewidth]{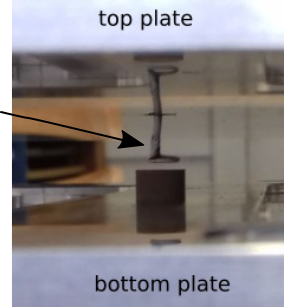}}}
    \caption{\label{fig:GUE_spectra}\label{fig:chiUE}
      Top: Reflection spectra for the chiUE for $N = 2, 3, 4$ dielectric
      cylinders. 
      The symbols denote the positions of the resonances extracted by the harmonic inversion technique.
      The solid lines in dark colors correspond to the measurement, the superimposed dashed lines in light colors to the reconstruction. 
      Bottom: Eigenfrequencies for a two resonator system, coupled by a circulator with an open-end side port terminator in dependence of the distance.
      The dashed line denotes the center of gravity of the two eigenfrequencies.
      The vertical green line denotes the lower limit of the distances used for the histograms in the left column of Fig.~\ref{fig:chiGUE_hist}.
      The lower inset shows the set-up for $N=4$ with the top plate removed. 
      The smaller inset shows the loop antenna (indicated by the black arrow) above the resonator between bottom and top plates.
      (See also Fig.~3 in \cite{reh20}.)
    }
  \end{center}
\end{figure}

For the realization of the chiUE time reversal symmetry has to be broken.
This is achieved by means of circulators, which had been already used for this purpose previously \cite{reh16}. 
A circulator is a microwave device with three ports, where waves entering via ports 1, 2, 3 exit only through ports 2, 3, 1, respectively. 
Figure \ref{fig:chiUE}~(bottom, lower inset) shows the set-up. 
Now the disks are at distances of approximately \(9\,\mathrm{cm}\), too large for a direct coupling. 
Instead, the coupling is achieved by circulators with two attached monopole antennas oriented horizontally, i.e., perpendicular to the cylindrical surface of the resonators. 
The distance of the monopole antennas from the disk surfaces can be varied by moving the circulators back and forth, where the two resonators have the same distance $d$ to the monopole antennas (see inset of Fig.~\ref{fig:GUE_spectra}. 
Since the circulators introduce directionality, the coupling constants are complex, $a=a_\mathrm{R}+\mathrm{i}a_\mathrm{I}=|a| e^{\mathrm{i}\varphi}$. 
In contrast to the chiOE situation now two parameters have to be varied, real and imaginary part of $a$, or, alternatively, modulus and phase. 
A Gaussian distribution of $a_R$ and $a_I$ results in a distribution 
\begin{equation} \label{eq:pUE}
    p(a)= \frac{a}{\sigma^2}e^{-\frac{a^2}{2\sigma^2}}
\end{equation}
for the modulus, and a uniform distribution for the phase. 

In the chiUE experiment, we only altered the modulus $|a|$ as the eigenvalues of the Hamiltonian do not depend on the phase. 
The phase was fixed by choosing standard open-end terminators for all circulators.
For details on other types of terminators and phases, see Appendix~\ref{sec:two-reson-coupl}. 
The distribution (\ref{eq:pUE}) has been realized, just as in the chiOE case, by a corresponding distribution of the distance between the disks and the circulators. 

\begin{figure}
  \begin{center}
    \includegraphics[width=\linewidth]{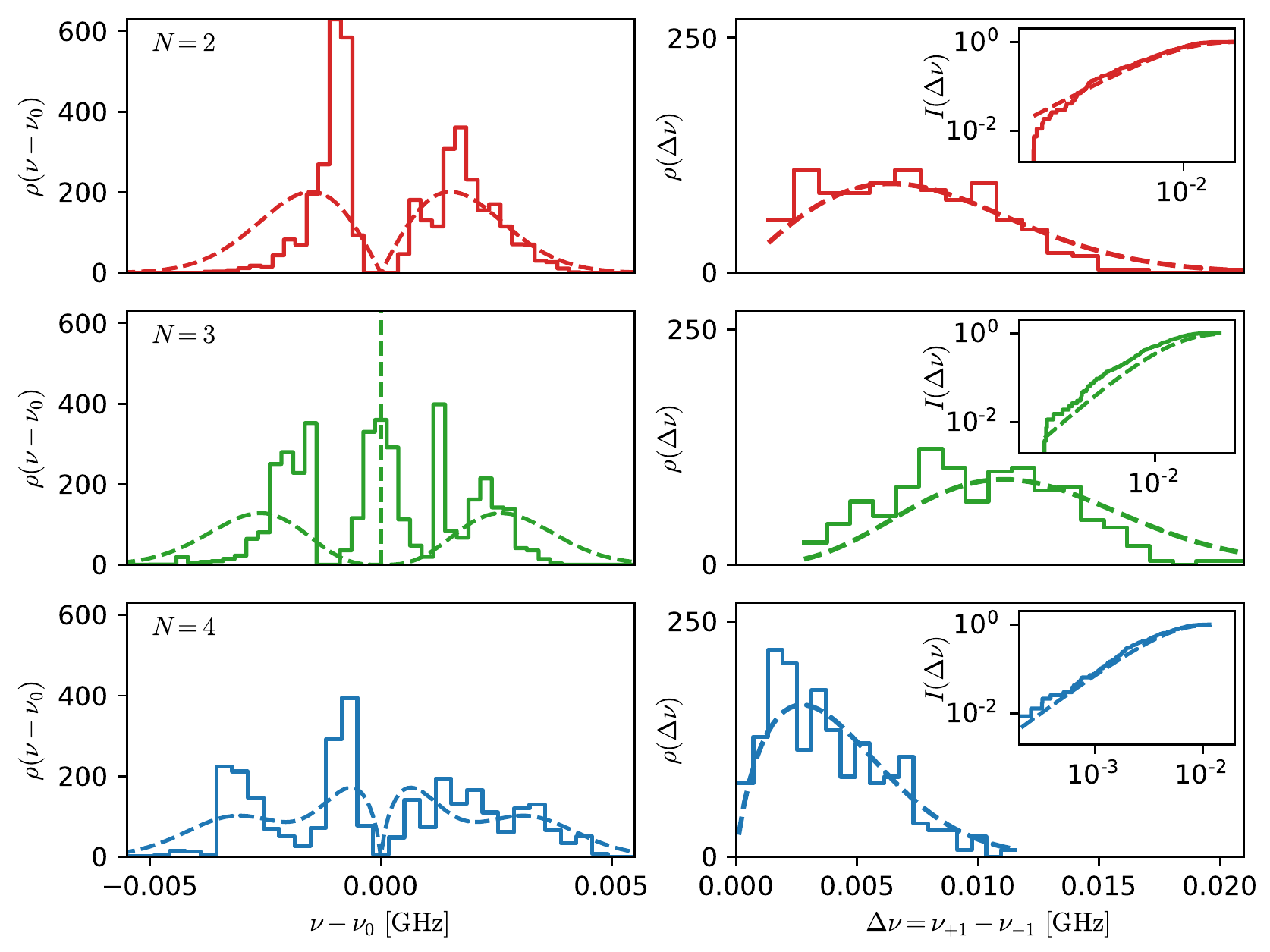}\\
    \caption{ \label{fig:chiGUE_hist}
      As Fig.~\ref{fig:chiOE_spectra}, but for the chiUE. 
      Left: Ensemble averaged density of states of the chiUE for $N=2$,  $3$, $4$. 
      Left: Ensemble averaged density of states of the chiUE for $N=2, 3, 4$. 
      The dashed lines correspond to the theoretical expectation. 
      Right: Two-point correlation function $\corr(\Delta\nu)$ for the chiUE. 
      The dashed lines show the theoretical prediction from Eq.~\eqref{eq:dos_general_3dim_with_abbrevs}. 
      The inset shows the integrated two-point correlation function $\ncorr(\Delta\nu)$ in a log-log plot together with the theoretical expectation, see Eq.~\eqref{eq:cdfTPcorrfunc}. 
      (See also Fig.~4 in \cite{reh20}.)
    }
  \end{center}
\end{figure}

Unfortunately, the circulators introduce strong absorption which is evident from Fig.~\ref{fig:chiUE} showing typical spectra for $N=2,3$ and 4 disks. 
In contrast to the chiOE, where the resonances are narrow and well separated, see Fig.~\ref{fig:chiOE_spectra}, now a considerable broadening is observed, making an analysis difficult. 
Therefore, this time the harmonic inversion technique has been used, allowing for a analysis of the spectra also for overlapping resonances, see e.g.\ Ref.~\cite{kuh08b}. 
The larger the number of resonators is, the more often a small wave function amplitude is obtained at the first resonator, where the antenna is placed above (see inset of Fig.~\ref{fig:chiUE}). 
For $N>4$ the number of these resonances which could not be detected via the harmonic inversion lead to a non-negligible number of missing resonances in the ensemble average. 
In the orthogonal case we could evaluate up to $N=5$, which is not possible for the unitary case, due to the additional broadening introduced by the circulators. 

Figure~\ref{fig:chiGUE_hist} shows the resultant ensemble averaged density of states for $N=2$, $3$, $4$. 
Qualitatively the features of chirality are still found, a ``positron'' spectrum for lower energies, an ``electron'' spectrum for the higher ones, and a central peak for $N=3$. 
But now a strong left-right asymmetry is found, showing that the chiral symmetry is severely disturbed.

Responsible for this fact are the circulators. 
The eigenfrequencies of the separated disks are identical up to 10~percent of the line width, but the presence of the circulators modify the electric field between top and bottom plate resulting in a detuning of the eigenfrequencies by several MHz, comparable to the distances between the resonances. 
To check this assumption, simulations have been performed taking the detuning into account, see Appendix~\ref{sec:simul-pert-chir}. 
A good qualitative over-all correspondence was found. 
In particular the left-right asymmetry was reproduced correctly. 
Fortunately the perturbations of the chiral symmetry drop out in first order for the two-point correlation function $\corr(E)$, which is shown in Fig.~\ref{fig:chiGUE_hist}~(right), again with the integrated pair correlation function $\ncorr(E)$ in a log-log plot. 
The dashed lines correspond to the theoretical expectations for the unperturbed system and match the experimental data nicely.

\subsection{The chiral symplectic ensemble}
\label{sec:chir-symp-ensemble}

In a recent paper we succeeded in the realization of the Gaussian symplectic ensemble in a microwave graph mimicking a spin 1/2 system~\cite{reh16}. 
The main ingredients had been two subgraphs, complex conjugate to each other, coupled by a pair of bonds at inversionally symmetric points with a phase shift of $\Delta\phi = \pi$ in one of the bonds and no phase shift in the other one. 
These ideas may be taken over to the present situation of coupled resonators. 

\begin{figure}
  \begin{center}
    \includegraphics[width=.55\linewidth]{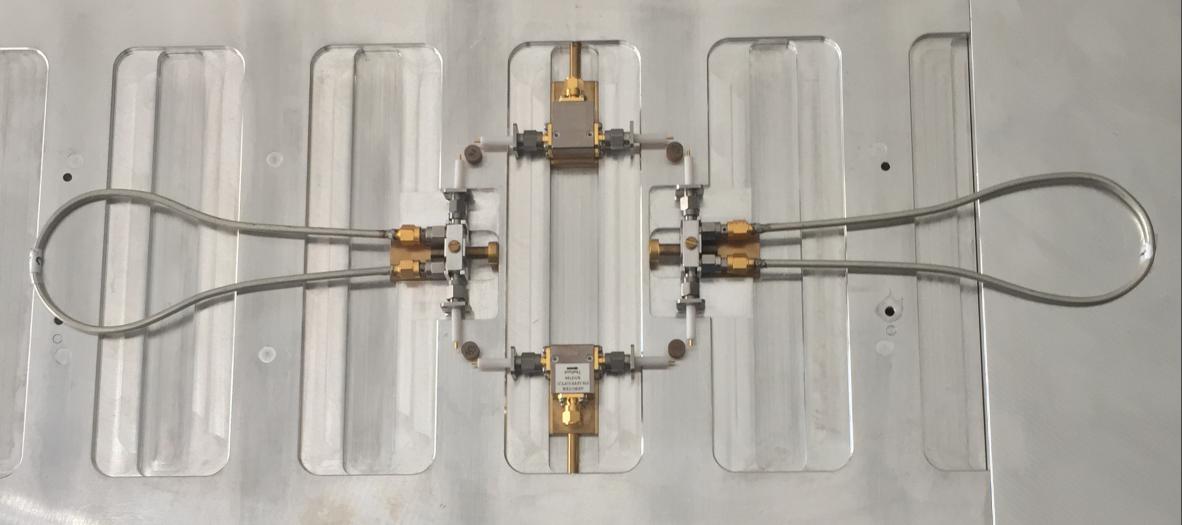}
    \includegraphics[width=.42\linewidth]{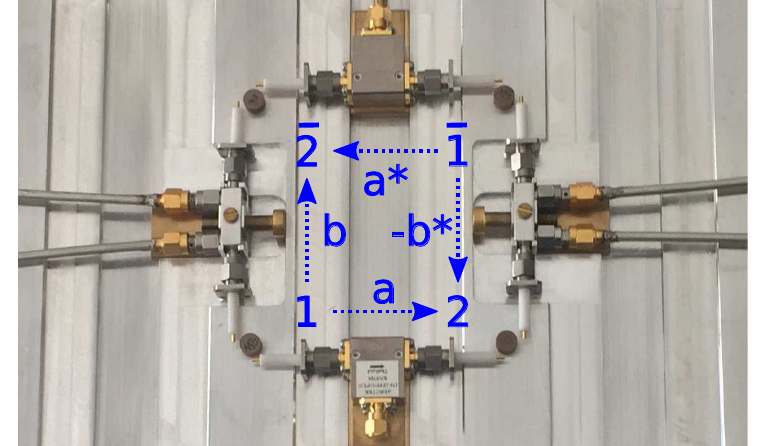}
  \end{center}
    \caption{\label{fig:exp-setup-chSE}
      Set-up for the realization of the chiSE for $N=2$. 
      Left: Photograph of the total set-up without top plate including the two cables connecting the two GUE subsystems which impose a phase difference of $\Delta\phi = \pi$ at the single-resonator resonance-frequency $\nu_0$. 
      Right: Zoom into the region with the 4 coupled resonators introducing also the notations for the resonator and the coupling constants for the chiSE used in Eq.~\eqref{eq:H_S}. 
      (For the right see also Fig.~5(left) in \cite{reh20}.)
    }
\end{figure}
A photograph of this setup is shown in Fig.~\ref{fig:exp-setup-chSE}~(left). 
The right hand side shows a zoom into the region where the resonators $1$, $\bar{1}$, $2$, and $\bar{2}$ are placed. 
The lower part corresponds to the GUE setup (shown on the inset of Fig.~\ref{fig:GUE_spectra}), i.e., the resonators $1$ and $2$ are coupled with coupling constant $a$ via two monopole antennas attached to a circulator where the third port of the circulator is either open or closed by a short terminator. 
The upper part is symmetric apart from the fact that the circulator has been inverted, thus the submatrix describing the system $\bar{1}$ and $\bar{2}$ is the complex conjugate of the upper one. 
The two GUE sub-systems are now coupled on the left and the right side via two cables of lengths $L_1$ and $L_2$, respectively. 
The length difference results in a phase difference $\Delta \phi=k (L_1-L_2)$ for the propagating waves, where $k$ is the wave number. 
For a symplectic symmetry to exist this phase difference must be $\Delta\phi = \pi$. 
In order to check that the system indeed has a symplectic symmetry, we performed further test measurements. 
They are shown in Appendix~\ref{sec:testing-sympl-sym} and test, for example, the presence of Kramers doublets similar to Ref.~\cite{reh18}. 

The Hamiltonian of this system can be represented by a $4 \times 4$ matrix. 
Written with rows and columns in the order $1, \bar{1}, 2, \bar{2}$, it reads 
\begin{equation} \label{eq:H_S}
H=\left(
\begin{array}{cccc}
\cdot & \cdot & a & b \\
\cdot & \cdot & -b^* & a^* \\
a^* & -b & \cdot & \cdot \\
b^* & a & \cdot & \cdot \\
\end{array}
\right)\,.
\end{equation}
Here, the \(2\times 2\) sub-blocks in the off-diagonal elements represent the couplings between the two subsystems and are chosen according to Eq.~\eqref{eq:quaternions_via_complex_nums}. 

To realize the ensemble average, a calibration measurement similar to Figs.~\ref{fig:schematic_chain}~(bottom) and~\ref{fig:chiUE}~(bottom) was done. 
Because of the symmetry, only two couplings are free parameters, namely $a$ and $b$.
In order to ensure that all couplings are such that they obey Eq.~\eqref{eq:H_S}, several calibrations needed to be done before the measurements for the ensemble averages could commence. 

In total, there are 4 distances to take into account: 
(i) the vertical distance of the bottom circulator to resonators $1$ and $2$, 
(ii) the vertical distance of the top circulator to $\bar{1}$ and $\bar{2}$, 
(iii) the horizontal distance of the left wire to resonators $1$ and $\bar{2}$, and 
(iv) the horizontal distance of the right wire to resonators $\bar{1}$ and $2$. 
(i) and (ii) give rise to the coupling strength $|a|$ and (iii) and (iv) to the coupling strength \(|b|\) in Eq.~\eqref{eq:H_S}. 
Like in the chiUE case, the complex parameters only enter the eigenvalues by their magnitude, see Eq.~\eqref{eq:evals_chSE} below. 
Therefore, a distribution of the form~\eqref{eq:pUE} is chosen here as well. 

The characteristic polynomial for Hamiltonian~\eqref{eq:H_S} is given by
\begin{equation} \label{eq:charpoly_chSE}
  \chi(E)=|E\cdot{\bf 1} -H|= \left(E^2-|a|^2-|b|^2\right)^2\,.
\end{equation}
Hence, there are doubly degenerate eigenvalues at the two positions
\begin{equation} \label{eq:evals_chSE}
  E_{+1,-1}=\pm\sqrt{|a|^2+|b|^2}\,.
\end{equation}
The four-disk system thus show both the chiral symmetry, with $E_{+1}$ and $E_{-1}=-E_{+1}$ coming in pairs, and the symplectic symmetry, with the characteristic Kramers doublet structure of the spectrum. 
Because the eigenvalues are given by Eq.~\eqref{eq:evals_chSE}, the density of states \(\dos_{+}\) has the same form as in Eq.~\eqref{eq:dos_N3_chGOE_chGUE_partial} as $a$ and $b$ are both complex numbers in~\eqref{eq:evals_chSE}. 
Therefore, Eqs~\eqref{eq:corr_func_Nthree} and~\eqref{eq:ncorr_func_Nthree} originally obtained for $N=3, \beta = 2$ are also valid for $N=2, \beta = 4$. 

\begin{figure}
	\begin{center}
		\includegraphics[width=\linewidth]{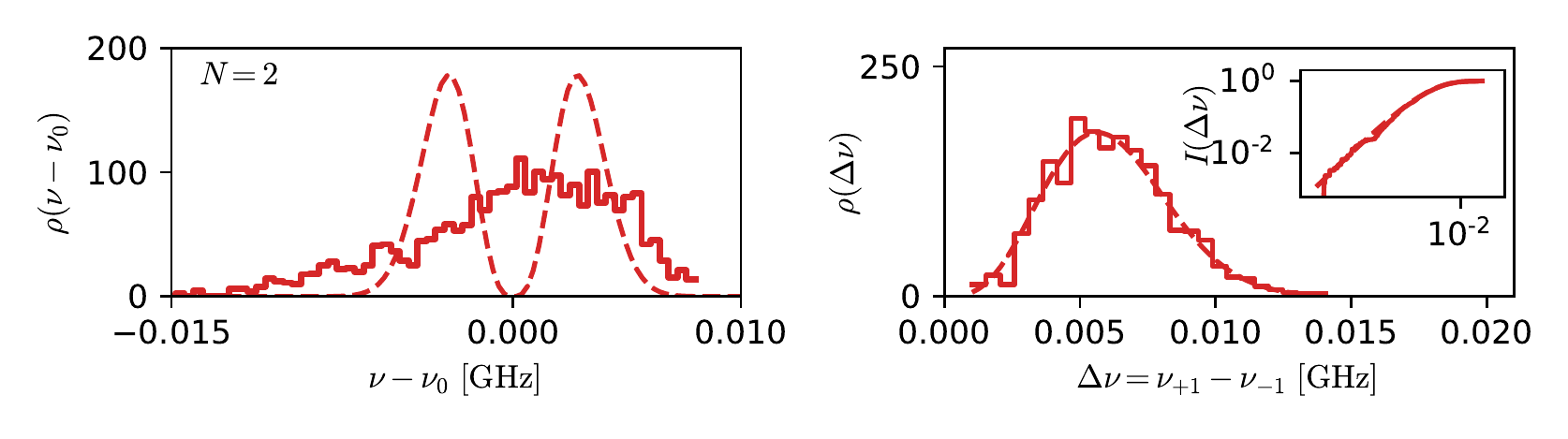}
	\end{center}
	\caption{ \label{fig:chiGSE}
		Left: Ensemble averaged density of states of the chiSE for $N=2$. 
		The dashed lines correspond to the theoretical expectation. 
		Right: Two-point correlation function $\corr(\Delta\nu)$ for the chiSE.
		The dashed lines show the theoretical prediction from Eq.~\eqref{eq:dos_general_3dim_with_abbrevs}. 
		The inset shows the integrated two-point correlation function $\ncorr(\Delta\nu)$ in a log-log plot together with Eq.~\eqref{eq:cdfTPcorrfunc}. 
		(For the right see also Fig.~10(right) in \cite{reh20}.)
	}
\end{figure}

In total, 200 measurements with open-end circulators were done and 170 measurements with short-end ones allowing to extract the resonances in a similar fashion to Sec.~\ref{sec:chir-unit-ensemble} using the harmonic inversion technique. 
Fig.~\ref{fig:chiGSE} shows the density of states \dos{} (left) and the correlation function \corr{} (right). 
Just as before for the chiUE we see that the density of states is distorted. 
This is again due to the presence of the circulators. 
Using the correlation function instead overcomes the problem and shows good agreement with the predictions just like before for the chiUE. 
Hence, we see the expected cubic repulsion of \(E_{+1}\) and \(E_{-1}\) in accordance with Tab.~\ref{tab:repulsion_exponents_ncorr}. 
An extension to larger \(N\) values seem hardly feasible. 
Each additional pair of resonators would mean four more bonds and a corresponding increase of absorption. 
Therefore, we have to be content with this demonstration for \(N = 2\). 

\section{Conclusions}

We present an experimental realization of the orthogonal, unitary, and symplectic version of the chiral ensembles BDI, AIII, CII, or, as in this paper simplified to chiOE, chiUE, and chiSE. 
All realizations are based on tight-binding chains created by coupled microwave resonators. 
The key ingredient are small resonators with a high index of refraction placed between two conducting plates. 
Different types of couplings between the resonators were then used to create the different ensembles. 
These are presented in Sections~\ref{sec:chir-orth-ensemble}, \ref{sec:chir-unit-ensemble}, and~\ref{sec:chir-symp-ensemble}, respectively. 

While the orthogonal ensemble could be created up to a length of $N=5$ resonators, only $N \le 4$ and $N=2$ were possible for the unitary and symplectic case, respectively.
This is mainly due to the more complex set-up of these cases using microwave circulators.
Furthermore, the presence of these circulators creates a perturbation which distorts the density of states. 
However, focusing on the correlation function, \(\corr\), and its integrated version, \(\ncorr\),  allows to clearly see the expected RMT repulsion  between positive and negative eigenvalues close to $E=0$. 
The sparsity of the coupling matrices for $N\ge 4$, however, leads to logarithmic corrections, too small to be seen in our experiment. 

\begin{acknowledgments}
This work was funded by the Deutsche Forschungsgemeinschaft via the individual grants STO 157/16-2 and KU 1525/3-1 as well as the European Commission through the H2020 programm by the Open Future Emerging Technology ``NEMF21'' Project (664828). 
One of the authors (MR) gratefully acknowledges the support through EPSRC grant EP/R012008/1.
\end{acknowledgments}

\appendix
\section{Density of States for \(N=4\) and \(N=5\)}
\label{sec:density-states-n=4}

Following the derivation of Eq.~\eqref{eq:dos_N3_chGOE_chGUE}, we can calculate the corresponding densities and correlation functions for $N=4$ and $N=5$. 
For $N=4$ $H$ reads
\begin{equation} \label{eq:ham4}
  H = \left(\begin{array}{cccc}
      \cdot & \cdot & a     & \cdot \\
      \cdot & \cdot & b^*   & c \\
      a^*   & b     & \cdot & \cdot \\
      \cdot & c^*   & \cdot & \cdot \\
    \end{array}\right)\,,
\end{equation}
where we ordered rows an columns just as in Eq.~\eqref{eq:ham3}, first all odd and then all even sites. 
Thus the characteristic polynomial becomes
\begin{equation}
  \chi(E) = E^4 - E^2\left(|a|^2 + |b|^2 + |c|^2\right) + |a|^2|c|^2\,.
\end{equation}
It is helpful to introduce polar coordinates where \(r^2 = |a|^2 + |b|^2 + |c|^2\). 
The specific choice of angles $a = r\cdot f_a(\vartheta, \varphi, \dots)$, $b = r\cdot f_b(\vartheta, \varphi, \dots)$ etc.\ depends on the dimensionality of the problem, i.e., whether $\beta=1$ or $\beta=2$ is considered. 
With this the eigenvalues can be written as
\begin{equation} \label{eq:eigenvals_squared_N4} 
  \left(2E_\pm^2\right) = r^2\left(1 \pm \sqrt{1 - f^2(\vartheta, \varphi, \dots)}\right)\,,
\end{equation}
where the angular part, \(f(\vartheta, \varphi, \dots) = 2 \frac{|a| |c|}{r^2}\) depends on the specific choice of coordinates. 
The sign \(\pm\) indicates the larger and smaller of the two positive eigenvalues. 
The whole spectrum is given by $(-E_+, -E_-, E_-, E_+)$ and therefore centered around zero. 
For later convenience, we introduce the function 
\begin{equation}
  h_\pm(x) = \frac{1}{\sqrt{1 \pm\sqrt{1 - x^2}}}
\end{equation}
such that we have $\sqrt{2} E_\pm = r / h_\pm(f(\vartheta, \varphi, \dots))$. 

The density of states for the values $E_\pm$ follows from the averages over the Gaussian degrees of freedom in \(a, b, c\). 
This density is identical to zero for $E < 0$ such that the argument $E$ can be assumed positive in the following equations. 
For $\beta = 1$ we have to calculate the average in three dimensions, for $\beta = 2$ in six dimensions. 

Hence, for $\beta = 1$ the 3-dimensional integral becomes a 2-dimensional integral
\begin{equation} \label{eq:dos_general_3dim}
  \dos_\pm(E) =
  \int\frac{\mathrm{d}\Omega}{{\pi}^{3/2}} h_\pm(f) 
  \left(E\!\cdot\! h_\pm(f)\right)^2 
  \mathrm{e}^{-(E\cdot h_\pm(f))^2}\,,
\end{equation}
where $\int\mathrm{d}\Omega$ indicates the integration over the angles including the Jacobian, e.g. for the $2$-sphere in the usual spherical coordinates
$\int\mathrm{d}\Omega{\dots} = \int_{0}^{2\pi}\,\mathrm{d}\varphi\, \int_{0}^{\pi}\,\mathrm{d}\vartheta\sin\vartheta \dots$. 
This can be used to introduce a distribution function for the quantity $f$ over the sphere via
\begin{equation} \label{eq:def_pf_func_general}
  \pf(r) = \frac{1}{A}\int\mathrm{d}\Omega\,\delta\left(r - f(\vartheta, \varphi, \dots)\right)\,, 
\end{equation}
where we set the constant $A = \int\mathrm{d}\Omega$ such that $\pf$ is normalized. 
Note that the value of $A$ depends on the choice of the polar coordinates used to parameterize the integrals. 
Furthermore, we can introduce the abbreviation
\begin{equation} \label{eq:pr_func_def_general}
  \pr(r) = \frac{r^2}{c} \mathrm{e}^{-r^2}
\end{equation}
which is normalized over \(r\in [0, \infty)\) by setting $c = \sqrt{\pi} / 4$. 

With the help of these abbreviations we can formulate the densities of states \dos{} via
\begin{equation} \label{eq:dos_general_3dim_with_abbrevs}
  \dos_\pm(E) = \frac{A c}{{\pi}^{n/2}} \int\limits_{0}^{1}\mathrm{d}f\,\pf(f)\,h_\pm(f)\,\,\pr
  \left(E\cdot h_\pm(f)\right)\,.
\end{equation}
The value $n$ in the prefactor reflects the dimensionality of the integral and is, for example, $n=3$ from Eq.~\eqref{eq:dos_general_3dim}. 
The integrated density of states follows from integrating Eq.~\eqref{eq:pr_func_def_general} to
\begin{equation}
  \nr(r) = \mathrm{Erf}(r) - \frac{2r}{\sqrt{\pi}}\mathrm{e}^{-r^2}
\end{equation}
and yields
\begin{equation} \label{eq:cdfTPcorrfunc}
  \intdos_\pm(E) = \frac{A c}{{\pi}^{n/2}}
    \int\limits_{0}^{1}\!\mathrm{d}f\,\,\pf(f)\,\,
    \nr\!\left(E\cdot h_\pm(f)\right)
\end{equation}
together with $n=3$ and $c = \sqrt{\pi}/4$ from above. 
In what follows we will give the necessary expressions for $A$, $c$, $n$, \pf, and \pr\ also for $N=4$ and $N=5$ for the different ensembles.

For $N = 4$ and $\beta = 1$ we use ordinary spherical coordinates, i.e.,\ $a = \sin\vartheta \cos\varphi$, $b = \sin\vartheta \sin\varphi$, and $c = \cos\vartheta$. 
This yields $A = 4\pi$ and from Eq.~\eqref{eq:def_pf_func_general}
\begin{eqnarray}
  \pf(f)
  &=& \frac{1}{4\pi}\!\!
      \int\limits_{0}^{2\pi}\!\!\mathrm{d}\varphi\!\!
      \int\limits_{0}^{\pi}\!\!\mathrm{d}\vartheta\,
      \sin\vartheta\,\times \\
  & & \qquad\notag \delta\left(f - \sin^2\vartheta|\sin(2\varphi)|\right) \\
  &=& \label{eq:pf_N4_beta1}
      \frac{4}{\pi\sqrt{1 + f}} K\left(\sqrt{\frac{1 - f}{1 + f}}\right)\,,
\end{eqnarray}
where $K$ is the complete elliptic integral. 
For later reference, we also state the behavior of this function for small values of $f$~\citep[see page 359]{mag66}
\begin{equation} \label{eq:pf_N4_beta1_small_f}
  \pf(f) = \frac{2}{\pi}\left(3\ln(2) + \ln\frac{1}{f}\right) + \frac{4}{3\pi}f + \mathcal{O}\left(f^2\ln\frac{1}{f}\right).
\end{equation}

For $\beta = 2$ the integral in Eq.~\eqref{eq:dos_general_3dim} becomes six-dimensional and
\begin{eqnarray} \label{eq:pr_func_def_5}
  \pr(r) &=& \frac{r^5}{c}\mathrm{e}^{-r^2} \\
  \nr(r) &=& 1 - \frac{r^4 + 2r^2 + 2}{2}\mathrm{e}^{-r^2}
\end{eqnarray}
with $c = 4$ to normalize~\eqref{eq:pr_func_def_5}. 
Choosing the six-dimensional coordinates in the following form
\begin{eqnarray}
  a &=& r\cos\zeta \sin\psi \mathrm{e}^{\mathrm{i}\varphi_1} \\
  b &=& r\cos\zeta \cos\psi \mathrm{e}^{\mathrm{i}\varphi_3} \\
  c &=& r\sin\zeta \mathrm{e}^{\mathrm{i}\varphi_2}
\end{eqnarray}
leads to $A = \pi^3$ and we obtain
\begin{equation} \label{eq:pf_N4_beta2}
  \pf(f) = f\ln\left(\frac{1 + \sqrt{1 - f^2}}{1 - \sqrt{1 - f^2}}\right)\,.
\end{equation}
The limiting behavior is
\begin{equation} \label{eq:pf_N4_beta2_small_f}
  \pf(f) = 2\left(\ln(2) + \ln\frac{1}{f}\right)\cdot f + \mathcal{O}\left(f^2\right)\,.
\end{equation}

The same steps yield for $N = 5$
\begin{equation} \label{eq:ham5}
  H=\left(
    \begin{array}{ccccc}
      \cdot & \cdot & \cdot & a     & \cdot \\
      \cdot & \cdot & \cdot & b^*   & c \\
      \cdot & \cdot & \cdot & \cdot & d^* \\
      a^*   & b     & \cdot & \cdot & \cdot \\
      \cdot & c^*   & d   & \cdot & \cdot \\
    \end{array}
  \right)
\end{equation}
with eigenvalues
\begin{equation} \label{eq:eigenvals_squared_N5}
  \left(2E^2\right)_\pm = r^2\left(1 \pm \sqrt{1 - f^2(\vartheta, \varphi, \chi, \dots)}\right)\,,
\end{equation}
where \(f(\vartheta, \varphi, \dots) = 2 \sqrt{|a|^2|c|^2 + |a|^2|d|^2 + |b|^2|d|^2} / r^2\). 
The four-dimensional integral leads to prefactor $Ac / {\pi}^{4/2} = Ac / \pi^2$ in Eq.~\eqref{eq:dos_general_3dim_with_abbrevs} and
\begin{eqnarray} \label{eq:pr_func_def_4}
  \pr(r) &=& \frac{r^3}{c}\mathrm{e}^{-r^2} \\
  \nr(r) &=& 1 - (1 + r^2)\,\mathrm{e}^{-r^2}
\end{eqnarray}
with $c = 2$ to normalize~\eqref{eq:pr_func_def_4}. 
Choosing the four-dimensional coordinates in the following form
\begin{eqnarray}
  a + \mathrm{i}b &=& r \cos\psi \mathrm{e}^{\mathrm{i}\varphi_1}\\
  c + \mathrm{i}d &=& r \sin\psi \mathrm{e}^{\mathrm{i}\varphi_2}
\end{eqnarray}
leads to $A = 2 \pi^2$ and we obtain~\cite[using formula (6)]{gla76}
\begin{equation} \label{eq:pf_N5_beta1}
  \pf(f) = \frac{8 f}{\pi^2 (1\! +\! f)}
    K\!\left(\sqrt{\frac{1\! -\! f}{1\! +\! f}}\right)
    K\!\left(\sqrt{\frac{2 f}{1\! +\! f}}\right)
\end{equation}
with the limiting behavior
\begin{align}
  \nonumber
  \pf(f) &=
           \frac{30\ln(2) - 1}{3\pi^2}\left(
           3\ln(2) + \ln\frac{1}{f}\right)\cdot f\, +
  \\  \label{eq:pf_N5_beta1_small_f}
         &\qquad\qquad \mathcal{O}\left(\ln\frac{1}{f}\cdot f^2\right).
\end{align}

For completeness, we also state the following results: 
For $N = 5$ and $\beta = 2$, we have $n = 8$, $c=3$ in
\begin{eqnarray} \label{eq:pr_func_def_8}
  \pr(r) &=& \frac{r^7}{c}\mathrm{e}^{-r^2} \\
  \nr(r) &=& \frac{e^{- x^{2}}}{2 \pi^{4}}
              \left(- x^{6} - 3 x^{4} - 6 x^{2} + 6 e^{x^{2}} - 6\right)
\end{eqnarray}
and we get $A = \pi^4 / 3$ and
\begin{align}
  \nonumber
  \pf(f)
  &= 3 f \left(\sqrt{1 - f^2}\ln\frac{f^2}{4(1 - f^2)}\,+ \right.
  \\
  \label{eq:pf_N5_beta2}
  & \quad \left.\rule[0pt]{5ex}{0pt}\ln\frac{1 + \sqrt{1 - f^2}}{1 - \sqrt{1 - f^2}}\right) \\
  \label{eq:pf_N5_beta2_small_f}
  &= 3\left(\ln 2 + \frac{1}{2} + \ln\frac{1}{f}\right)\cdot f^3 + \mathcal{O}\left(f^4\right)\,.
\end{align}
Expressions for the case $N = 4$ and $\beta = 4$ can be done in a similar way and yield
\begin{align} \label{eq:pf_N4_beta4}
  \pf(f) &= 30 f^3\left(\ln\left(\sqrt{1 - f^2} + 1\right) -\right. \nonumber \\
         & \quad\ \left.\rule[0pt]{6ex}{0pt} \ln{(f)} - \sqrt{1 - f^2}\right) \\
  \label{eq:pf_N4_beta4_small_f}
         &= 30\cdot\left(\ln(2) - 1 + \ln\frac{1}{f}\right)\cdot f^3 + \mathcal{O}\left(f^4\right) \\
  \pr(r) &= \frac{1}{\pi^6}r^{11} e^{- r^{2}} \\
  \nr(r) &= \frac{1}{2 \pi^{6}} 
            \left(\rule[-0.6ex]{0cm}{3ex} 120\! -\! \left(r^{10} + 5 r^{8} + 20 r^{6}\,+ \right.\right. \nonumber \\
         & \quad \left.\left.\rule[0pt]{6ex}{0pt}
            60 r^{4} +  120 r^{2} + 120\right)
             \rule[-0.6ex]{0cm}{3ex}
           e^{- r^{2}} \right)\,.
\end{align}
For $N = 5$ and $\beta = 4$ we obtain
\begin{align} \label{eq:pf_N5_beta4}
  \pf(f) &=
            \frac{8!}{2^9}\!\cdot f^7\! \int\limits_{f^2}^{1}\mathrm{d}\beta
            \frac{(1 - \beta)\left( - (2 - \beta)\ln(1 - \beta)- 2\beta\right)}{
            \beta^{7/2}\sqrt{\beta - f^2}} \\ 
         &= \label{eq:pf_N5_beta4_small_f}
            \mathrm{const}\cdot f^7\left(1 + \text{logarithmic corrections}\right) \\
  \pr(r) &= \frac{1}{2520}r^{15} e^{- r^{2}} \\
  \nr(r) &= \frac{1}{2 \pi^{8}}
           \left(\rule[-0.6ex]{0cm}{3ex} 5040 - \left(r^{14} + 7 r^{12} + 42 r^{10}\, + \right.\right. \nonumber \\ 
         & \rule[0pt]{6ex}{0pt} 
           \quad 210 r^{8} + 840 r^{6} + 2520 r^{4} + 5040 r^{2}\,+ \nonumber \\ 
         & \left.\left.\rule[0pt]{6ex}{0pt} \quad 5040\right)e^{- r^{2}}\right).
\end{align}

These expression provide the densities of states and the integrated density of states for the positive values $E_\pm$ via Eqs~\eqref{eq:dos_general_3dim_with_abbrevs} and~\eqref{eq:cdfTPcorrfunc}. 
The overall expressions for the density of states is then given by the sum of the expressions above. 
For $N = 4$ we get
\begin{equation}
  \dos(E) = \frac{1}{4}\left(\rule[0pt]{0ex}{3ex} \dos_{-}(|E|) + \dos_{+}(|E|)\right)
\end{equation}
and for $N = 5$ we have
\begin{equation}
  \dos(E) = \frac{1}{5}\left(
              \rule[0pt]{0ex}{3ex}
              \delta(E) + \dos_{-}(|E|) +
              \dos_{+}(|E|)\right)
\end{equation}
where the \(\delta(E)\) accounts for the symmetry-protected central eigenenergy. 

Like Eq.~\eqref{eq:dos_N3_chGOE_chGUE} in the main text, these results are valid for the specific value of $N$ they were derived for. 
This is due to the fact that the matrices $A$ are sparse for $N > 3$. 

\section{Repulsion behavior for the correlation function \corr{} for \(N=4\) and \(N=5\)}
\label{sec:repulsion_exponents_appendix}

As mentioned in the main text in Sec.~\ref{sec:ensemble-aver-dens}, the analytical results can be used to derive the repulsion behavior. 
For $N=2$ and $N=3$ the results are trivial. 
For $N=4$ and $N=5$ they can be obtained from the analytic expressions of $\corr$ as follows. 

The smallest eigenvalue is given by using the minus sign in Eqs~\eqref{eq:eigenvals_squared_N4} and~\eqref{eq:eigenvals_squared_N5}. 
For small values of the energy, we have
\begin{equation} \label{eq:small_E_formula}
  E_{-} = \frac{r}{\sqrt{2}}\sqrt{1 - \sqrt{1 - f^2}} \approx \frac{r\cdot f}{2}.
\end{equation}
and the density is given by
\begin{align} \label{eq:dos_small_E}
  \dos(E) &= \left\langle\delta\left(E - \frac{r\cdot f}{2}\right)\right\rangle_{r, f} \\
          &= \int\limits_0^\infty\mathrm{d}r\,\, \pr(r)
             \int\limits_0^1\mathrm{d}f\,\, \pf(f)\,
             \delta\left(E - \frac{r\cdot f}{2}\right)\,.
\end{align}
For small $E$ values there are hence two contributions from the delta function to the integral, one from the region of small $r$ of the integrand, the other one from the region of small $f$.
Because of the asymptotic behavior \(\pr(r) \sim r^{n-1}\) the contribution from the small $r$ is negligible, and we are left with
\begin{equation}
  \dos(E) = \int\limits_0^\infty\mathrm{d}r  \frac{2 \pr(r)}{r} \pf\left(\frac{2E}{r}\right)
\end{equation}
The repulsion exponents of \(\dos_{-}(E) \sim E^{\alpha'}\) follow directly from the limiting behavior of \(\pf(f) \sim f^{\alpha'}\). 
The exponent for $\ncorr(E) \sim E^{\alpha' + 1}$ is larger by 1 due to the additional integration. 
These exponents \(\alpha' + 1\) are used to create the dashed lines in Fig.~\ref{fig:rep_ncorr_small_energies} for comparison with the integrated correlations \ncorr. 

\section{Two Resonator Couplings}
\label{sec:two-reson-coupl}

\begin{figure*}
    \mbox{
        \includegraphics[width=.325\textwidth]{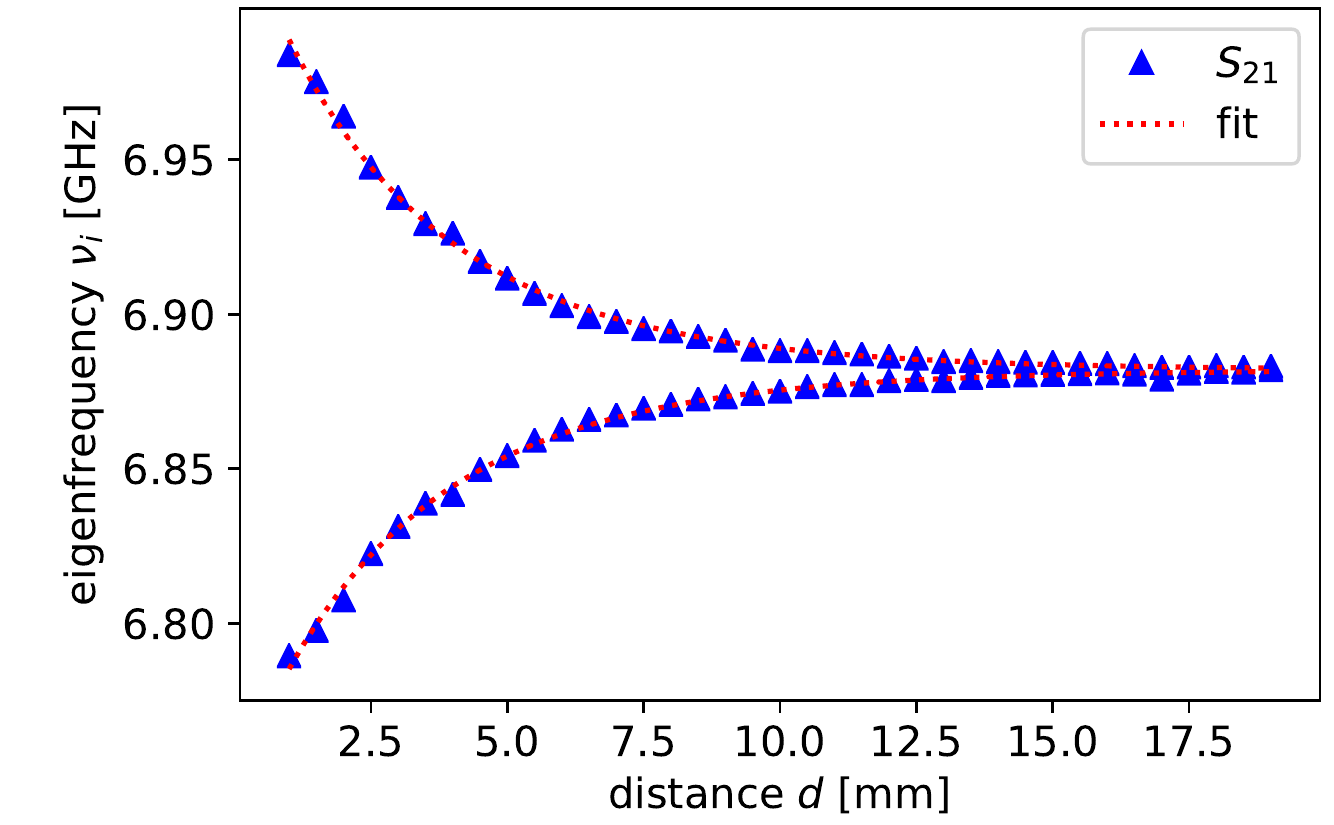}
        \includegraphics[width=.325\textwidth]{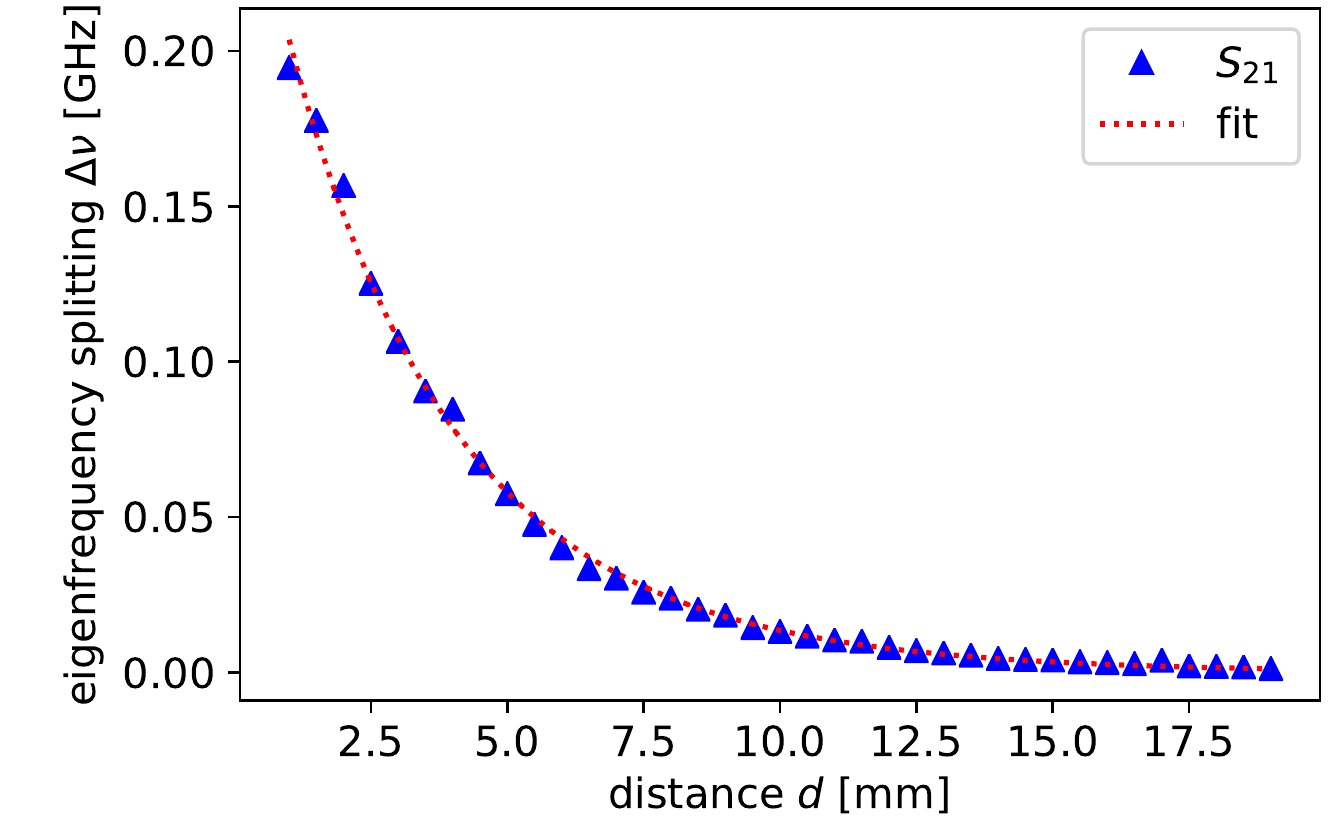}
        \includegraphics[width=.325\textwidth]{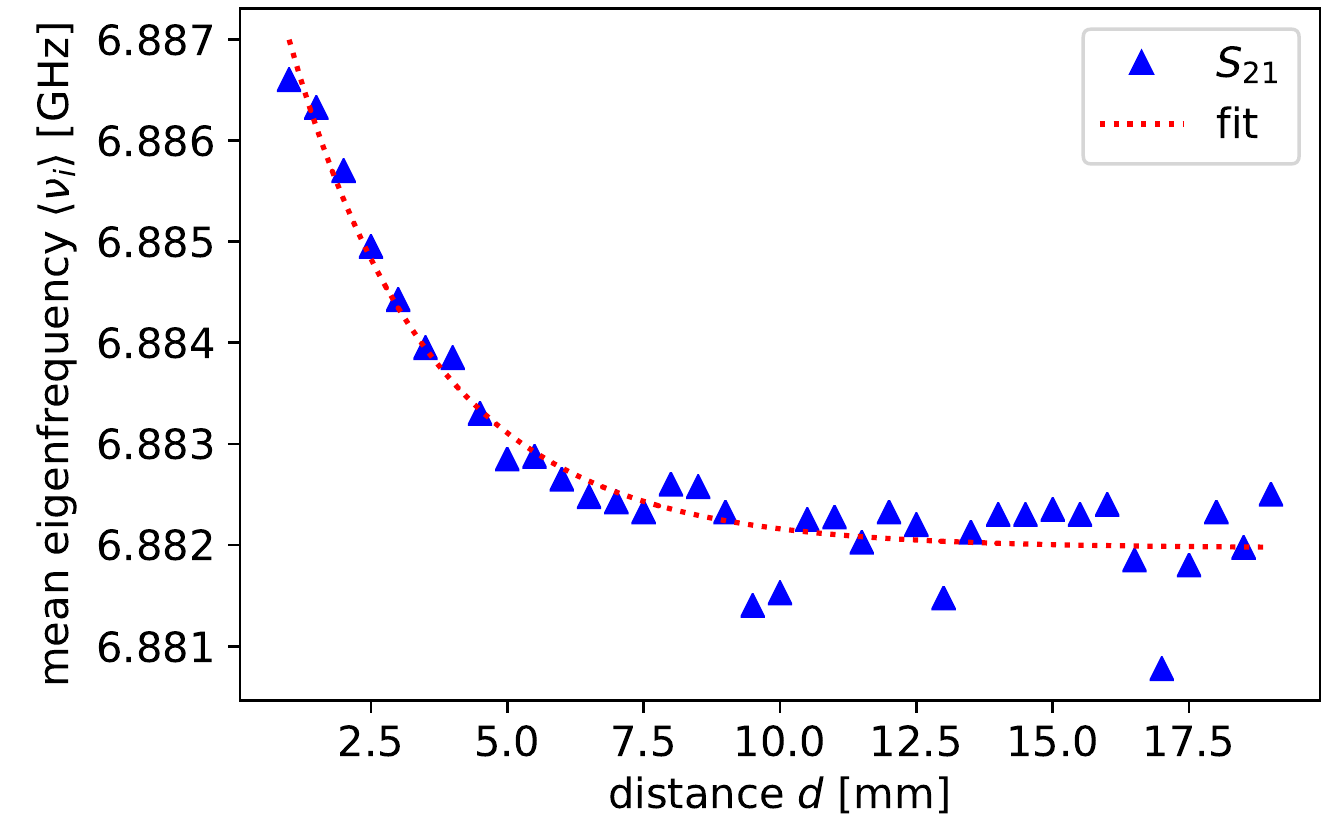}}\\
    \caption{ \label{fig_annex:couplings_chiOE}
        The extracted resonances $\nu_-$ (lower curve) and $\nu_+$ (upper curve), using Lorentzian fits, are shown by blue triangles for the 2-resonator chiOE setup (left). 
        The resonances have been extracted from a transmission ($S_{21}$) measurement. 
        The corresponding splittings $\Delta\nu=\nu_+-\nu_-$ (center) and the mean frequency $\langle\nu\rangle=(\nu_++\nu_-)/2$ (right) are presented. 
        The dotted lines present the fits to Eq~\eqref{eq:fit_func_coupling} (center), Eq.~\eqref{eq_annex:meanfitGOE} (right), and their sum (left). 
        The fit values are given in the text.
    }
\end{figure*}

\begin{figure*}
    \mbox{
        \includegraphics[width=.325\textwidth]{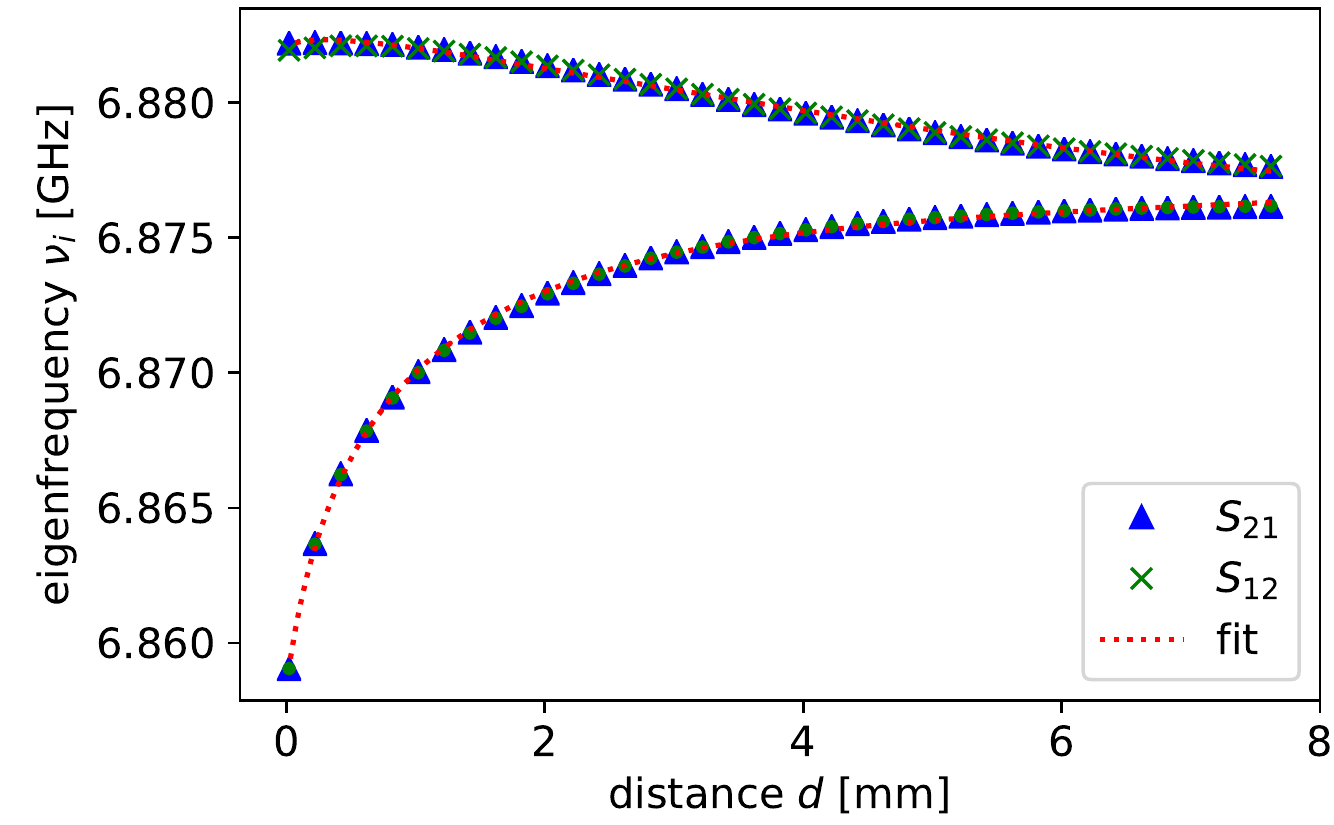}
        \includegraphics[width=.325\textwidth]{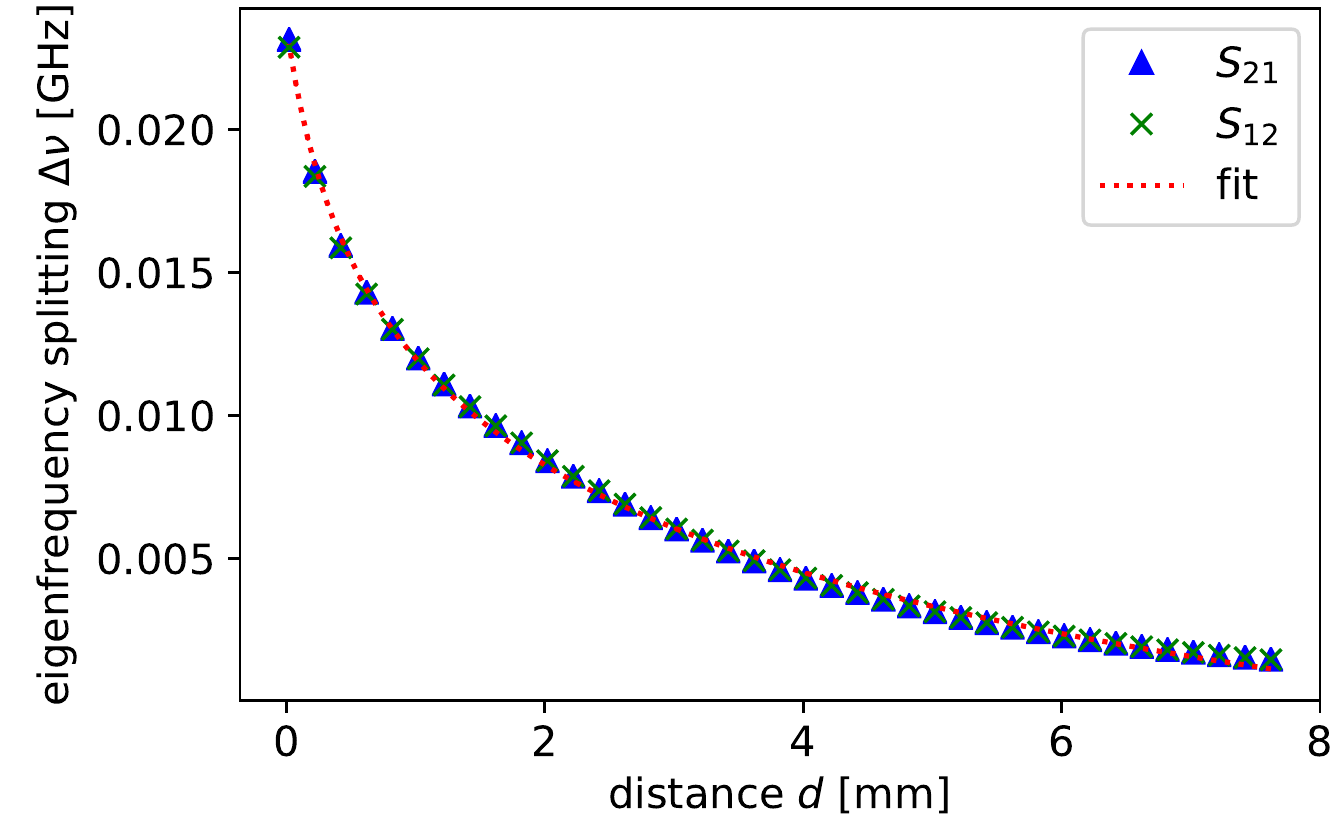}
        \includegraphics[width=.325\textwidth]{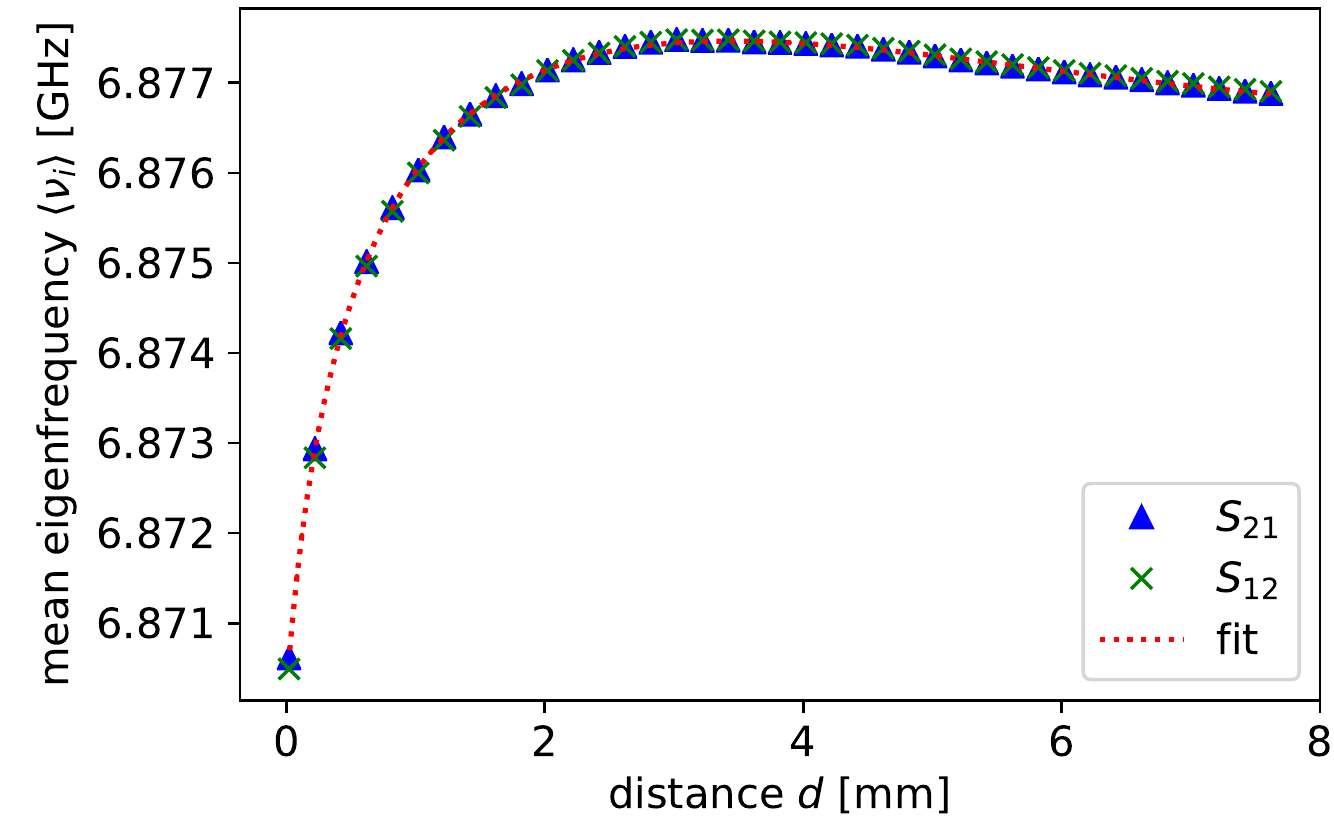}}\\
    \caption{ \label{fig_annex:couplings_chiUE_open}
        As Fig.~\ref{fig_annex:couplings_chiOE} but for the chiral GUE case where an open termination was used to close the third port of the circulator.
        Additionally, eigenvalues extracted from the inverse transmission $S_{12}$ (crosses) are shown.
    }
\end{figure*}

\begin{figure*}
    \mbox{
        \includegraphics[width=.325\textwidth]{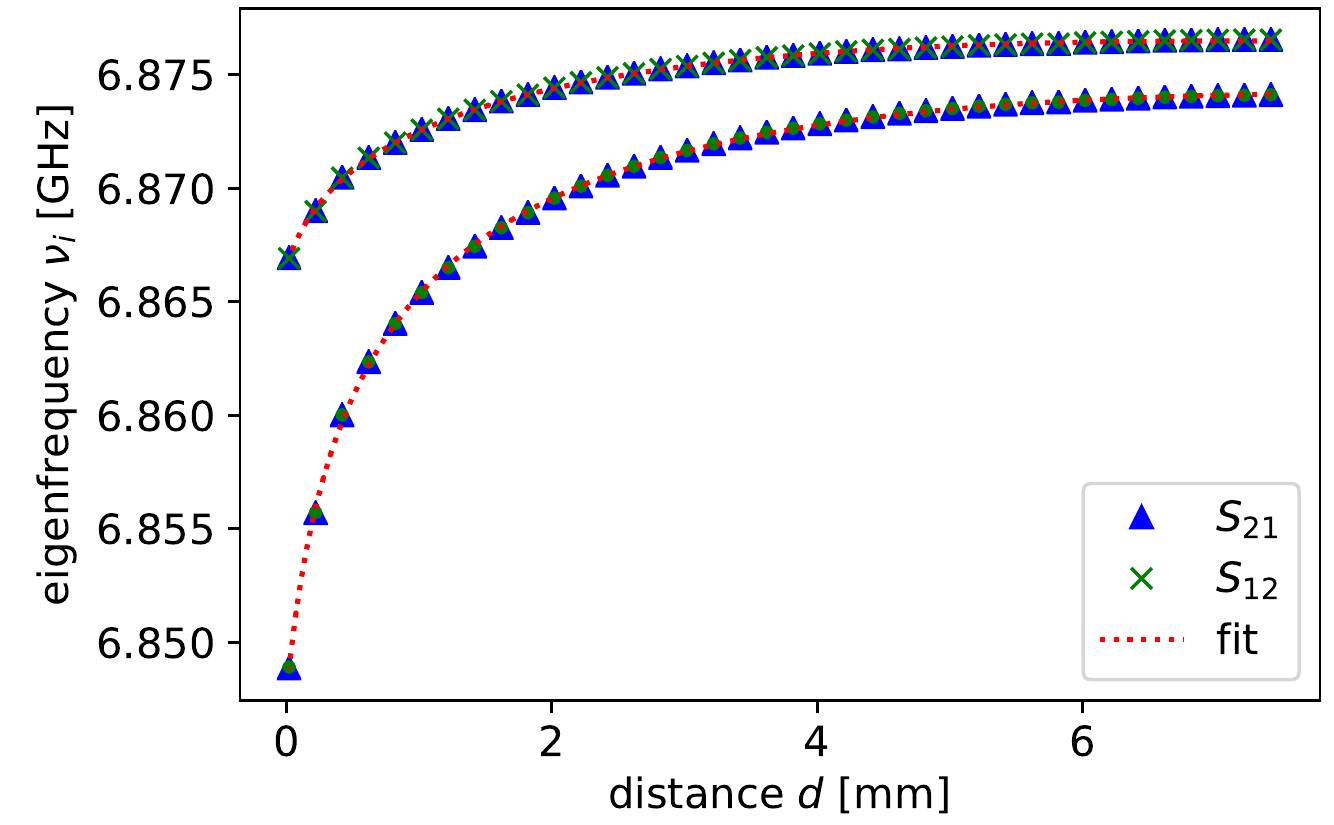}
        \includegraphics[width=.325\textwidth]{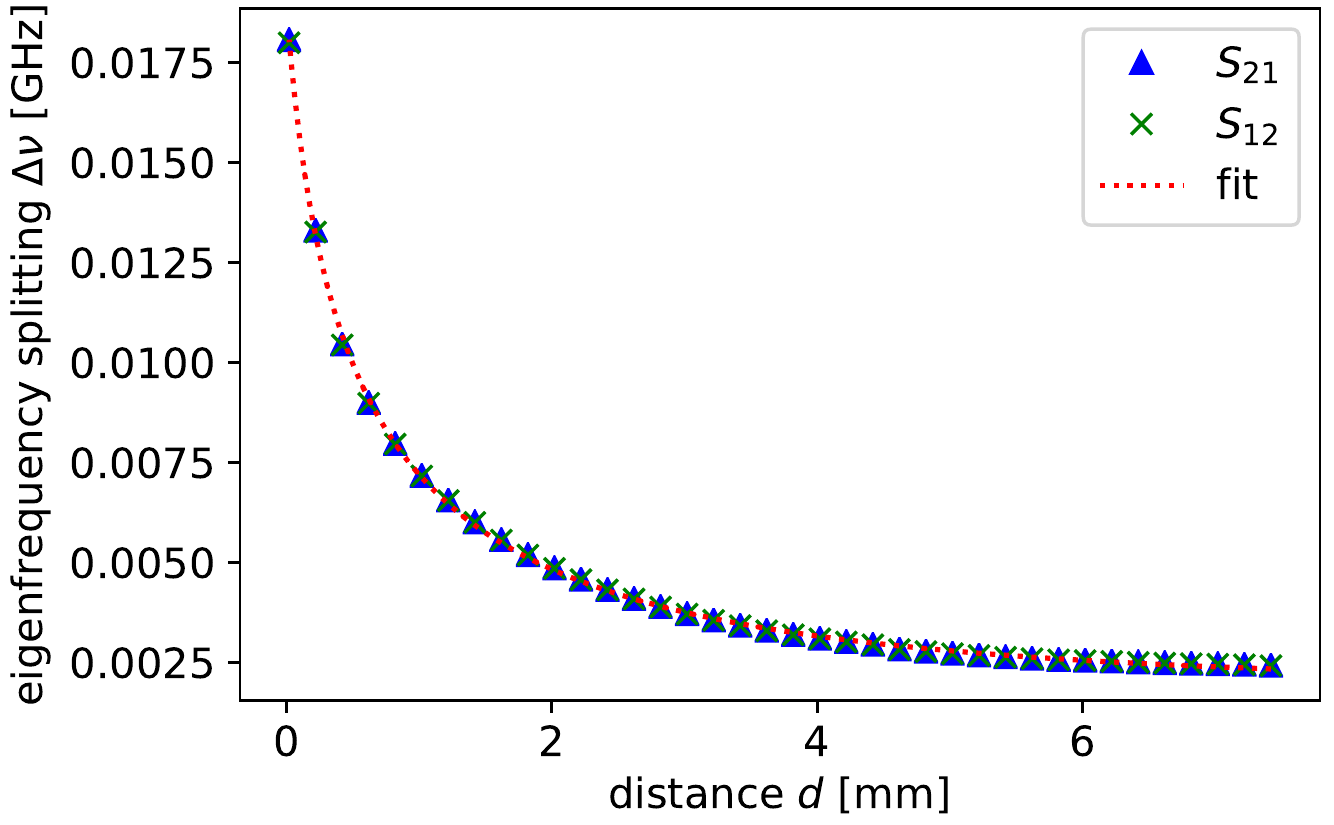}
        \includegraphics[width=.325\textwidth]{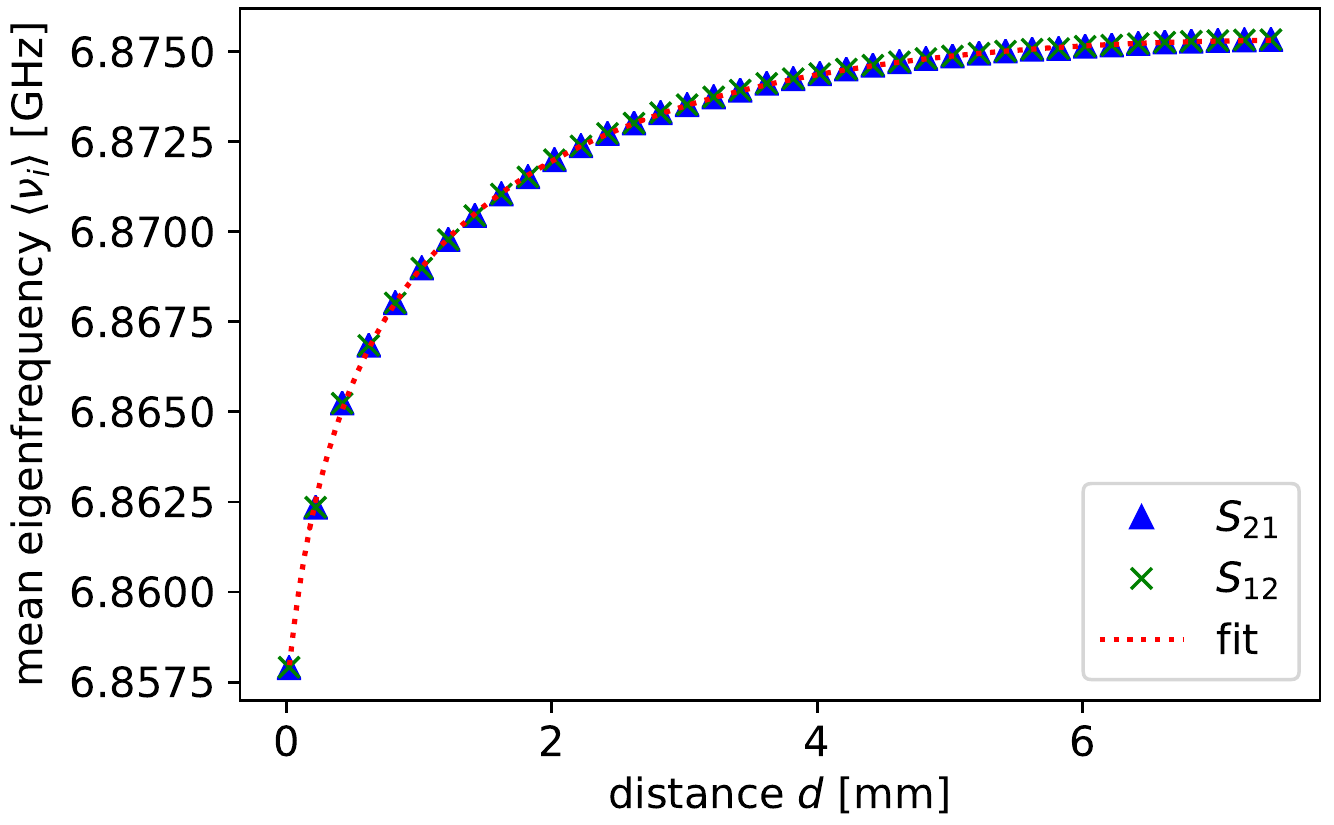}}\\
    \mbox{
        \includegraphics[width=.325\textwidth]{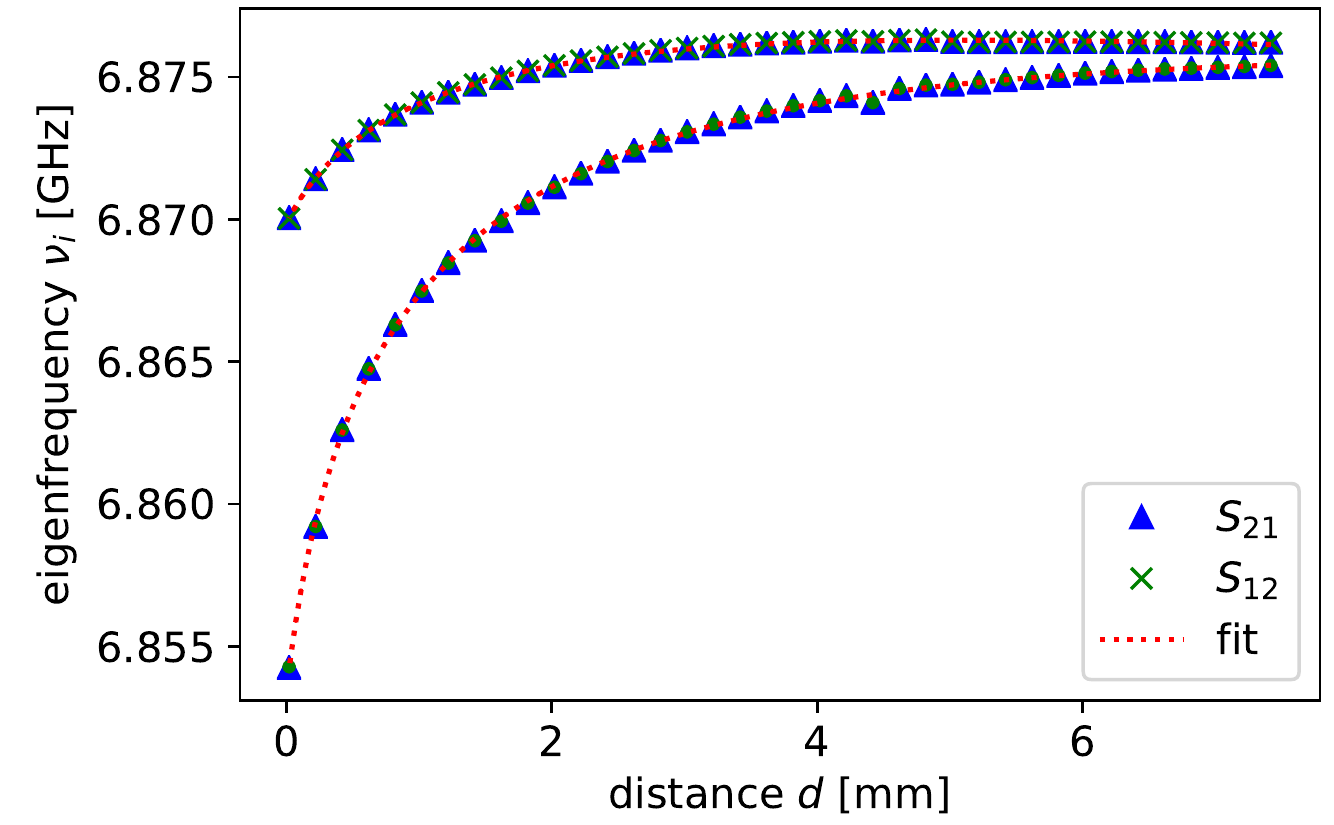}
        \includegraphics[width=.325\textwidth]{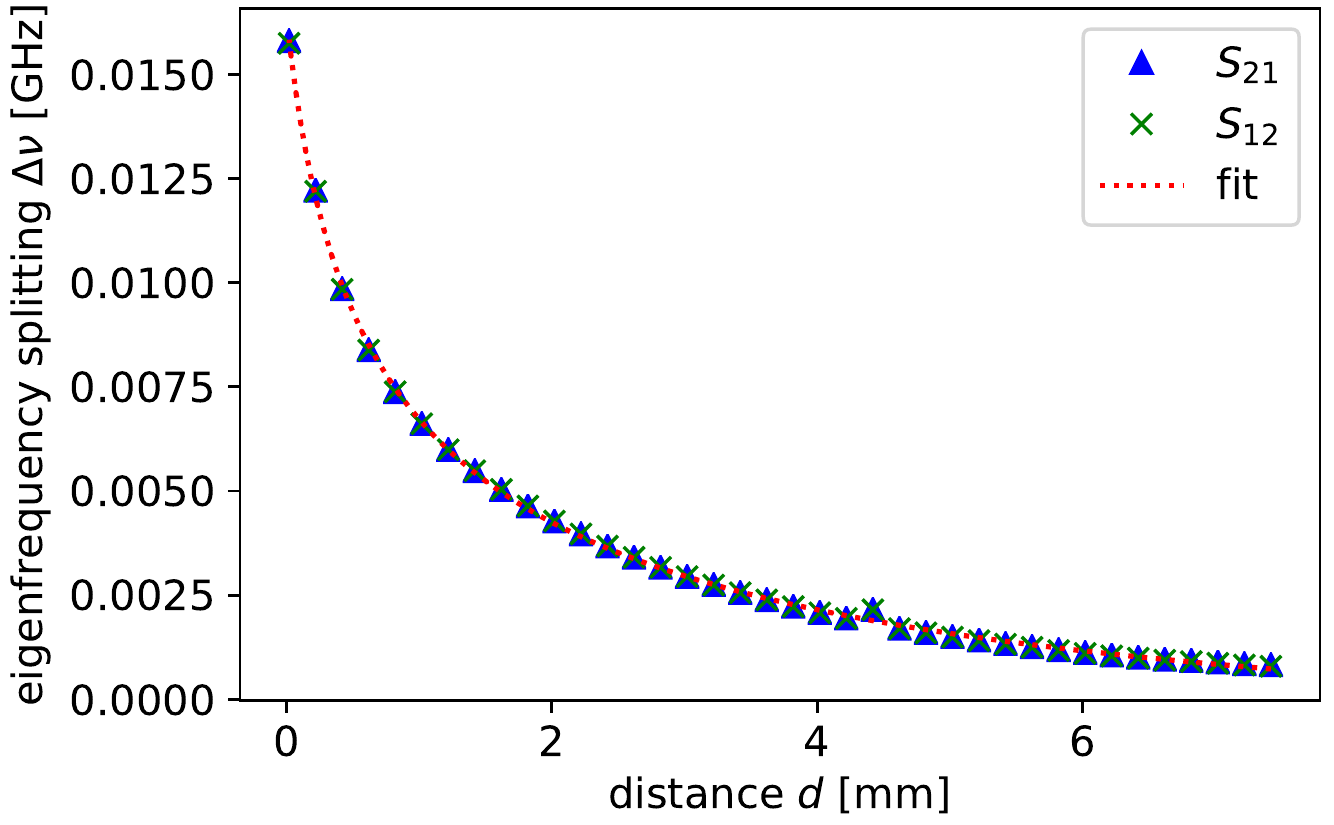}
        \includegraphics[width=.325\textwidth]{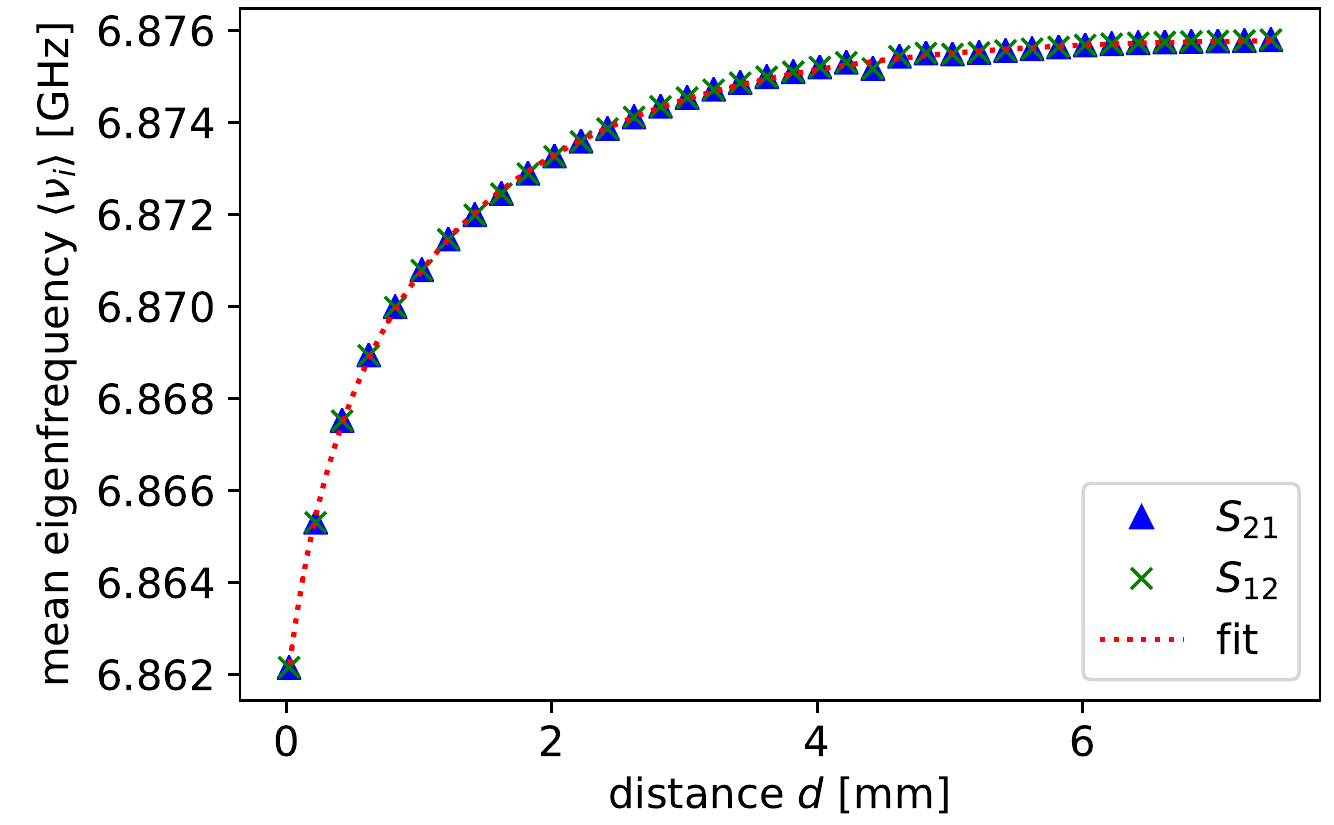}}\\
    \mbox{
        \includegraphics[width=.325\textwidth]{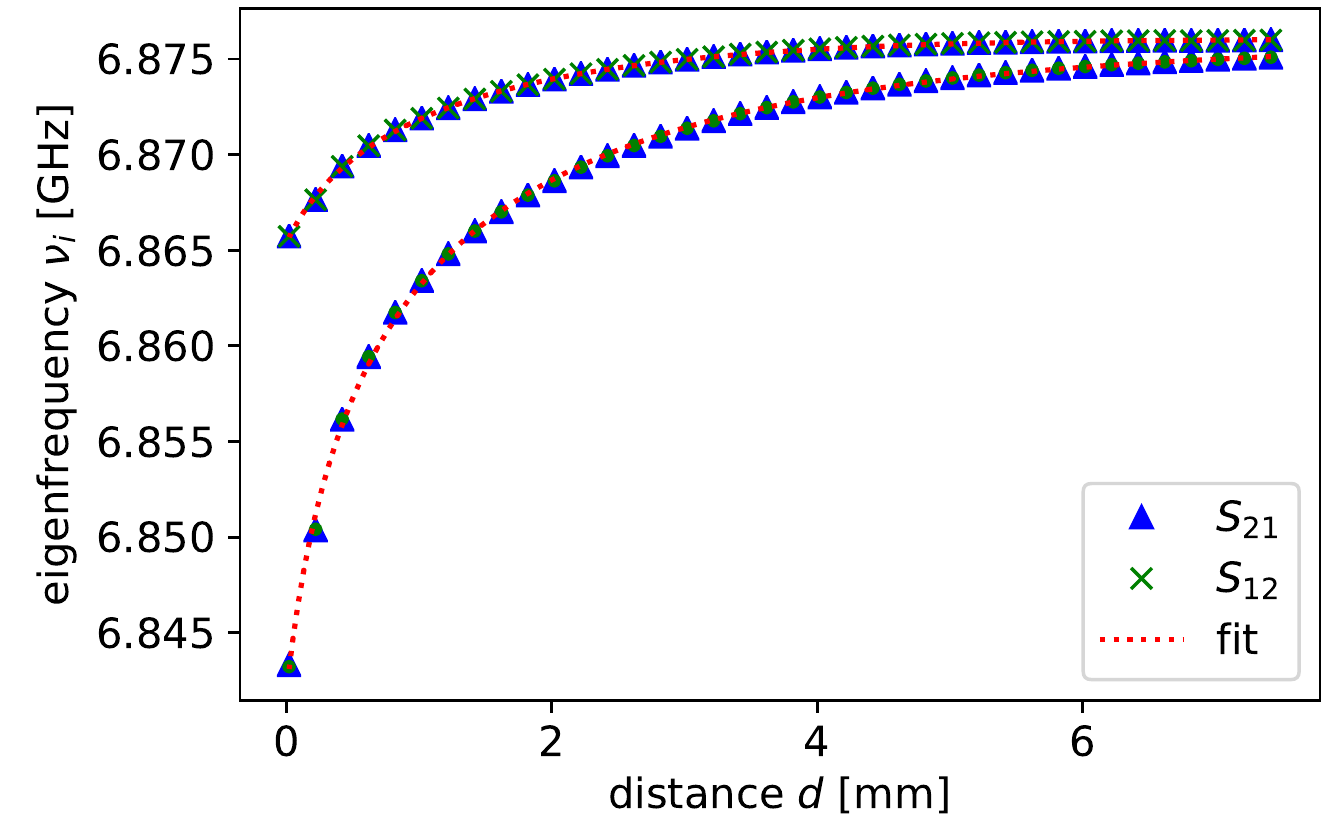}
        \includegraphics[width=.325\textwidth]{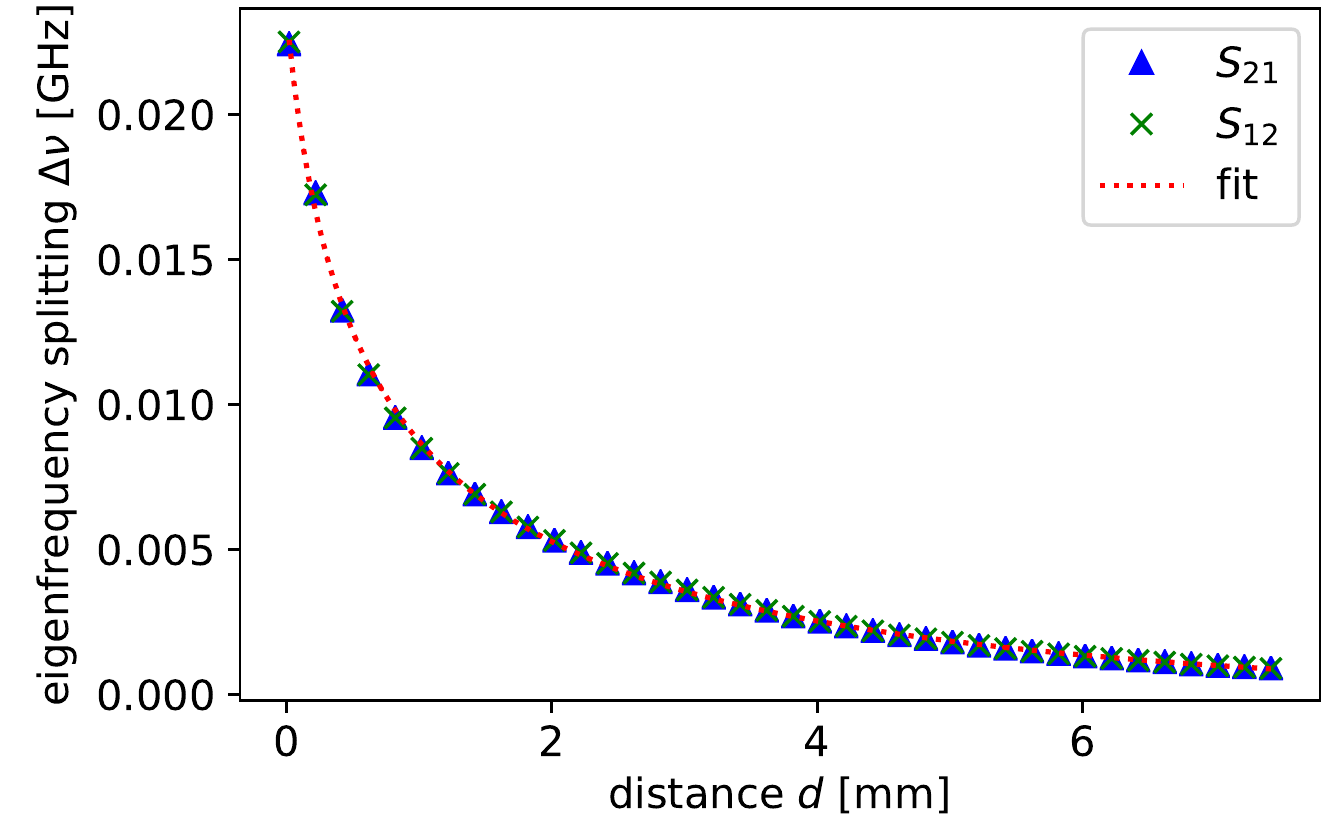}
        \includegraphics[width=.325\textwidth]{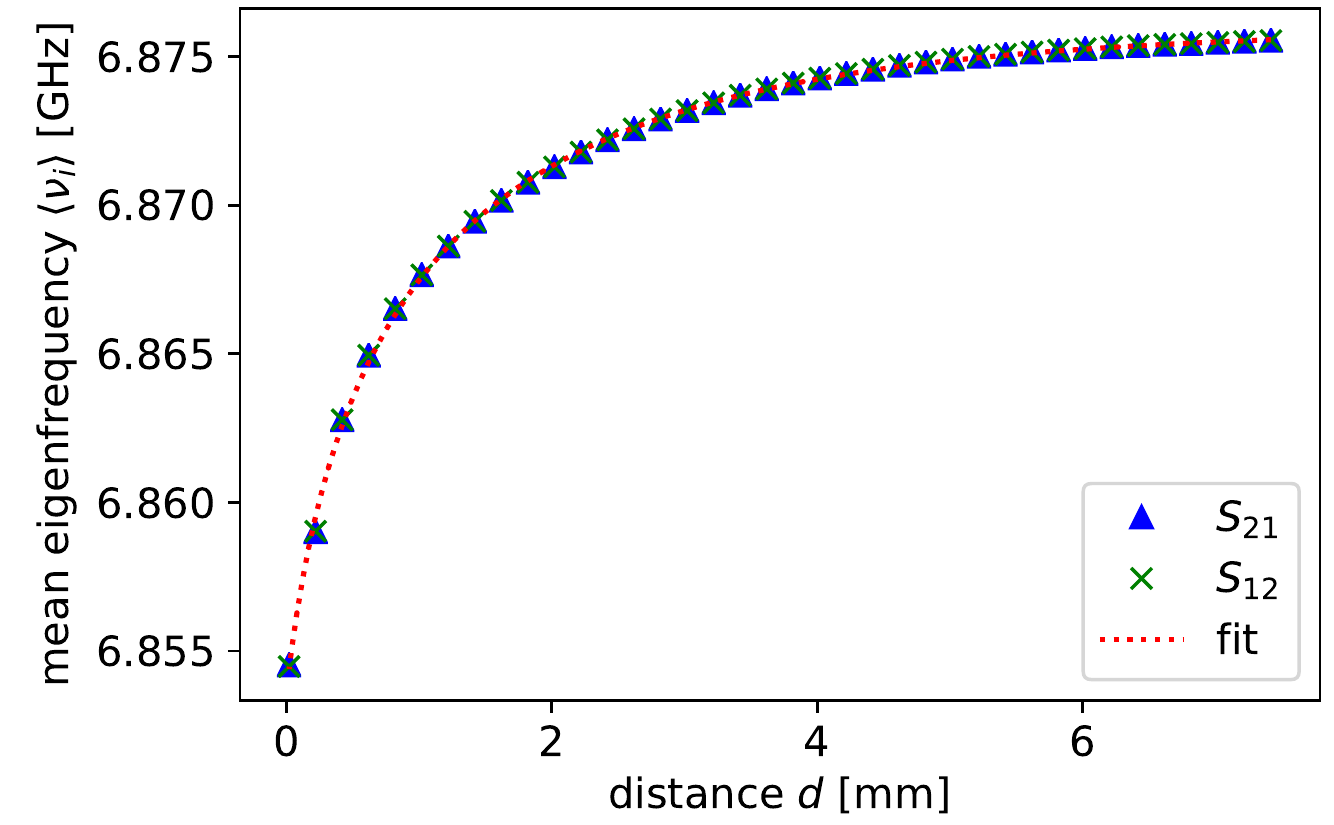}}\\
    \caption{ \label{fig_annex:couplings_chiUE_rest}
        As Fig.~\ref{fig_annex:couplings_chiOE} where instead of an open termination three different terminations corresponding to $\varphi_T = \pi$ (short), $\pi/2$, and $3\pi/2$ were used. 
    }
\end{figure*}

When considering the ensemble averages for the different ensembles, the couplings must be put in relation to a parameter one can control. 
In the case of the chiOE, this is the distance between the resonators and in case of the chiUE and chSE, it is the distance between the resonators and the two monopole antennas which are attached to the circulators since those mediate the coupling (see Fig.~\ref{fig:chiUE}). 
Furthermore, to get different complex couplings different phase shifts can be introduced to the circulators in this case. 

In this appendix, we want to present how we extract the coupling-distance relationship for the chiOE and chiUE case (the chiSE case is done analogously, see Sec.~\ref{sec:chir-symp-ensemble}). 
The results for the chiUE coupling experiment lead to the decision to only use the open termination for all circulators to avoid the effects of on-site shifts discussed in Appendix~\ref{sec:simul-pert-chir}. 

For the chiOE case, two resonators were set up similar to the inset in Fig.~\ref{fig:schematic_chain}~(bottom). 
For different distances $d$ between the resonator borders the transmission $S_{21}$ has been measured and the two resonance frequencies $\nu_-$ and $\nu_+$ have been extracted. 
The corresponding eigenfrequencies (left), their splitting $\Delta\nu$ (center), and their mean $\langle\nu\rangle$ are shown in Fig.~\ref{fig_annex:couplings_chiOE}. 
The two resonance frequencies \(\nu_\pm\) for every distance $d$ can be used to extract the coupling \(a\) via Eq.~\eqref{eq:definition_coupling_from_splitting}. 
The dashed lines in Fig.~\ref{fig_annex:couplings_chiOE} shows a fit to the experimental data based on Eq.~\eqref{eq:fit_func_coupling} for the splitting with $a_0=1.32$\,GHz and $\gamma=0.233$\,mm$^{-1}$. 
The mean frequency shift $\langle \nu\rangle=(\nu_+ - \nu_-)/2$ in the chOE case can be approximated by
\begin{equation}\label{eq_annex:meanfitGOE}
\langle \nu^{OE}\rangle=\nu_0+a_m K_0\left[\gamma_m(2 r_D + d)\right]\,,
\end{equation}
as it is related to the wavefunction at the position of the second resonator. 
Fitting this equation to the data gives $\nu_0=6.882$\,GHz, $a_m=0.137$\,GHz and $\gamma_m=0.327$\,mm$^{-1}$. 
As $\gamma_m$ and $\gamma$ describe the decay of the wavefunction outside the resonator they should be approximately the same. 
Note, that the $K_0$ dependence is an effective description as in the experiment the air gap above the resonators leads to excitation of higher modes outside the resonator (for details see \cite{bel13b}) provoking a difference in $\gamma$ and $\gamma_m$.

To check the assumption of constant \(\nu_0\), i.e., no on-site detuning, we take a look on the right hand side plot of Fig.~\ref{fig_annex:couplings_chiOE} showing the center of mass of the two resonances on the left.
While the scale on this plot of the shift is much smaller than the scale for the extracted resonances, one can still see, that there is a detuning present for small distances. 
Instead of being constant, the mean of the extracted eigenfrequencies varies of the order of $\pm 2\,\mathrm{MHz}$. 
This detuning violates the chiral symmetry as mentioned in the main text. 
However, it is sufficiently small to be negligible in the chiOE case against the splitting $\Delta\nu$, typically larger than 10\,MHz. 
For this reason Fig.~\ref{fig:chiGOE_hist_TP} shows a good agreement for the density of states between experiment and theory. 

In case of the chiUE experiment the description using Eqs~\eqref{eq:fit_func_coupling} and~\eqref{eq_annex:meanfitGOE} is not valid anymore as the coupling is mediated by the monopole antennas, the circulator and the closing of the third port of the circulator.
The phases acquired by the wave going through the circulator in the two directions and the distortion of the electromagnetic field by the presence of the metallic case of the circulator lead to additional dependencies of the mean frequency and the splitting.
In lack of any theoretical model we assume the splitting $\Delta\nu$ can be described by
\begin{equation}\label{eq_annex:deltafitGUE}
\Delta\nu(d)= c_1 \left|K_0\left(c_2[d-c_4]\right)\right|^2 + c_3
\end{equation}
to account for the uncertainties regarding the distances and for the mean frequency $\langle\nu\rangle$ an additional linear and quadratic term were added
\begin{equation}\label{eq_annex:meanfitGUE}
\langle\nu\rangle(d)=a_1 K_0\left(a_2[d-a_4]\right) + a_3 + a_5 (d-a_4)+a_6(\nu-a_4)^2\,.
\end{equation}
As can be seen in Fig.~\ref{fig_annex:couplings_chiUE_open} and Fig.~\ref{fig_annex:couplings_chiUE_rest} these dependencies are sufficient to describe our experimental data and the difference between the fit and the experimentally measured points is smaller than the symbol size for the data. 
For completeness the extracted fit values of the frequency splittings and the mean frequencies are given in Tab.~\ref{tab_annex:resosplit} and Tab.~\ref{tab_annex:resomean}, correspondingly for the different terminators. 

\begin{table}
    \begin{tabular}{|c|c|c|c|c|}\hline
        \parbox{2.5cm}{\vspace*{7pt}Terminator\vspace*{7pt}} & \mbox{$\displaystyle\frac{c_1}{\mathrm{GHz}}$} & $\displaystyle\frac{c_2}{\mathrm{mm}^{-1}}$ & $\displaystyle\frac{c_3}{\mathrm{GHz}}$ & $\displaystyle\frac{c_4}{\mathrm{mm}}$ \\\hline
        Open ($\varphi_T=0$)    & 0.0003 &  0.000 & -0.0203 &  0.23 \\\hline
        Short ($\varphi_T=\pi$) & 0.0010 &  0.099 &  0.0020 &  0.19 \\\hline
        $\varphi_T=\pi/2$       & 0.0007 &  0.033 & -0.0009 &  0.23 \\\hline
        $\varphi_T=3\pi/2$      & 0.0012 &  0.057 & -0.0004 &  0.21 \\\hline
    \end{tabular}
    \caption{ \label{tab_annex:resosplit}
        Fit values for the resonance splitting $\Delta\nu$ obtained by Eq.~\eqref{eq_annex:deltafitGUE} for the chiUE case for different termination of the third port of the circulator. 
    }
\end{table}

\begin{table}
    \begin{tabular}{|c|c|c|c|c|c|c|}\hline
        \parbox{2.5cm}{\vspace*{7pt}Terminator\vspace*{7pt}} & $\displaystyle\frac{a_1}{\mathrm{GHz}}$ & $\displaystyle\frac{a_2}{\mathrm{mm}^{-1}}$ & $\displaystyle\frac{a_3}{\mathrm{GHz}}$ & $\displaystyle\frac{a_4}{\mathrm{mm}}$ & $\displaystyle\frac{a_5}{\mathrm{m}^{-1}}$ & $\displaystyle\frac{a_5}{\mathrm{m}^{-2}}$\\\hline
        Open ($\varphi_T=0$)     & -0.0037 &  0.246 &  6.882 &  0.18 & -1.0 &  46 \\\hline
        Short ($\varphi_T=\pi$)  & -0.0062 &  0.033 &  6.891 &  0.15 & -0.9 &  9 \\\hline
        $\varphi_T=\pi/2$       & -0.0061 &  0.112 &  6.885 &  0.24 & -1.0 &  34 \\\hline
        $\varphi_T=3\pi/2$      & -0.0091 &  0.002 &  6.927 &  0.22 & -1.8 &  50 \\\hline
    \end{tabular}
    \caption{ \label{tab_annex:resomean}
        As Tab.~\ref{tab_annex:resosplit} but for the mean resonance $\langle\nu\rangle$ obtained by Eq.~\eqref{eq_annex:meanfitGUE}. 
        Note that $a_5$ and $a_6$ are given in meter scales. 
    }
\end{table}

In Fig.~\ref{fig_annex:couplings_chiUE_open}, we show the results for open-end circulators, i.e., the setup used to generate the histograms in Fig.~\ref{fig:chiGUE_hist}. 
The left plot shows the two extracted resonances $\nu_\pm$. 
As one can already observe here the behavior of these two resonances  is far from being symmetric especially for distances $d$ less than 2\,mm. 
From this figure we observe that for the whole distance range the variation of the mean frequency (around 4\,MHz) is of the order of the splitting (10\,MHz). 
Taking only distances into account for $d\ge 2$\,mm the mean fluctuates less than 0.5\,MHz compared to splittings of the order of 5\,MHz. 
Therefore we decided to adjust the distribution of couplings taking into account only distances $d\ge 2$\,mm (see discussion in the main text). 

For completeness we want to address the variation of the phases of the couplings $a=|a| e^{\mathrm{i}\varphi}$. 
Because the eigenvalues do not depend on $\varphi$ the resulting histograms in Fig.~\ref{fig:chiGUE_hist} should be unchanged when different phases $\varphi$ are used. 
For this we consider now different terminators for the free port of the circulator as mentioned in Sec.~\ref{sec:chir-unit-ensemble}. 
In a preliminary investigation, the value of $\varphi_T$ was restricted to three more choices in addition to the already mentioned open-end termination (giving $\varphi_T=0$ upon reflection): 
standard short terminators were used to realize phases shifts of $\pi$ upon reflection. 
It is important to note that the phase of the terminator reflection $\varphi_T$ equals $\varphi$ only approximately as some additional lengths are introduced by the open and short terminator. 
However, these additional lengths are small compared to the wavelength used in our setup. 
In addition, phase shifts of $\pi/2$, and $3\pi/2$ were realized by introducing short cables of appropriate lengths between the circulator exit and the terminator.
For these three additional terminations, Fig.~\ref{fig_annex:couplings_chiUE_rest} shows the 2-resonator couplings and extracted fits. 
We never observe a reasonably symmetric coupling behavior for any distance range but both of the resonances shift always to larger frequencies with increasing distance.
Additionally, the shift has increased compared to the open termination roughly by a factor of 3 whereas the splitting is similar or even smaller.

To avoid problems due to on-site variations we therefore only used the open-end setup for the histogram in the chiUE case. 
Note, again, that a variation of the phase is not really needed since the eigenvalues of the Hamiltonian depend on the modulus of the coupling constants only. 

\section{Simulation of the perturbation of the chiral symmetry}
\label{sec:simul-pert-chir}

The circulators used to break time-reversal symmetry disturb the chiral symmetry. 
This is visible in the densities of states in the right-hand side of Figs.~\ref{fig:chiGUE_hist} and~\ref{fig:chiGSE}. 
These perturbations result in spectra which do not follow the analytical predictions from Sec.~\ref{sec:ensemble-aver-dens}. 
In order to account for this, we also investigate perturbed systems theoretically. 
In particular, we show that perturbations of the model Hamiltonians will result in curves similar to the perturbed experimental data shown later in Sec.~\ref{sec:chir-unit-ensemble}. 
In order to achieve this, we calculate spectra numerically and show that the deviations from the expected formulas from Sec.~\ref{sec:ensemble-aver-dens} can be understood this way. 
The perturbation can occur in two ways: by an onsite detuning or a next-to-nearest neighbor coupling. 
Both introduce non-zero elements in the diagonal blocks of $H$, see Eq.~\eqref{eq:chiham}. 

The coupling between the disks does not only split their eigenfrequencies according to Eq.~(\ref{eq:split_eigenfreqs}) but also leads to a detuning of their unperturbed eigenfrequencies. 
This is already evident for Fig.~\ref{fig:2disks} showing the splitting of the eigenfrequencies of two coupled disk in dependence of the distance. 
This splitting is not symmetrical about the disk eigenfrequencies. 
There is an additional shift of the center of gravity of the two resonances shown as red dashed line in Fig.~\ref{fig:2disks}. 
This effect is moderate if the disks are coupled directly, as for the chiOE studies, but it becomes more pronounced for the chiUE studies where the coupling is accomplished by means of circulators.

For a better qualitative understanding of these findings we performed a series of simulations taking into account the eigenfrequency detuning. 
There is an obvious correlation between the coupling between two disks and the detuning of the eigenfrequencies of the involved disk. 
To describe the effect we applied the most simple assumption, namely that a disk experiences a shift of its eigenfrequency proportional to the coupling. 
For the example of $N=4$ we then obtain a Hamiltonian 
\begin{equation} \label{eq:ham4a}
  H=\left(
    \begin{array}{cccc}
      \varepsilon |a| & \cdot & a & \cdot \\
      \cdot & \varepsilon (|b| + |c|) & b^* & c \\
      a^* & b & \varepsilon (|a|+|b|) & \cdot \\
      \cdot & c^* & \cdot & \varepsilon |c|
    \end{array}
  \right)\,,
\end{equation}
where $\varepsilon$ is a free parameter representing the strength of the onsite detuning. 

\begin{figure}
	\centering
	\includegraphics[width=\linewidth]{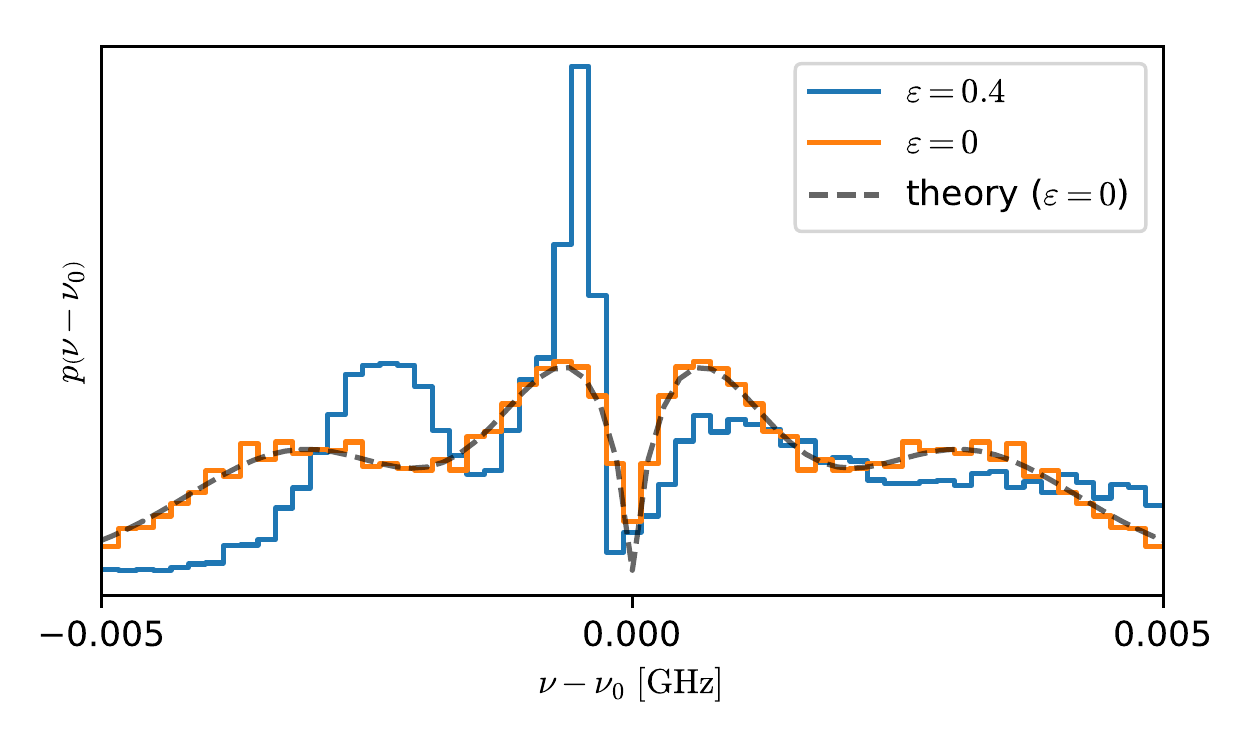}
	\caption{ \label{fig:asym_GUE}
		Histogram showing the influence of onsite detuning proportional to the coupling. 
	}
\end{figure}

In order to check the influence of this parameter we calculated an example spectrum based on 2000 random realizations of the Hamiltonian~\eqref{eq:ham4a}. 
For this the off-diagonal elements were chosen from a complex Gaussian distribution thereby creating a chiral unitary ensemble. 

The density of states from this calculation is shown in Fig.~\ref{fig:asym_GUE}. 
The orange histogram is a numerical calculation for $\varepsilon = 0$ and corresponds to the case with unbroken chiral symmetry, Eq.~\eqref{eq:ham4}. 
It is predicted by the theoretical curve for the density of states, Eq.~\eqref{eq:dos_general_3dim_with_abbrevs}. 
For $\epsilon=0.4$ (blue histogram) the obtained distribution resembles the one found in the experiment, see the bottom left panel of Fig.~\ref{fig:chiGUE_hist}, showing a clear asymmetric behavior. 
This supports the assumption that indeed a detuning of the eigenfrequencies is responsible for the experimental deviations from the theoretical curves.

While this explains the detuning, there is a better way to compare the experimental data with theoretical predictions by means of correlation functions, see Eqs~\eqref{eq:ncorr_func_Nthree} and~\eqref{eq:cdfTPcorrfunc}. 
This achieves that any shift of the spectra between realizations drops out of the comparison. 

\section{Testing the symplectic symmetry}
\label{sec:testing-sympl-sym}

\begin{figure}
	\begin{center}
		\includegraphics[width=\linewidth]{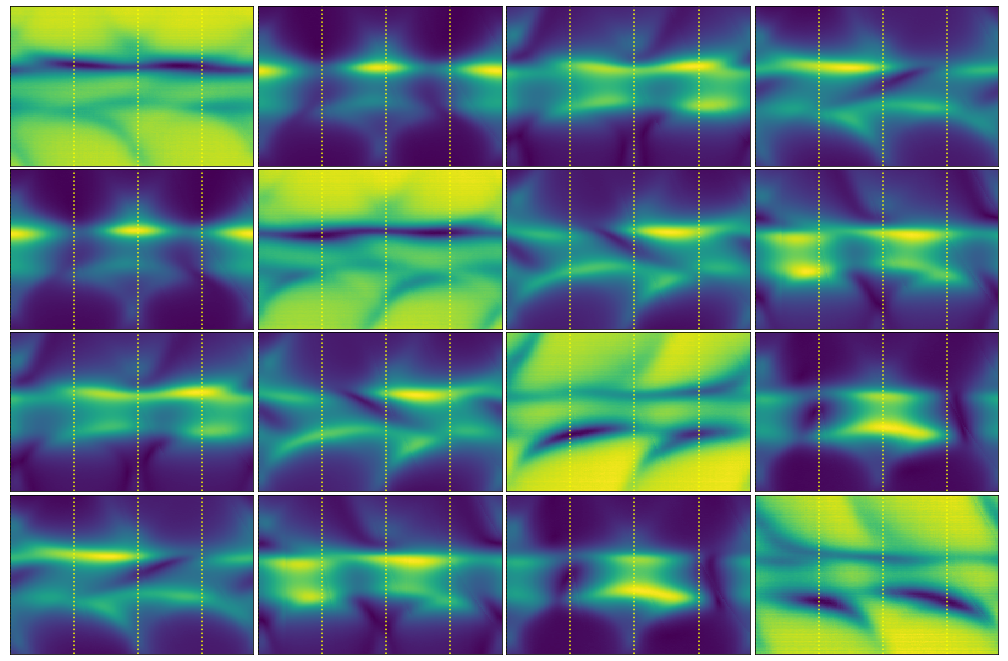}
		\caption{ \label{fig:chSE-test-allSpar}
			The different subfigures from top left to bottom right, read from the left to the right, correspond to the modulus of different measured S-matrix elements, $|S_{11}|$, $|S_{1\bar{1}}|$, $|S_{12}|$, $|S_{1\bar{2}}|$, $|S_{\bar{1}1}|$, $\dots$, $|S_{\bar{2}\bar{2}}|$. 
			The abscissa corresponds to the phase change \(\Delta \phi\) induced by a variation of length using a phase shifter in one of the connecting arms whereas the ordinate shows the frequency range between 6.835 and 6.882\,GHz.
			Color scale is adjusted for each subfigures where dark blue is the minimal and light green the maximal value.
			The dotted vertical lines emphasize the $\pi$, $2\pi$, and $3\pi$ phases.
		}
	\end{center}
\end{figure}

To verify that the symplectic symmetry is present we have performed a test measurement where instead of the cable $L_1$ an additional phaseshifter was attached thus the length of $L_1$ can be changed. 
200 measurements for different lengths of $L_1$ have been taken. 
This approach is similar to Ref.~\cite{reh18} where a graph of microwave cables was used to create the symplectic symmetry. 
In Fig.~\ref{fig:chSE-test-allSpar} the modulus of all four $S$-parameter are shown as a function of the phase difference $\Delta \phi$ (abscissa) and frequency (ordinate). 
Dark blue corresponds to the minimal value and bright yellow to the maximal value. 
Each subfigure is scaled individually. 
The figure shows that reflection (diagonal) and transmission (off-diagonal) behave differently as expected. 
Besides this, there is a $1$-$\bar{1}$ symmetry, i.e., $S_{1\bar{1}} = S_{\bar{1}1}$ etc. 
This is due to the time-reversal symmetry of the system. 
Furthermore, the elements of $S$ are roughly \(2\pi\) periodic, as suggested by the repeated bright spots in the transmissions or the dark spots in the reflections. 

For the phase differences $\Delta\phi=\pi$ and $3\pi$ the system should show a symplectic symmetry which we will detail below. 
At this this phase differences the four $|S_{11}|$, $|S_{1\bar{1}}|$ panels, $|S_{\bar{1}1}|$, $|S_{\bar{1}\bar{1}}|$ should be interpreted as a reflection quarternion and $2\times2$ subplot $|S_{12}|$, $|S_{1\bar{2}}|$, $|S_{\bar{1}2}|$, $|S_{\bar{1}\bar{2}}|$ as transmission quarternion. 
The same holds for the two other subplots correspondingly. 

\begin{figure}
    \begin{center}
        \includegraphics[width=\linewidth]{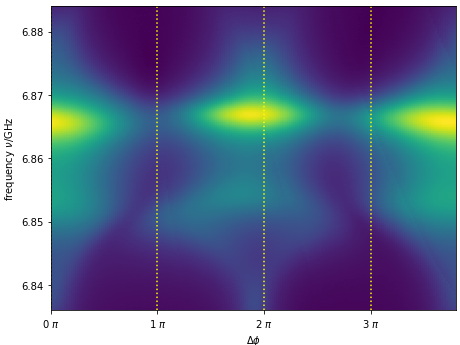}
        \includegraphics[width=\linewidth]{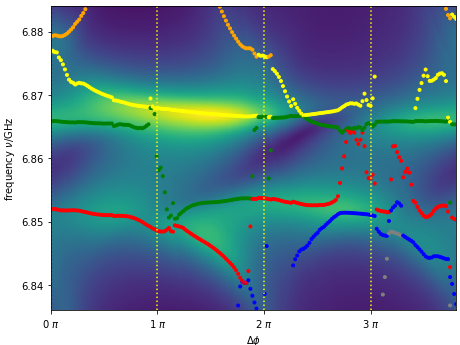}
    \end{center}
    \caption{\label{fig:chSE-test-trans}
        Two different transmission measurements for $|S_{1\bar{1}}|$ (top) and $|S_{1\bar{2}}|$ (bottom) as a function of phase difference between the two couplings and frequency.
        The vertical dotted lines correspond to a phase difference of $1\pi$, $2\pi$, and $3\pi$.
        They are enlargements from Fig.~\ref{fig:chSE-test-allSpar}.
        Additionally, the circles show the position of extracted resonances.
        The color scales have been adjusted individually, where bright yellow corresponds to the maximal value (top: 0.068, bottom: 0.053) and dark blue to 0.
    }
\end{figure}

In Fig.~\ref{fig:chSE-test-trans} two transmission $S_{1\bar{1}}$ (top) and $S_{1\bar{2}}$ (bottom) are shown enlarged. 
Additionally, vertical dotted lines indicate the phase-differences corresponding to multiples of $\pi$. 
In the upper graph of Fig.~\ref{fig:chSE-test-trans} shows $S_{1\bar{1}}$, i.e., a transmission to the "symmetric point" in the other subsystem. 
In the presence of a symplectic symmetry the transmission $S_{1\bar{1}}$ should vanish, a direct consequence of the quaternionic structure of the scattering matrix , see Refs~\cite{reh16,reh18} for details. 
This can indeed be seen in the reduced transmission close to the $1\pi$ and $3\pi$ lines. 
As one can see, the minimal transmission around 6.85 and 6.87\,GHz occur at slightly different $\Delta \phi$ values which comes from the fact that $\Delta \phi$ was determined using the two electrical lengths at the fixed eigenfrequency of a single resonator at
6.86,\,GHz and we did not take into account the phase variation induced by the frequency dependence. 
Also the minimal value $|S_{1\bar{1}}|^2\approx 10^{-4}$ (compared to a maximal transmission of $\approx 2.5^{-3}$) is not zero due to experimental imperfections: 
(i) two subsystems are not perfectly the complex conjugate of each other, 
(ii) the two coupling between the subsystems are not exactly the same in absolute (absorption and impedance mismatches) as well as the phase difference (deviations from $\pi$). 
Nevertheless, the dark vertical regions close to $\Delta\phi = \pi$ and $3\pi$ do indicate the chiral symmetry as there is a reduced amount of transmission $S_{1\bar{1}}$. 

The lower graph of Fig.~\ref{fig:chSE-test-trans} shows the transmission $S_{1\bar{2}}$, i.e., a transmission between the two different points in the subsystems. 
Instead of minima at the values of $\Delta\phi = 1\cdot\pi$ and $3\cdot\pi$, we see non-zero values (or small values) of the transmission allowing to extract the resonances and especially the eigenfrequencies of the system. 
The filled circles correspond to extracted resonances using first the harmonic inversion technique \cite{kuh08a} as initial parameter determination and then a fit of a sum of four Lorentzians plus a linear complex background. 
From the extracted resonances we can see a clear approximate degeneracy close to the $\Delta\phi=\pi$ for the two lower resonances and the upper ones are getting close as well and then only a single resonance can be fitted to the peak (green dots for $\Delta\phi$ slightly larger than $\pi$). 
This also shows that the symplectic symmetry is approximately present where we should see two degenerate eigenfrequencies corresponding to the two Kramer's doublets. 
Again the degeneracy is not perfect due to the same reasons as mentioned above. 
Increasing the phase difference further the lowest resonance separates from the central region and finally a new resonance is approaching from higher frequencies (yellow dots) when coming closer to $2\pi$ (similar to the phase difference at 0). 

To perform the final experiments on the chiGSE-system we constructed two different coaxial cables in total shorter than the setup with the phaseshifters by approximately 45\,cm on each connection to reduce the above mentioned discrepancy of $\Delta \phi$ with frequency. 
Also the setup has a slightly reduced absorption and removed additional reflections stemming from the insertion of the phase shifter in the connecting bond.


\begin{thebibliography}{37}%
	\makeatletter
	\providecommand \@ifxundefined [1]{%
		\@ifx{#1\undefined}
	}%
	\providecommand \@ifnum [1]{%
		\ifnum #1\expandafter \@firstoftwo
		\else \expandafter \@secondoftwo
		\fi
	}%
	\providecommand \@ifx [1]{%
		\ifx #1\expandafter \@firstoftwo
		\else \expandafter \@secondoftwo
		\fi
	}%
	\providecommand \natexlab [1]{#1}%
	\providecommand \enquote  [1]{``#1''}%
	\providecommand \bibnamefont  [1]{#1}%
	\providecommand \bibfnamefont [1]{#1}%
	\providecommand \citenamefont [1]{#1}%
	\providecommand \href@noop [0]{\@secondoftwo}%
	\providecommand \href [0]{\begingroup \@sanitize@url \@href}%
	\providecommand \@href[1]{\@@startlink{#1}\@@href}%
	\providecommand \@@href[1]{\endgroup#1\@@endlink}%
	\providecommand \@sanitize@url [0]{\catcode `\\12\catcode `\$12\catcode
		`\&12\catcode `\#12\catcode `\^12\catcode `\_12\catcode `\%12\relax}%
	\providecommand \@@startlink[1]{}%
	\providecommand \@@endlink[0]{}%
	\providecommand \url  [0]{\begingroup\@sanitize@url \@url }%
	\providecommand \@url [1]{\endgroup\@href {#1}{\urlprefix }}%
	\providecommand \urlprefix  [0]{URL }%
	\providecommand \Eprint [0]{\href }%
	\providecommand \doibase [0]{https://doi.org/}%
	\providecommand \selectlanguage [0]{\@gobble}%
	\providecommand \bibinfo  [0]{\@secondoftwo}%
	\providecommand \bibfield  [0]{\@secondoftwo}%
	\providecommand \translation [1]{[#1]}%
	\providecommand \BibitemOpen [0]{}%
	\providecommand \bibitemStop [0]{}%
	\providecommand \bibitemNoStop [0]{.\EOS\space}%
	\providecommand \EOS [0]{\spacefactor3000\relax}%
	\providecommand \BibitemShut  [1]{\csname bibitem#1\endcsname}%
	\let\auto@bib@innerbib\@empty
	\bibitem [{\citenamefont {Keating}\ and\ \citenamefont {Snaith}(2000)}]{kea00}%
	\BibitemOpen
	\bibfield  {author} {\bibinfo {author} {\bibfnamefont {J.}~\bibnamefont
			{Keating}}\ and\ \bibinfo {author} {\bibfnamefont {N.}~\bibnamefont
			{Snaith}},\ }\bibfield  {title} {\bibinfo {title} {Random matrix theory and
			$\zeta(1/2 + \mathrm{i}t)$},\ }\href {https://doi.org/10.1007/s002200000261}
	{\bibfield  {journal} {\bibinfo  {journal} {Commun. Math. Phys.}\ }\textbf
		{\bibinfo {volume} {214}},\ \bibinfo {pages} {57–89} (\bibinfo {year}
		{2000})}\BibitemShut {NoStop}%
	\bibitem [{\citenamefont {St{\"o}ckmann}(2007)}]{stoe07b}%
	\BibitemOpen
	\bibfield  {author} {\bibinfo {author} {\bibfnamefont {H.-J.}\ \bibnamefont
			{St{\"o}ckmann}},\ }\href {https://doi.org/10.2277/0521027152} {\emph
		{\bibinfo {title} {Quantum Chaos - An Introduction}}}\ (\bibinfo  {publisher}
	{University Press},\ \bibinfo {address} {Cambridge},\ \bibinfo {year}
	{2007})\BibitemShut {NoStop}%
	\bibitem [{\citenamefont {Haake}(2001)}]{haa01b}%
	\BibitemOpen
	\bibfield  {author} {\bibinfo {author} {\bibfnamefont {F.}~\bibnamefont
			{Haake}},\ }\href {https://doi.org/10.1007/978-3-642-05428-0} {\emph
		{\bibinfo {title} {Quantum Signatures of Chaos. 2nd edition}}}\ (\bibinfo
	{publisher} {Springer},\ \bibinfo {address} {Berlin},\ \bibinfo {year}
	{2001})\BibitemShut {NoStop}%
	\bibitem [{\citenamefont {Feist}\ \emph {et~al.}(2009)\citenamefont {Feist},
		\citenamefont {B{\"a}cker}, \citenamefont {Ketzmerick}, \citenamefont
		{Burgd{\"o}rfer},\ and\ \citenamefont {Rotter}}]{fei09}%
	\BibitemOpen
	\bibfield  {author} {\bibinfo {author} {\bibfnamefont {J.}~\bibnamefont
			{Feist}}, \bibinfo {author} {\bibfnamefont {A.}~\bibnamefont {B{\"a}cker}},
		\bibinfo {author} {\bibfnamefont {R.}~\bibnamefont {Ketzmerick}}, \bibinfo
		{author} {\bibfnamefont {J.}~\bibnamefont {Burgd{\"o}rfer}},\ and\ \bibinfo
		{author} {\bibfnamefont {S.}~\bibnamefont {Rotter}},\ }\bibfield  {title}
	{\bibinfo {title} {Nanowires with surface disorder: Giant localization length
			and dynamical tunneling in the presence of directed chaos},\ }\href
	{https://doi.org/10.1103/PhysRevB.80.245322} {\bibfield  {journal} {\bibinfo
			{journal} {Phys. Rev. B}\ }\textbf {\bibinfo {volume} {80}},\ \bibinfo
		{pages} {245322} (\bibinfo {year} {2009})}\BibitemShut {NoStop}%
	\bibitem [{\citenamefont {Tulino}\ and\ \citenamefont
		{Verd{\'u}}(2004)}]{tul04}%
	\BibitemOpen
	\bibfield  {author} {\bibinfo {author} {\bibfnamefont {A.~M.}\ \bibnamefont
			{Tulino}}\ and\ \bibinfo {author} {\bibfnamefont {S.}~\bibnamefont
			{Verd{\'u}}},\ }\bibfield  {title} {\bibinfo {title} {Random matrix theory
			and wireless communications},\ }\href {https://doi.org/10.1561/0100000001}
	{\bibfield  {journal} {\bibinfo  {journal} {Foundations and Trends in
				Communications and Information Theory}\ }\textbf {\bibinfo {volume} {1}},\
		\bibinfo {pages} {1} (\bibinfo {year} {2004})}\BibitemShut {NoStop}%
	\bibitem [{\citenamefont {Couillet}\ and\ \citenamefont
		{Debbah}(2011)}]{cou11}%
	\BibitemOpen
	\bibfield  {author} {\bibinfo {author} {\bibfnamefont {R.}~\bibnamefont
			{Couillet}}\ and\ \bibinfo {author} {\bibfnamefont {M.}~\bibnamefont
			{Debbah}},\ }\href@noop {} {\emph {\bibinfo {title} {Random Matrix Methods
				for Wireless Communications}}}\ (\bibinfo  {publisher} {Cambridge University
		Press},\ \bibinfo {address} {Cambridge},\ \bibinfo {year} {2011})\BibitemShut
	{NoStop}%
	\bibitem [{\citenamefont {Sch{\"a}fer}\ \emph {et~al.}(2010)\citenamefont
		{Sch{\"a}fer}, \citenamefont {Nilsson},\ and\ \citenamefont {Guhr}}]{sch10b}%
	\BibitemOpen
	\bibfield  {author} {\bibinfo {author} {\bibfnamefont {R.}~\bibnamefont
			{Sch{\"a}fer}}, \bibinfo {author} {\bibfnamefont {N.~F.}\ \bibnamefont
			{Nilsson}},\ and\ \bibinfo {author} {\bibfnamefont {T.}~\bibnamefont
			{Guhr}},\ }\bibfield  {title} {\bibinfo {title} {Power mapping with dynamical
			adjustment for improved portfolio optimization},\ }\href
	{https://doi.org/10.1080/14697680902748498} {\bibfield  {journal} {\bibinfo
			{journal} {Quantitative Finance}\ }\textbf {\bibinfo {volume} {10}},\
		\bibinfo {pages} {107} (\bibinfo {year} {2010})}\BibitemShut {NoStop}%
	\bibitem [{\citenamefont {M{\"u}nnix}\ \emph {et~al.}(2010)\citenamefont
		{M{\"u}nnix}, \citenamefont {Sch{\"a}fer},\ and\ \citenamefont
		{Guhr}}]{mun10}%
	\BibitemOpen
	\bibfield  {author} {\bibinfo {author} {\bibfnamefont {M.~C.}\ \bibnamefont
			{M{\"u}nnix}}, \bibinfo {author} {\bibfnamefont {R.}~\bibnamefont
			{Sch{\"a}fer}},\ and\ \bibinfo {author} {\bibfnamefont {T.}~\bibnamefont
			{Guhr}},\ }\bibfield  {title} {\bibinfo {title} {Compensating asynchrony
			effects in the calculation of financial correlations},\ }\href
	{https://doi.org/10.1016/j.physa.2009.10.033} {\bibfield  {journal} {\bibinfo
			{journal} {Physica A}\ }\textbf {\bibinfo {volume} {389}},\ \bibinfo {pages}
		{767} (\bibinfo {year} {2010})}\BibitemShut {NoStop}%
	\bibitem [{\citenamefont {Krbalek}\ and\ \citenamefont {Seba}(2002)}]{krb02}%
	\BibitemOpen
	\bibfield  {author} {\bibinfo {author} {\bibfnamefont {M.}~\bibnamefont
			{Krbalek}}\ and\ \bibinfo {author} {\bibfnamefont {P.}~\bibnamefont {Seba}},\
	}\bibfield  {title} {\bibinfo {title} {Headway statistics of public transport
			in mexican cities},\ }\href {https://doi.org/10.1088/0305-4470/36/1/102}
	{\bibfield  {journal} {\bibinfo  {journal} {J. Phys. A}\ }\textbf {\bibinfo
			{volume} {36}},\ \bibinfo {pages} {L7} (\bibinfo {year} {2002})}\BibitemShut
	{NoStop}%
	\bibitem [{\citenamefont {Jagannath}\ and\ \citenamefont
		{Trogdon}(2017)}]{jag17}%
	\BibitemOpen
	\bibfield  {author} {\bibinfo {author} {\bibfnamefont {A.}~\bibnamefont
			{Jagannath}}\ and\ \bibinfo {author} {\bibfnamefont {T.}~\bibnamefont
			{Trogdon}},\ }\bibfield  {title} {\bibinfo {title} {Random matrices and the
			{New York City} subway system},\ }\href
	{https://doi.org/10.1103/PhysRevE.96.030101} {\bibfield  {journal} {\bibinfo
			{journal} {Phys. Rev. E}\ }\textbf {\bibinfo {volume} {96}},\ \bibinfo
		{pages} {030101} (\bibinfo {year} {2017})}\BibitemShut {NoStop}%
	\bibitem [{\citenamefont {Dyson}(1962)}]{dys62c}%
	\BibitemOpen
	\bibfield  {author} {\bibinfo {author} {\bibfnamefont {F.~J.}\ \bibnamefont
			{Dyson}},\ }\bibfield  {title} {\bibinfo {title} {The threefold way.
			{A}lgebraic structure of symmetry groups and ensembles in quantum
			mechanics},\ }\href {https://doi.org/10.1063/1.1703863} {\bibfield  {journal}
		{\bibinfo  {journal} {J. Math. Phys.}\ }\textbf {\bibinfo {volume} {3}},\
		\bibinfo {pages} {1199} (\bibinfo {year} {1962})}\BibitemShut {NoStop}%
	\bibitem [{\citenamefont {Seligman}\ and\ \citenamefont
		{Nishioka}(1986)}]{sel86a}%
	\BibitemOpen
	\bibinfo {editor} {\bibfnamefont {T.~H.}\ \bibnamefont {Seligman}}\ and\
	\bibinfo {editor} {\bibfnamefont {H.}~\bibnamefont {Nishioka}},\ eds.,\
	\href@noop {} {\emph {\bibinfo {title} {Quantum Chaos and Statistical Nuclear
				Physics}}},\ Lect. Notes Phys. 263\ (\bibinfo  {publisher} {Springer},\
	\bibinfo {address} {Berlin},\ \bibinfo {year} {1986})\BibitemShut {NoStop}%
	\bibitem [{\citenamefont {Mitchell}\ \emph {et~al.}(2010)\citenamefont
		{Mitchell}, \citenamefont {Richter},\ and\ \citenamefont
		{Weidenm{\"u}ller}}]{mit10}%
	\BibitemOpen
	\bibfield  {author} {\bibinfo {author} {\bibfnamefont {G.~E.}\ \bibnamefont
			{Mitchell}}, \bibinfo {author} {\bibfnamefont {A.}~\bibnamefont {Richter}},\
		and\ \bibinfo {author} {\bibfnamefont {H.~A.}\ \bibnamefont
			{Weidenm{\"u}ller}},\ }\bibfield  {title} {\bibinfo {title} {Random matrices
			and chaos in nuclear physics: Nuclear reactions},\ }\href
	{https://doi.org/10.1103/RevModPhys.82.2845} {\bibfield  {journal} {\bibinfo
			{journal} {Rev. Mod. Phys.}\ }\textbf {\bibinfo {volume} {82}},\ \bibinfo
		{pages} {2845} (\bibinfo {year} {2010})}\BibitemShut {NoStop}%
	\bibitem [{\citenamefont {Casati}\ \emph {et~al.}(1980)\citenamefont {Casati},
		\citenamefont {Valz-Gris},\ and\ \citenamefont {Guarnieri}}]{cas80}%
	\BibitemOpen
	\bibfield  {author} {\bibinfo {author} {\bibfnamefont {G.}~\bibnamefont
			{Casati}}, \bibinfo {author} {\bibfnamefont {F.}~\bibnamefont {Valz-Gris}},\
		and\ \bibinfo {author} {\bibfnamefont {I.}~\bibnamefont {Guarnieri}},\
	}\bibfield  {title} {\bibinfo {title} {On the connection between quantization
			of nonintegrable systems and statistical theory of spectra},\ }\href
	{https://doi.org/10.1007/BF02798790} {\bibfield  {journal} {\bibinfo
			{journal} {Lett. Nuov. Cim.}\ }\textbf {\bibinfo {volume} {28}},\ \bibinfo
		{pages} {279} (\bibinfo {year} {1980})}\BibitemShut {NoStop}%
	\bibitem [{\citenamefont {Bohigas}\ \emph {et~al.}(1984)\citenamefont
		{Bohigas}, \citenamefont {Giannoni},\ and\ \citenamefont {Schmit}}]{boh84c}%
	\BibitemOpen
	\bibfield  {author} {\bibinfo {author} {\bibfnamefont {O.}~\bibnamefont
			{Bohigas}}, \bibinfo {author} {\bibfnamefont {M.~J.}\ \bibnamefont
			{Giannoni}},\ and\ \bibinfo {author} {\bibfnamefont {C.}~\bibnamefont
			{Schmit}},\ }\bibfield  {title} {\bibinfo {title} {Spectral properties of the
			{L}aplacian and random matrix theories},\ }\href
	{https://doi.org/10.1051/jphyslet:0198400450210101500} {\bibfield  {journal}
		{\bibinfo  {journal} {J. Physique Lett.}\ }\textbf {\bibinfo {volume} {45}},\
		\bibinfo {pages} {L} (\bibinfo {year} {1984})}\BibitemShut {NoStop}%
	\bibitem [{\citenamefont {M{\"u}ller}\ \emph {et~al.}(2004)\citenamefont
		{M{\"u}ller}, \citenamefont {Heusler}, \citenamefont {Braun}, \citenamefont
		{Haake},\ and\ \citenamefont {Altland}}]{muel04}%
	\BibitemOpen
	\bibfield  {author} {\bibinfo {author} {\bibfnamefont {S.}~\bibnamefont
			{M{\"u}ller}}, \bibinfo {author} {\bibfnamefont {S.}~\bibnamefont {Heusler}},
		\bibinfo {author} {\bibfnamefont {P.}~\bibnamefont {Braun}}, \bibinfo
		{author} {\bibfnamefont {F.}~\bibnamefont {Haake}},\ and\ \bibinfo {author}
		{\bibfnamefont {A.}~\bibnamefont {Altland}},\ }\bibfield  {title} {\bibinfo
		{title} {Semiclassical foundation of universality in quantum chaos},\ }\href
	{https://doi.org/10.1103/PhysRevLett.93.014103} {\bibfield  {journal}
		{\bibinfo  {journal} {Phys. Rev. Lett.}\ }\textbf {\bibinfo {volume} {93}},\
		\bibinfo {pages} {014103} (\bibinfo {year} {2004})}\BibitemShut {NoStop}%
	\bibitem [{\citenamefont {Zirnbauer}(1996)}]{zir96}%
	\BibitemOpen
	\bibfield  {author} {\bibinfo {author} {\bibfnamefont {M.~R.}\ \bibnamefont
			{Zirnbauer}},\ }\bibfield  {title} {\bibinfo {title} {Riemannian symmetric
			superspaces and their origin in random-matrix theory},\ }\href
	{https://doi.org/10.1063/1.531675} {\bibfield  {journal} {\bibinfo  {journal}
			{J. Math. Phys.}\ }\textbf {\bibinfo {volume} {37}},\ \bibinfo {pages} {4986}
		(\bibinfo {year} {1996})}\BibitemShut {NoStop}%
	\bibitem [{\citenamefont {Bohigas}\ \emph {et~al.}(1991)\citenamefont
		{Bohigas}, \citenamefont {Legrand}, \citenamefont {Schmit},\ and\
		\citenamefont {Sornette}}]{boh91b}%
	\BibitemOpen
	\bibfield  {author} {\bibinfo {author} {\bibfnamefont {O.}~\bibnamefont
			{Bohigas}}, \bibinfo {author} {\bibfnamefont {O.}~\bibnamefont {Legrand}},
		\bibinfo {author} {\bibfnamefont {C.}~\bibnamefont {Schmit}},\ and\ \bibinfo
		{author} {\bibfnamefont {D.}~\bibnamefont {Sornette}},\ }\bibfield  {title}
	{\bibinfo {title} {Comment on spectral statistics in elastodynamics},\ }\href
	{https://doi.org/10.1121/1.400662} {\bibfield  {journal} {\bibinfo  {journal}
			{J. Acoust. Soc. Am.}\ }\textbf {\bibinfo {volume} {89}},\ \bibinfo {pages}
		{1456} (\bibinfo {year} {1991})}\BibitemShut {NoStop}%
	\bibitem [{\citenamefont {Reynders}\ \emph {et~al.}(2014)\citenamefont
		{Reynders}, \citenamefont {Legault},\ and\ \citenamefont {Langley}}]{rey14}%
	\BibitemOpen
	\bibfield  {author} {\bibinfo {author} {\bibfnamefont {E.}~\bibnamefont
			{Reynders}}, \bibinfo {author} {\bibfnamefont {J.}~\bibnamefont {Legault}},\
		and\ \bibinfo {author} {\bibfnamefont {R.~S.}\ \bibnamefont {Langley}},\
	}\bibfield  {title} {\bibinfo {title} {An efficient probabilistic approach to
			vibro-acoustic analysis based on the {G}aussian orthogonal ensemble},\ }\href
	{https://doi.org/10.1121/1.4881930} {\bibfield  {journal} {\bibinfo
			{journal} {J. Acoust. Soc. Am.}\ }\textbf {\bibinfo {volume} {136}},\
		\bibinfo {pages} {201} (\bibinfo {year} {2014})}\BibitemShut {NoStop}%
	\bibitem [{\citenamefont {St{\"o}ckmann}(1999)}]{stoe99}%
	\BibitemOpen
	\bibfield  {author} {\bibinfo {author} {\bibfnamefont {H.-J.}\ \bibnamefont
			{St{\"o}ckmann}},\ }\href {https://doi.org/10.2277/0521592844} {\emph
		{\bibinfo {title} {Quantum Chaos - An Introduction}}}\ (\bibinfo  {publisher}
	{University Press},\ \bibinfo {address} {Cambridge},\ \bibinfo {year}
	{1999})\BibitemShut {NoStop}%
	\bibitem [{\citenamefont {So}\ \emph {et~al.}(1995)\citenamefont {So},
		\citenamefont {Anlage}, \citenamefont {Ott},\ and\ \citenamefont
		{Oerter}}]{so95}%
	\BibitemOpen
	\bibfield  {author} {\bibinfo {author} {\bibfnamefont {P.}~\bibnamefont
			{So}}, \bibinfo {author} {\bibfnamefont {S.~M.}\ \bibnamefont {Anlage}},
		\bibinfo {author} {\bibfnamefont {E.}~\bibnamefont {Ott}},\ and\ \bibinfo
		{author} {\bibfnamefont {R.~N.}\ \bibnamefont {Oerter}},\ }\bibfield  {title}
	{\bibinfo {title} {Wave chaos experiments with and without time reversal
			symmetry: {GUE} and {GOE} statistics},\ }\href
	{https://doi.org/10.1103/PhysRevLett.74.2662} {\bibfield  {journal} {\bibinfo
			{journal} {Phys. Rev. Lett.}\ }\textbf {\bibinfo {volume} {74}},\ \bibinfo
		{pages} {2662} (\bibinfo {year} {1995})}\BibitemShut {NoStop}%
	\bibitem [{\citenamefont {Stoffregen}\ \emph {et~al.}(1995)\citenamefont
		{Stoffregen}, \citenamefont {Stein}, \citenamefont {St{\"o}ckmann},
		\citenamefont {Ku\'s},\ and\ \citenamefont {Haake}}]{sto95b}%
	\BibitemOpen
	\bibfield  {author} {\bibinfo {author} {\bibfnamefont {U.}~\bibnamefont
			{Stoffregen}}, \bibinfo {author} {\bibfnamefont {J.}~\bibnamefont {Stein}},
		\bibinfo {author} {\bibfnamefont {H.-J.}\ \bibnamefont {St{\"o}ckmann}},
		\bibinfo {author} {\bibfnamefont {M.}~\bibnamefont {Ku\'s}},\ and\ \bibinfo
		{author} {\bibfnamefont {F.}~\bibnamefont {Haake}},\ }\bibfield  {title}
	{\bibinfo {title} {Microwave billiards with broken time reversal symmetry},\
	}\href {https://doi.org/10.1103/PhysRevLett.74.2666} {\bibfield  {journal}
		{\bibinfo  {journal} {Phys. Rev. Lett.}\ }\textbf {\bibinfo {volume} {74}},\
		\bibinfo {pages} {2666} (\bibinfo {year} {1995})}\BibitemShut {NoStop}%
	\bibitem [{\citenamefont {{\L}awniczak}\ \emph {et~al.}(2010)\citenamefont
		{{\L}awniczak}, \citenamefont {Bauch}, \citenamefont {Hul},\ and\
		\citenamefont {Sirko}}]{law10}%
	\BibitemOpen
	\bibfield  {author} {\bibinfo {author} {\bibfnamefont {M.}~\bibnamefont
			{{\L}awniczak}}, \bibinfo {author} {\bibfnamefont {S.}~\bibnamefont {Bauch}},
		\bibinfo {author} {\bibfnamefont {O.}~\bibnamefont {Hul}},\ and\ \bibinfo
		{author} {\bibfnamefont {L.}~\bibnamefont {Sirko}},\ }\bibfield  {title}
	{\bibinfo {title} {Experimental investigation of the enhancement factor for
			microwave irregular networks with preserved and broken time reversal symmetry
			in the presence of absorption},\ }\href
	{https://doi.org/10.1103/PhysRevE.81.046204} {\bibfield  {journal} {\bibinfo
			{journal} {Phys. Rev. E}\ }\textbf {\bibinfo {volume} {81}},\ \bibinfo
		{pages} {046204} (\bibinfo {year} {2010})}\BibitemShut {NoStop}%
	\bibitem [{\citenamefont {Joyner}\ \emph {et~al.}(2014)\citenamefont {Joyner},
		\citenamefont {M{\"u}ller},\ and\ \citenamefont {Sieber}}]{joy14}%
	\BibitemOpen
	\bibfield  {author} {\bibinfo {author} {\bibfnamefont {C.~H.}\ \bibnamefont
			{Joyner}}, \bibinfo {author} {\bibfnamefont {S.}~\bibnamefont {M{\"u}ller}},\
		and\ \bibinfo {author} {\bibfnamefont {M.}~\bibnamefont {Sieber}},\
	}\bibfield  {title} {\bibinfo {title} {{GSE} statistics without spin},\
	}\href {https://doi.org/10.1209/0295-5075/107/50004} {\bibfield  {journal}
		{\bibinfo  {journal} {Europhys. Lett.}\ }\textbf {\bibinfo {volume} {107}},\
		\bibinfo {pages} {50004} (\bibinfo {year} {2014})}\BibitemShut {NoStop}%
	\bibitem [{\citenamefont {Rehemanjiang}\ \emph {et~al.}(2016)\citenamefont
		{Rehemanjiang}, \citenamefont {Allgaier}, \citenamefont {Joyner},
		\citenamefont {M{\"u}ller}, \citenamefont {Sieber}, \citenamefont {Kuhl},\
		and\ \citenamefont {St{\"o}ckmann}}]{reh16}%
	\BibitemOpen
	\bibfield  {author} {\bibinfo {author} {\bibfnamefont {A.}~\bibnamefont
			{Rehemanjiang}}, \bibinfo {author} {\bibfnamefont {M.}~\bibnamefont
			{Allgaier}}, \bibinfo {author} {\bibfnamefont {C.~H.}\ \bibnamefont
			{Joyner}}, \bibinfo {author} {\bibfnamefont {S.}~\bibnamefont {M{\"u}ller}},
		\bibinfo {author} {\bibfnamefont {M.}~\bibnamefont {Sieber}}, \bibinfo
		{author} {\bibfnamefont {U.}~\bibnamefont {Kuhl}},\ and\ \bibinfo {author}
		{\bibfnamefont {H.-J.}\ \bibnamefont {St{\"o}ckmann}},\ }\bibfield  {title}
	{\bibinfo {title} {Microwave realization of the {G}aussian symplectic
			ensemble},\ }\href {https://doi.org/10.1103/PhysRevLett.117.064101}
	{\bibfield  {journal} {\bibinfo  {journal} {Phys. Rev. Lett.}\ }\textbf
		{\bibinfo {volume} {117}},\ \bibinfo {pages} {064101} (\bibinfo {year}
		{2016})}\BibitemShut {NoStop}%
	\bibitem [{\citenamefont {Rehemanjiang}\ \emph {et~al.}(2018)\citenamefont
		{Rehemanjiang}, \citenamefont {Richter}, \citenamefont {Kuhl},\ and\
		\citenamefont {St{\"o}ckmann}}]{reh18}%
	\BibitemOpen
	\bibfield  {author} {\bibinfo {author} {\bibfnamefont {A.}~\bibnamefont
			{Rehemanjiang}}, \bibinfo {author} {\bibfnamefont {M.}~\bibnamefont
			{Richter}}, \bibinfo {author} {\bibfnamefont {U.}~\bibnamefont {Kuhl}},\ and\
		\bibinfo {author} {\bibfnamefont {H.-J.}\ \bibnamefont {St{\"o}ckmann}},\
	}\bibfield  {title} {\bibinfo {title} {Spectra and spectral correlations of
			microwave graphs with symplectic symmetry},\ }\href
	{https://doi.org/10.1103/PhysRevE.97.022204} {\bibfield  {journal} {\bibinfo
			{journal} {Phys. Rev. E}\ }\textbf {\bibinfo {volume} {97}},\ \bibinfo
		{pages} {022204} (\bibinfo {year} {2018})}\BibitemShut {NoStop}%
	\bibitem [{\citenamefont {Beenakker}(2015)}]{bee15}%
	\BibitemOpen
	\bibfield  {author} {\bibinfo {author} {\bibfnamefont {C.~W.~J.}\
			\bibnamefont {Beenakker}},\ }\bibfield  {title} {\bibinfo {title}
		{Random-matrix theory of {M}ajorana fermions and topological
			superconductors},\ }\href {https://doi.org/10.1103/RevModPhys.87.1037}
	{\bibfield  {journal} {\bibinfo  {journal} {Rev. Mod. Phys.}\ }\textbf
		{\bibinfo {volume} {87}},\ \bibinfo {pages} {1037} (\bibinfo {year}
		{2015})}\BibitemShut {NoStop}%
	\bibitem [{\citenamefont {Rehemanjiang}\ \emph {et~al.}(2020)\citenamefont
		{Rehemanjiang}, \citenamefont {Richter}, \citenamefont {Kuhl},\ and\
		\citenamefont {St{\"o}ckmann}}]{reh20}%
	\BibitemOpen
	\bibfield  {author} {\bibinfo {author} {\bibfnamefont {A.}~\bibnamefont
			{Rehemanjiang}}, \bibinfo {author} {\bibfnamefont {M.}~\bibnamefont
			{Richter}}, \bibinfo {author} {\bibfnamefont {U.}~\bibnamefont {Kuhl}},\ and\
		\bibinfo {author} {\bibfnamefont {H.-J.}\ \bibnamefont {St{\"o}ckmann}},\
	}\bibfield  {title} {\bibinfo {title} {Microwave realization of the chiral
			orthogonal, unitary, and symplectic ensembles},\ }\href
	{https://doi.org/10.1103/PhysRevLett.124.116801} {\bibfield  {journal}
		{\bibinfo  {journal} {Phys. Rev. Lett.}\ }\textbf {\bibinfo {volume} {124}},\
		\bibinfo {pages} {116801} (\bibinfo {year} {2020})}\BibitemShut {NoStop}%
	\bibitem [{\citenamefont {Verbaarschot}\ and\ \citenamefont
		{Zahed}(1993)}]{ver93}%
	\BibitemOpen
	\bibfield  {author} {\bibinfo {author} {\bibfnamefont {J.~J.~M.}\
			\bibnamefont {Verbaarschot}}\ and\ \bibinfo {author} {\bibfnamefont
			{I.}~\bibnamefont {Zahed}},\ }\bibfield  {title} {\bibinfo {title} {Spectral
			density of the {QCD} {D}irac operator near zero virtuality},\ }\href
	{https://doi.org/10.1103/PhysRevLett.70.3852} {\bibfield  {journal} {\bibinfo
			{journal} {Phys. Rev. Lett.}\ }\textbf {\bibinfo {volume} {70}},\ \bibinfo
		{pages} {3852} (\bibinfo {year} {1993})}\BibitemShut {NoStop}%
	\bibitem [{\citenamefont {Ivanov}(2002)}]{iva02}%
	\BibitemOpen
	\bibfield  {author} {\bibinfo {author} {\bibfnamefont {D.~A.}\ \bibnamefont
			{Ivanov}},\ }\bibfield  {title} {\bibinfo {title} {The supersymmetric
			technique for random-matrix ensembles with zero eigenvalues},\ }\href
	{https://doi.org/10.1063/1.1423765} {\bibfield  {journal} {\bibinfo
			{journal} {J. Math. Phys.}\ }\textbf {\bibinfo {volume} {43}},\ \bibinfo
		{pages} {126} (\bibinfo {year} {2002})}\BibitemShut {NoStop}%
	\bibitem [{\citenamefont {Kuhl}\ \emph {et~al.}(2010)\citenamefont {Kuhl},
		\citenamefont {Barkhofen}, \citenamefont {Tudorovskiy}, \citenamefont
		{St{\"o}ckmann}, \citenamefont {Hossain}, \citenamefont {de~Forges~de
			Parny},\ and\ \citenamefont {Mortessagne}}]{kuh10a}%
	\BibitemOpen
	\bibfield  {author} {\bibinfo {author} {\bibfnamefont {U.}~\bibnamefont
			{Kuhl}}, \bibinfo {author} {\bibfnamefont {S.}~\bibnamefont {Barkhofen}},
		\bibinfo {author} {\bibfnamefont {T.}~\bibnamefont {Tudorovskiy}}, \bibinfo
		{author} {\bibfnamefont {H.-J.}\ \bibnamefont {St{\"o}ckmann}}, \bibinfo
		{author} {\bibfnamefont {T.}~\bibnamefont {Hossain}}, \bibinfo {author}
		{\bibfnamefont {L.}~\bibnamefont {de~Forges~de Parny}},\ and\ \bibinfo
		{author} {\bibfnamefont {F.}~\bibnamefont {Mortessagne}},\ }\bibfield
	{title} {\bibinfo {title} {Dirac point and edge states in a microwave
			realization of tight-binding graphene-like structures},\ }\href
	{https://doi.org/10.1103/PhysRevB.82.094308} {\bibfield  {journal} {\bibinfo
			{journal} {Phys. Rev. B}\ }\textbf {\bibinfo {volume} {82}},\ \bibinfo
		{pages} {094308} (\bibinfo {year} {2010})}\BibitemShut {NoStop}%
	\bibitem [{\citenamefont {Barkhofen}\ \emph {et~al.}(2013)\citenamefont
		{Barkhofen}, \citenamefont {Bellec}, \citenamefont {Kuhl},\ and\
		\citenamefont {Mortessagne}}]{bar13a}%
	\BibitemOpen
	\bibfield  {author} {\bibinfo {author} {\bibfnamefont {S.}~\bibnamefont
			{Barkhofen}}, \bibinfo {author} {\bibfnamefont {M.}~\bibnamefont {Bellec}},
		\bibinfo {author} {\bibfnamefont {U.}~\bibnamefont {Kuhl}},\ and\ \bibinfo
		{author} {\bibfnamefont {F.}~\bibnamefont {Mortessagne}},\ }\bibfield
	{title} {\bibinfo {title} {Disordered graphene and boron nitride in a
			microwave tight-binding analog},\ }\href
	{https://doi.org/10.1103/PhysRevB.87.035101} {\bibfield  {journal} {\bibinfo
			{journal} {Phys. Rev. B}\ }\textbf {\bibinfo {volume} {87}},\ \bibinfo
		{pages} {035101} (\bibinfo {year} {2013})}\BibitemShut {NoStop}%
	\bibitem [{\citenamefont {Kuhl}\ \emph
		{et~al.}(2008{\natexlab{a}})\citenamefont {Kuhl}, \citenamefont
		{H{\"o}hmann}, \citenamefont {Main},\ and\ \citenamefont
		{St{\"o}ckmann}}]{kuh08b}%
	\BibitemOpen
	\bibfield  {author} {\bibinfo {author} {\bibfnamefont {U.}~\bibnamefont
			{Kuhl}}, \bibinfo {author} {\bibfnamefont {R.}~\bibnamefont {H{\"o}hmann}},
		\bibinfo {author} {\bibfnamefont {J.}~\bibnamefont {Main}},\ and\ \bibinfo
		{author} {\bibfnamefont {H.-J.}\ \bibnamefont {St{\"o}ckmann}},\ }\bibfield
	{title} {\bibinfo {title} {Resonance widths in open microwave cavities
			studied by harmonic inversion},\ }\href
	{https://doi.org/10.1103/PhysRevLett.100.254101} {\bibfield  {journal}
		{\bibinfo  {journal} {Phys. Rev. Lett.}\ }\textbf {\bibinfo {volume} {100}},\
		\bibinfo {pages} {254101} (\bibinfo {year} {2008}{\natexlab{a}})}\BibitemShut
	{NoStop}%
	\bibitem [{\citenamefont {Magnus}\ \emph {et~al.}(1966)\citenamefont {Magnus},
		\citenamefont {Oberhettinger},\ and\ \citenamefont {Soni}}]{mag66}%
	\BibitemOpen
	\bibfield  {author} {\bibinfo {author} {\bibfnamefont {W.}~\bibnamefont
			{Magnus}}, \bibinfo {author} {\bibfnamefont {F.}~\bibnamefont
			{Oberhettinger}},\ and\ \bibinfo {author} {\bibfnamefont {R.~P.}\
			\bibnamefont {Soni}},\ }\href@noop {} {\emph {\bibinfo {title} {Formulas and
				Theorems for the Special Functions of Mathematical Physics}}},\ \bibinfo
	{edition} {3rd}\ ed.\ (\bibinfo  {publisher} {Springer},\ \bibinfo {address}
	{New York},\ \bibinfo {year} {1966})\BibitemShut {NoStop}%
	\bibitem [{\citenamefont {Glasser}(1976)}]{gla76}%
	\BibitemOpen
	\bibfield  {author} {\bibinfo {author} {\bibfnamefont {M.~L.}\ \bibnamefont
			{Glasser}},\ }\bibfield  {title} {\bibinfo {title} {Definite integrals of the
			complete elliptic integral {K}},\ }\href
	{https://doi.org/10.6028/jres.080B.032} {\bibfield  {journal} {\bibinfo
			{journal} {J. Res. Nat. Bur. Standards Sect. B}\ }\textbf {\bibinfo {volume}
			{80B}},\ \bibinfo {pages} {313} (\bibinfo {year} {1976})}\BibitemShut
	{NoStop}%
	\bibitem [{\citenamefont {Bellec}\ \emph {et~al.}(2013)\citenamefont {Bellec},
		\citenamefont {Kuhl}, \citenamefont {Montambaux},\ and\ \citenamefont
		{Mortessagne}}]{bel13b}%
	\BibitemOpen
	\bibfield  {author} {\bibinfo {author} {\bibfnamefont {M.}~\bibnamefont
			{Bellec}}, \bibinfo {author} {\bibfnamefont {U.}~\bibnamefont {Kuhl}},
		\bibinfo {author} {\bibfnamefont {G.}~\bibnamefont {Montambaux}},\ and\
		\bibinfo {author} {\bibfnamefont {F.}~\bibnamefont {Mortessagne}},\
	}\bibfield  {title} {\bibinfo {title} {Tight-binding couplings in microwave
			artificial graphene},\ }\href {https://doi.org/10.1103/PhysRevB.88.115437}
	{\bibfield  {journal} {\bibinfo  {journal} {Phys. Rev. B}\ }\textbf {\bibinfo
			{volume} {88}},\ \bibinfo {pages} {115437} (\bibinfo {year}
		{2013})}\BibitemShut {NoStop}%
	\bibitem [{\citenamefont {Kuhl}\ \emph
		{et~al.}(2008{\natexlab{b}})\citenamefont {Kuhl}, \citenamefont {Izrailev},\
		and\ \citenamefont {Krokhin}}]{kuh08a}%
	\BibitemOpen
	\bibfield  {author} {\bibinfo {author} {\bibfnamefont {U.}~\bibnamefont
			{Kuhl}}, \bibinfo {author} {\bibfnamefont {F.~M.}\ \bibnamefont {Izrailev}},\
		and\ \bibinfo {author} {\bibfnamefont {A.~A.}\ \bibnamefont {Krokhin}},\
	}\bibfield  {title} {\bibinfo {title} {Enhancement of localization in
			one-dimensional random potentials with long-range correlations},\ }\href
	{https://doi.org/10.1103/PhysRevLett.100.126402} {\bibfield  {journal}
		{\bibinfo  {journal} {Phys. Rev. Lett.}\ }\textbf {\bibinfo {volume} {100}},\
		\bibinfo {pages} {126402} (\bibinfo {year} {2008}{\natexlab{b}})}\BibitemShut
	{NoStop}%
\end{thebibliography}
\end{document}